%% file: FL.tex
\UseRawInputEncoding  

\documentclass[times,twocolumn,final]{elsarticle}

\usepackage{lineno}
\usepackage{framed,multirow}
\usepackage{amssymb}
\usepackage{latexsym}

\usepackage{url}
\usepackage{xcolor}

\usepackage{graphicx}
\usepackage{amsmath}
\usepackage{amssymb}
\usepackage{verbatim}
\usepackage{booktabs}
\usepackage{bm}
\usepackage{array}
\usepackage{epstopdf}
\usepackage{makecell}
\usepackage{tabularx} 
\usepackage{booktabs}   
\usepackage{mathrsfs} 
\usepackage{framed, multirow} 
\usepackage{amsfonts} 
\usepackage{wrapfig}  
\usepackage{cite}
\usepackage{footnote}

\usepackage{latexsym}
\usepackage[caption = false]{subfig}
\usepackage{graphicx}

\usepackage{tablefootnote}
\usepackage{float}
\usepackage[hidelinks,colorlinks=true,linkcolor=blue,citecolor=blue]{hyperref}

\input{defs}

\usepackage{geometry}
\begin{document}
	
	\begin{frontmatter}
		\title{Data efficient deep learning for medical image analysis: A survey}

\author[mymainaddress]{Suruchi Kumari}\ead{suruchi\_k@cs.iitr.ac.in}
\author[mymainaddress]{Pravendra~Singh\corref{mycorrespondingauthor}}\ead{pravendra.singh@cs.iitr.ac.in}

\cortext[mycorrespondingauthor]{Corresponding author: Pravendra Singh }

\address[mymainaddress]{Department of Computer Science and Engineering, Indian Institute of Technology Roorkee, India}

\begin{abstract}
The rapid evolution of deep learning has significantly advanced the field of medical image analysis. However, despite these achievements, the further enhancement of deep learning models for medical image analysis faces a significant challenge due to the scarcity of large, well-annotated datasets. To address this issue, recent years have witnessed a growing emphasis on the development of data-efficient deep learning methods. This paper conducts a thorough review of data-efficient deep learning methods for medical image analysis. To this end, we categorize these methods based on the level of supervision they rely on, encompassing categories such as \emph{no supervision}, \emph{inexact supervision}, \emph{incomplete supervision}, \emph{inaccurate supervision}, and \emph{only limited supervision}. We further divide these categories into finer subcategories. For example, we categorize \emph{inexact supervision} into \emph{multiple instance learning} and \emph{learning with weak annotations}. Similarly, we categorize \emph{incomplete supervision} into \emph{semi-supervised learning}, \emph{active learning}, and \emph{domain-adaptive learning} and so on.
Furthermore, we systematically summarize commonly used datasets for data efficient deep learning in medical image analysis and investigate future research directions to conclude this survey.

\end{abstract}

\begin{keyword}
 Data efficient deep learning \sep Medical image analysis \sep Inexact supervision \sep Incomplete supervision \sep Inaccurate supervision \sep Only limited supervision \sep No supervision.
\end{keyword}
\end{frontmatter}

\section{Introduction}
\label{Intro}
Deep learning has significantly influenced various medical fields, particularly medical imaging, with its influence expected to further expand \citep{topol2019high}. In the context of medical image analysis (MIA), deep learning methods have demonstrated remarkable performance across various tasks, including disease classification \citep{hashimoto2020multi, shao2021transmil, xue2019robust, azizi2021big}, medical object detection \citep{9760421, campanella2019clinical}, ROI segmentation \citep{zhu2019pick, taleb20203d, hervella2020learning, chen2019synergistic}, and image registration \citep{litjens2017survey, chen2022recent, ronneberger2015u}. Initially, supervised learning was widely adopted in MIA. Despite its success in numerous applications, the broader use of supervised models faces a significant challenge due to the typically small size of most medical datasets. Medical image datasets are often considerably smaller than standard computer vision datasets. The initial amount of available data is limited, and obtaining additional data is hindered by factors such as patient confidentiality and institutional policies. Furthermore, in many instances, only a small fraction of the images are annotated by domain experts.

\begin{figure*}[htb!]
\begin{center}
\includegraphics[scale=0.47]{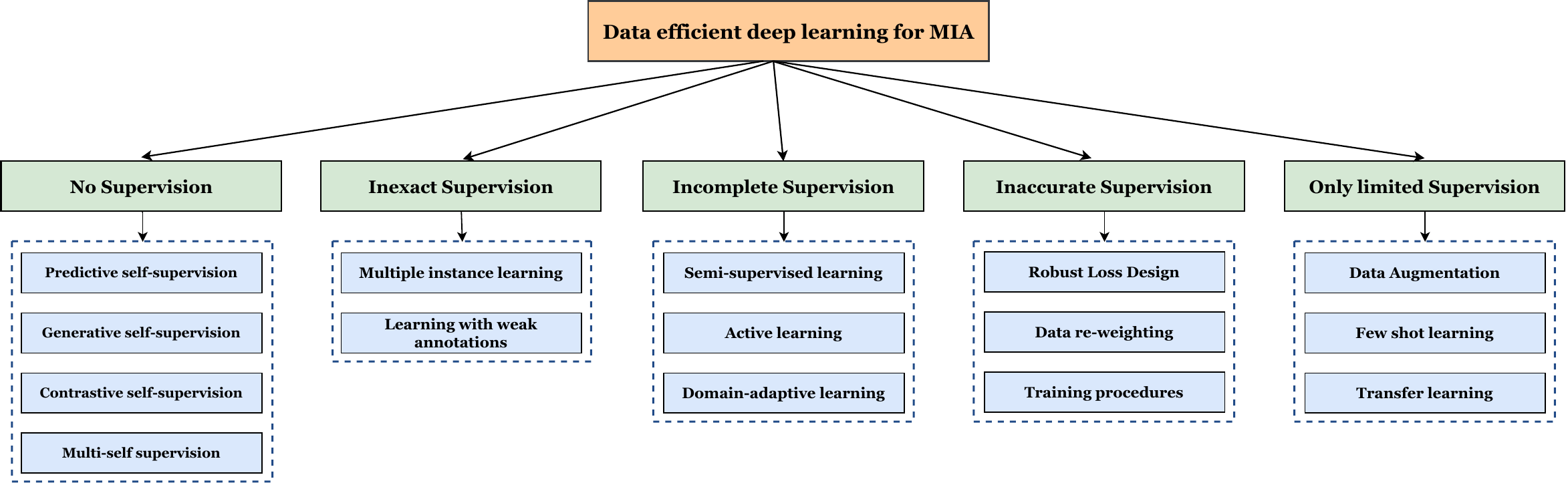}
\end{center}
\caption{Taxonomy of data efficient deep learning approaches for medical image analysis.}
\label{Taxonomy}
\end{figure*}

Typically, researchers rely on domain experts, such as radiologists or pathologists, to create task-specific annotations for image data. Labeling a sufficiently large dataset can be time-consuming \citep{willemink2020preparing}. For example, training deep learning systems for radiology, especially when involving 3D data, requires meticulous slice-by-slice annotations, which can be particularly time-intensive \citep{litjens2017survey}. Some research efforts have involved numerous experts in annotating extensive medical image datasets \citep{gulshan2016development, esteva2017dermatologist}. However, such initiatives demand substantial financial and logistical resources, which are often not readily available across various domains. Other investigations have resorted to crowd-sourcing approaches for obtaining labels from non-experts \citep{7046014, albarqouni2016aggnet, rajpurkar2022ai}. Although this method may have potential in specific cases, its applicability is limited because non-experts typically cannot provide meaningful labels for most medical applications. To overcome these limitations, there is a growing trend among researchers to develop data-efficient deep learning approaches for medical image analysis. We broadly categorize these approaches into the following groups: no supervision, inexact supervision, incomplete supervision, inaccurate supervision, and limited supervision, as shown in Figure~\ref{Taxonomy}.

This survey covers more than 250 papers, with the majority published in recent years (2020-2023). These papers span a diverse range of applications of deep learning in medical image analysis and have been presented in conference proceedings for MICCAI, EMBC, and ISBI, as well as various journals such as TMI, Medical Image Analysis, and Computers in Biology and Medicine, among others.

Several related review articles have already been published summarizing a few specific categories of data efficient learning in the domain of medical image analysis. Cheplygina et al. \citep{cheplygina2019not} provided an overview of semi-supervised learning, multiple instance learning, and transfer learning within the context of medical imaging, addressing both diagnostic and segmentation tasks. Meanwhile, Tajbakhsh et al. \citep{tajbakhsh2020embracing} explored numerous strategies for handling dataset limitations, such as cases involving scarce or weak annotations, with a particular focus on medical image segmentation. Chen et al. \citep{chen2022recent} present a summary of the latest developments in deep learning, encompassing supervised, unsupervised, and semi-supervised methodologies. More recently, Jin et al. \citep{jin2023label} provide an overview of semi-supervised, self-supervised, multi-instance learning, active learning, and annotation-efficient techniques. However, it's worth noting that their review does not delve into subjects such as domain-adaptive learning or few-shot learning, among others. Also, in the previously discussed surveys, their coverage is either restricted concerning data-efficient methods in MIA or not up to date with the current trends. To tackle this challenge, we undertake a systematic review of recent data-efficient methodologies, as outlined in Figure~\ref{Taxonomy}. Our goal is to offer a thorough review of data-efficient learning methods in medical image analysis and outline future challenges. We also provide an overview of several widely used available datasets in the field of medical imaging, as illustrated in Table~\ref{tab:datasets}. The major contributions of our work can be summarized as follows:
\begin{itemize}
    \item This is the first survey paper that summarizes recent advances in data efficient deep learning for medical image analysis. Specifically, we present a comprehensive overview of more than 250 relevant papers to cover the recent progress.
    \item We systematically categorize these methods into five distinct groups: incomplete supervision, no supervision, inaccurate supervision, inexact supervision, and only limited supervision.
    \item Lastly, we explore several potential future directions for further research and development for data-efficient deep learning methods in MIA. 
\end{itemize}

The remainder of this survey is organized as follows.  In Section~\ref{NS}, we delve into techniques falling under the category of No Supervision, which we further subdivide into Predictive Self-Supervision (Subsection~\ref{PSSL}), Generative Self-Supervision (Subsection~\ref{GSSL}), Contrastive Self-Supervision (Subsection~\ref{CSSL}), and Multi-Self Supervised Learning (Subsection~\ref{Multi-SSL}). In Section~\ref{InS}, we explore Inexact Supervision techniques, further classified into Multiple Instance Learning (Subsection~\ref{MIL}) and Learning with Weak Annotations (Subsection~\ref{LWA}). Section~\ref{Incomplete} is dedicated to Incomplete Supervision methods, which we further categorize as Semi-Supervised Learning (Subsection~\ref{semi}), Active Learning (Subsection~\ref{active}), and Domain-Adaptive Learning (Subsection~\ref{UDA}). Similarly, Section~\ref{Inaccurate} deals with Inaccurate Supervision techniques, which we further categorize as Robust Loss Design (Subsection~\ref{RLD}), Data reweighting (Subsection~\ref{DR}), and Training procedures (Subsection~\ref{TP}). Moving on to Section~\ref{Only}, we focus on Only Limited Supervision techniques, which are classified into Data Augmentation (Subsection~\ref{DA}), Few-Shot Learning (Subsection~\ref{FSL}), and Transfer Learning (Subsection~\ref{TL}). Additionally, we outline potential future research directions in Section~\ref{Future} before concluding this survey in Section~\ref{conclusion}. The structural overview of this survey is presented in Figure~\ref{Taxonomy}.

\begin{table*}[]
    \centering
    \scriptsize
    \begin{center} 
    \caption{Commonly used datasets for data efficient deep learning in medical image analysis.}
    \label{tab:datasets}
   \resizebox{\textwidth}{!}{
    \begin{tabular}{p{2.5cm}p{1cm}p{2cm}p{3cm}p{5cm}p{4cm}}
    \toprule
  Dataset & \multicolumn{1}{l}{Organ} & \multicolumn{1}{l}{Types} & \multicolumn{1}{l}{Task}  & \multicolumn{1}{l}{Description} & \multicolumn{1}{l}{Link}\\ \midrule 

JSRT Database (2000) \citep{shiraishi2000development} & Brain & Chest radiographs & Classification & The database includes 154 conventional chest radiographs with a lung nodule (100 malignant and 54 benign nodules) and 93 radiographs without a nodule. &  \url{http://db.jsrt.or.jp/eng.php}  \\ \midrule

ADNI-3 dataset \citep{jack2008alzheimer, weiner2017alzheimer} & Brain & MRI, PET, fMRI, etc.. & Alzheimer’s Disease identification & 697 subjects from ADNI-2 and additional 133 CN, 151 amnestic MCI and 87 AD subjects were added (371 total new subjects) & \url{https://adni.loni.usc.edu/adni-3/} \\ \midrule

BraTS 2012 \citep{menze2014multimodal} & Brain & MR images & Brain tumor segmentation & Training: 30 datasets(pre- and post-therapy images) Synthetic data: 50 simulated datasets; Test: 15 clinical and 15 simulated datasets & \url{http://www.imm.dtu.dk/projects/BRATS2012/data.html} \\ \midrule

BraTS 2013 \citep{menze2014multimodal} & Brain & MR images & Brain tumor segmentation & Training: Clinical dataset from BraTS12 training data; Test: 15 clinical test images from BraTS12 and 10 new test dataset & \url{https://www.smir.ch/BRATS/Start2013\#!\#download}  \\ \midrule

BraTS 2014 \citep{menze2014multimodal} & Brain & MR images & Brain tumor segmentation & Training: 200 datasets from both BraTS12 and BraTS13 and TCIA [16] including longitudinal datasets; Test: 38 unseen datasets from both BraTS12 and BraTS13 test datasets and TCIA  & \url{https://www.smir.ch/BRATS/Start2014}  \\ \midrule

BraTS 2015 \citep{menze2014multimodal} & Brain & MR images & Brain tumor segmentation & Training: Identical to the BraTS14 training dataset; Test: 53 unseen datasets from both BraTS12 and BraTS13 test datasets and TCIA  & \url{https://www.smir.ch/BRATS/Start2015}  \\ \midrule

TCIA (2015) & Brain & MR images & Segmentation & 20 subjects with primary newly diagnosed glioblastoma who were treated with surgery and standard concomitant chemo-radiation therapy (CRT) followed by adjuvant chemotherapy. & \url{https://www.cancerimagingarchive.net/} \\ \midrule

BraTS 2016 \citep{menze2014multimodal} & Brain & MR images & Brain tumor segmentation & Training: Identical to the BraTS14 training dataset; Test: 191 unseen datasets from both BraTS12 and BraTS13 test datasets and TCIA  &  \url{https://www.smir.ch/BRATS/Start2016} \\ \midrule

ABIDE-II (2016) & Brain & fMRI sequences & Autism spectrum disorder classification & 1114 datasets from 521 individuals with ASD and 593 controls & \url{https://fcon 1000.projects.nitrc.org/indi/abide/}  \\ \midrule

BraTS 2017 \citep{menze2014multimodal} & Brain & MR images & Brain tumor segmentation & Training: 285 training datasets from BraTS12 and BraTS13 + pre-operative MRI scans from 19 institution; Validation: 6 unseen datasets from different institution; Test: 146 unseen datasets from both BraTS13 test datasets and different institutions  & \url{https://sites.google.com/site/braintumorsegmentation/}  \\ \midrule

BraTS 2018 \citep{menze2014multimodal} & Brain & MR images & Brain tumor segmentation & Training: Identical to the BraTS17 dataset; Validation: 6 unseen datasets from different institution; Test: 191 unseen datasets from both BraTS13 test datasets and different institutions  & \url{https://wiki.cancerimagingarchive.net/pages/viewpage.action?pageId=37224922}  \\ \midrule

dHCP 2018 \citep{makropoulos2018developing} & Brain & MRI  & Cortical and sub-cortical volume segmentation, cortical surface extraction, and inflation & 465 subjects ranging from 28 to 45 weeks post-menstrual age. & \url{http://www.developingconnectome.org/data-release/}  \\ \midrule

Calgary-Campinas-359 (CC-359) \citep{souza2018open} & Brain & MR images & Skull stripping or Brain segmentation  & 359 subjects on scanners from three different vendors (GE, Philips, and Siemens) and at two magnetic field strengths (1.5 T and 3 T) & \url{https://www.ccdataset.com/download} \\ \midrule

MICCAI WMH Challenge \citep{kuijf2019standardized}  & Brain & MR images & White matter hyperintensities (WMH) segmentation & Training: 60 images; Test: 110 images  & \url{https://wmh.isi.uu.nl/\#_Toc122355662}\\ \midrule

REST-meta-MDD Consortium \citep{yan2019reduced} & Brain & Resting-state functional MRI (R-fMRI) & Major Depressive Disorder (MDD) classification & Neuroimaging data of 1,300 depressed patients and 1,128 normal controls from 25 research groups & \url{http://rfmri.org/REST-meta-MDD}
  \\  \midrule

BraTS (2021)  & Brain & MR images & Segmentation; Classification & 2,000 cases (8,000 mpMRI scans) & \url{http://braintumorsegmentation.org/}  \\ \midrule

MM-WHS challenge dataset (2017) \citep{zhuang2016multi, zhuang2019evaluation} & Heart & MR and CT images & Whole heart segmentation  & 20 labeled and 40 unlabeled CT volumes; 20 labeled and 40 unlabeled MR volumes. & \url{https://zmiclab.github.io/zxh/0/mmwhs} \\ \midrule

ACDC (2018) \citep{bernard2018deep} & Heart & Cine MR images & Classification and segmentation & Training: 100 patients; Test: 50 patients & \url{https://www.creatis.insa-lyon.fr/Challenge/acdc/databases.html} \\ \midrule

Atrial LGE-MRI dataset (2018) \citep{xiong2021global} & Heart & Cardiac (LA) segmentation & Late gadolinium-enhanced magnetic resonance images (LGE-MRI) & Training: 100 LGE-MRI; Test: 54 LGE-MRI & \url{http://atriaseg2018.cardiacatlas.org}   \\ \midrule

MSCMRseg (2019) \citep{zhuang2016multivariate} & Heart & MR images & Cardiac(MYO, RV and LV) segmentation & Data was collected from 45 patients, who underwent cardiomyopathy. & \url{https://zmiclab.github.io/zxh/0/mscmrseg19} \\ \midrule

M\&Ms (2020) \citep{campello2021multi} & Heart & MR images & Cardiac segmentation & Training: 175; Validation: 40; Test: 160 MR images & \url{https://www.ub.edu/mnms/} \\ \midrule

STARE & Eye & Fundus images & Blood vessel segmentation &  20 equal-sized (700×605) color fundus images & \url{https://cecas.clemson.edu/~ahoover/stare/} \\ \midrule

DRIVE (2004) & Eye & Images captured withCanon CR5 non-mydriatic 3CCD camera & Vasculature segmentation & Training: 20 images; Test: 20 images & \url{https://drive.grand-challenge.org/} \\ \midrule

DRISHTI-GS (2014) \citep{sivaswamy2014drishti} & Eye & Fundus images & Optic disc (OD) and (OC) cup segmentation
& Training: 50 images; Test: 51 images & \url{https://ieeexplore.ieee.org/document/6867807} \\ 
 \bottomrule

\multicolumn{6}{r}{\footnotesize\textit{continued on the next page}} \\

\end{tabular}
}
\end{center}
\end{table*}

\begin{table*}[]
\ContinuedFloat
    \centering
    \scriptsize
    \begin{center} 
    \caption{Commonly used datasets for data efficient deep learning in medical image analysis (continued).}
    \label{tab:datasets}
   \resizebox{\textwidth}{!}{
    \begin{tabular}{p{2.5cm}p{1cm}p{2cm}p{3cm}p{5cm}p{4cm}}
    \toprule
  Dataset & \multicolumn{1}{l}{Organ} & \multicolumn{1}{l}{Types} & \multicolumn{1}{l}{Task}  & \multicolumn{1}{l}{Description} & \multicolumn{1}{l}{Link}\\ \midrule 

ReTOUCH (2017) \citep{bogunovic2019retouch} & Eye & OCT volumes & Fluid detection and fluid segmentation & Training: 70 OCT volumes; Test: 42 OCT volumes & \url{https://retouch.grand-challenge.org}  \\ \midrule

RetinalOCT (2018) \citep{kermany2018identifying} & Eye & Optical Coherence Tomography (OCT) Images & Classification & 207,130 OCT images & \url{https://www.kaggle.com/datasets/paultimothymooney/kermany2018}    \\ \midrule

LDLOCTCXR (2018) \citep{kermany2018identifying} & Eye & OCT and Chest X-Ray images & Classification & 108,312 images(37,206 with choroidal neovascularization, 11,349 with diabetic macular edema, 8,617 with drusen, and 51,140 normal) from 4,686 patient & \url{https://data.mendeley.com/datasets/rscbjbr9sj/3} \\ \midrule

PALM (2019) \citep{Palm2019} & Eye & Images captured with Zeiss Visucam 500 & Classification of normal and myopia fundus; lesion segmentation in pathologic myopia. & Training: 400 images, Validation: 400 images; Test: 400 images & \url{https://palm.grand-challenge.org}  \\ \midrule

REFUGE challenge dataset \citep{orlando2020refuge} & Eye & Fundus images & Classification of clinical Glaucoma; OD and OC segmentation; Localization of Fovea & 1200 fundus images with ground truth segmentations and clinical glaucoma labels & 
\url{https://refuge.grand-challenge.org/} \\ \midrule

ADAM (2020) \citep{fang2022adam} & Eye & Fundus images captured using a Zeiss Visucam 500 fundus camera & Classification; Optic disc detection and segmentation; Fovea localization and Lesion detection and segmentation & 1200 retinal fundus images & \url{https://amd.grand-challenge.org/}  \\ \midrule

RIGA+ dataset (2022) \citep{hu2022domain} & Eye & Fundus images & Segmentation of Optic Disc (OD) and Cup (OC) & 744 labeled samples and 717 Unlabeled samples & \url{https://zenodo.org/record/6325549} \\ \midrule

ISIC (2016) & Skin & Dermoscopic lesion images & 1.Lesion Segmentation; 2.Dermoscopic Feature Classification and segmentation; 3.Disease Classification & 1.Training:900, Test:379 images; 2.Training:807, Test:335 images; 3.Training:900, Test:379 images & \url{https://challenge.isic-archive.com/data/\#2016}  \\ \midrule

HAM10000 (2018) & Skin & Dermatoscopic images & Lesion classification and segmentation & 10000 training images & \url{https://dataverse.harvard.edu/dataset.xhtml?persistentId=doi:10.7910/DVN/DBW86T} \\ \midrule

MITOS12 \citep{ludovic2013mitosis} & Breast & Histological Images & Breast cancer grading  & 50 high power fields (HPF) coming from 5 different slides scanned at ×40 magnification & \url{http://ludo17.free.fr/mitos_2012/dataset.html}  \\ \midrule

MITOS14  & Breast & Histological Images & Breast cancer grading & Training data set there are 284 frames at X20 magnification and 1,136 frames at X40 magnification. & \url{https://mitos-atypia-14.grand-challenge.org/Dataset/}  \\ \midrule

MIAS (2015) & Breast & Mammograms & Detection; Classification & 322 images (161 pairs) at 50 micron resolution in \emph{Portable Gray Map} format & \url{https://www.kaggle.com/datasets/kmader/mias-mammography} \\ \midrule

TUPAC (2016) \citep{veta2019predicting} & Breast & Whole-slide histopathology images & Automatic prediction of tumor proliferation scores of breast tumors & Training: 500 WSIs; Test: 321 WSIs & \url{https://github.com/CODAIT/deep-histopath}  \\ \midrule

CAMELYON (2016) \citep{litjens20181399} & Breast & Whole-slide images (WSIs) & Detection and classification of breast cancer metastases & Training: 270 WSI; Test: 130 WSI  & \url{https://camelyon16.grand-challenge.org/Data} \\ \midrule

CAMELYON (2017) \citep{litjens20181399} & Breast & Whole-slide images (WSIs) & Detection and classification of breast cancer metastases & Training: 500 WSI; Test: 500 WSI  & \url{https://camelyon17.grand-challenge.org/Data} \\ \midrule

CBIS-DDSM (2017) & Breast & Mammograms & Segmentation & Data set contains 753 calcification cases and 891 mass cases & 
\url{https://www.kaggle.com/datasets/awsaf49/cbis-ddsm-breast-cancer-image-dataset} \\ \midrule

BACH (2018) \citep{aresta2019bach} & Breast & Microscopy and Whole-slide images & Breast cancer classification & Microscopy: 400 images; WSI: 30 images & \url{https://iciar2018-challenge.grand-challenge.org/Dataset/}  \\ \midrule

TNBC (2018)  & Breast & Histopathology images stained with H\&E & Nuclei segmentation & Data Set1: 50 images with a total of 4022 annotated cells; Data Set2: 30 images from 7 different organs with a total of 21 623 annotated nuclei & \url{https://ega-archive.org/datasets/EGAD00001000063}  \\ \midrule

FNAC (2019) \citep{saikia2019comparative} & Breast & Cytology images & Classification & 212 images in two classes: benign (99) and malignant (113) & \url{https://1drv.ms/u/s!Al-T6d-_ENf6axsEbvhbEc2gUFs}  \\ \midrule

NYUBCS (2019) & Breast & Mammograms & Segmentation & 29,426 digital screening mammography
exams (1,001,093 images) from 141,473 patients  & \url{https://cs.nyu.edu/~kgeras/reports/datav1.0.pdf} \\ \midrule

BreastPathQ (2019) \citep{petrick2021spie} & Breast & Whole slide images stained with H\&E & Estimation of tumor cellularity (TC) & Training: 2,579 patches extracted from 69 WSIs; Test: 1,121 patches extracted from 25 WSIs & \url{https://breastpathq.grand-challenge.org/Overview/} \\ \midrule

CERVIX93 (2018) \citep{phoulady2018new} & Cervix & Cytology images & Classification; detection & 93 stacks of images (2705 nuclei) &  \url{https://github.com/parhamap/cytology_dataset}  \\ \midrule

LBC (2020) \citep{hussain2020liquid} & Cervix &  Cytology images & Classification & 963 LBC images in classes of NILM, LSIL, HSIL, and SCC & \url{https://data.mendeley.com/datasets/zddtpgzv63/4}  \\ \midrule

CHAOS (2021) \citep{kavur2021chaos} & Abdomen & CT and MR images & Liver and Abdominal segmentation & CT: 40 images; MRI: 120 DICOM data sets & \url{https://chaos.grand-challenge.org/}  \\ \midrule

KiTS (2023) & Kidney & CT scan & Kidney Tumor Segmentation & Training: 489 cases; Test: 110 cases &  \url{https://kits-challenge.org/kits23/} \\ \midrule

LiTS (2017) & Liver & CT scans & Liver lesions segmentation & Training: 130 CT scans; Test: 70 CT scans & \url{https://competitions.codalab.org/competitions/17094} \\ \midrule

Asciteps (2020) \citep{su2020development} & Stomach & Classification; detection & Cytology images & 487 images for classification: malignant(18,558) and benign(6089); 176 images for detection (6573 bounding boxes) & \url{https://pan.baidu.com/s/1r0cd0PVm5DiUmaNozMSxgg} \\ \bottomrule

\multicolumn{6}{r}{\footnotesize\textit{continued on the next page}}\\

\end{tabular}
}
\end{center}
\end{table*}

\begin{table*}[]
\ContinuedFloat
    \centering
    \scriptsize
    \begin{center} 
    \caption{Commonly used datasets for data efficient deep learning in medical image analysis (continued).}
    \label{tab:datasets}
   \resizebox{\textwidth}{!}{
    \begin{tabular}{p{2.5cm}p{1cm}p{2cm}p{3cm}p{5cm}p{4cm}}
    \toprule
  Dataset & \multicolumn{1}{l}{Organ} & \multicolumn{1}{l}{Types} & \multicolumn{1}{l}{Task}  & \multicolumn{1}{l}{Description} & \multicolumn{1}{l}{Link}\\ \midrule

MoNuSeg (2017) \citep{kumar2017dataset}  & Multi-organ & H\&E stained tissue images & Nuclei segmentation & Training: 30 images and around 22,000 nuclear boundary annotations; Test: 7000 nuclear boundary annotations & \url{https://monuseg.grand-challenge.org/}  \\ \midrule

BTCV (2017) \citep{gibson2018automatic} & Multi-organ & CT images & Multi-organ segmentation & 90 abdominal CT images & \url{n https://zenodo.org/record/1169361\#.Y8Ud-OxBwUE}  \\ \midrule

DeepLesion (2018) \citep{yan2018deeplesion} & Multi-organ & CT slices & For different applications & 32,735 lesions in 32,120 CT slices & \url{https://nihcc.app.box.com/v/DeepLesion}  \\ \midrule

DECATHLON (2019) & Multi-organ & CT and MRI & Segmentation & Brain: 750 MRI; Heart: 30 MRI; Liver: 201 CT images; Hippocampus: 195 MRI; Prostate: 48 MRI; Lung: 96 CT scans; Pancreas: 420 CT scans; HepaticVessel: 443 CT scans; Spleen: 61 CT scans; Colon: 190 CT scans & \url{http://medicaldecathlon.com/}  \\ \midrule

MIDOG \citep{aubreville2023mitosis} & Multi-organ & Whole Slide Images & Segmentation & Canine Lung Cancer: 44 cases; Human Breast Cancer: 150 cases; Canine Lymphoma: 55 cases; Human neuroendocrine tumor: 55 cases; Canine Cutaneous Mast Cell Tumor: 50 cases; Human melanoma: 49 cases & \url{https://imig.science/midog/the-dataset/}  \\ \midrule

CRCHistoPhenotypes (2016) \citep{sirinukunwattana2016locality} & Colon & Histology images & Cancer classification & 100 H\&E stained histology images of colorectal adenocarcinomas & \url{https://warwick.ac.uk/fac/cross fac/tia/data/crchistolabelednucleihe}  \\ \midrule

KATHER (2018) \citep{kather2019predicting} & Colon & Histological images & Cancer classification & 100,000 histological images of human colorectal cancer and healthy tissue & \url{https://zenodo.org/record/1214456\#.Y8fgV-zP1hE} \\ \midrule

PROMISE12 challenge dataset \citep{litjens2014evaluation} & Prostate & MR images & Prostate segmentation & Training: 50; Test: 30; Live challenge: 20 datasets & \url{promise12.grand-challenge.org/}   \\ \midrule

TMA-Zurich (2018) \citep{arvaniti2018automated} & Prostate & Histopathology images & Gleason grading of prostate cancer & Training: 641 patients; Test: 245 patients & 
\url{https://www.nature.com/articles/s41598-018-30535-1?source=app\#data-availability} \\ \midrule

The Cancer Genome Atlas (TCGA) dataset & Prostate & Histopathology WSIs & Cancer tumour classification based on gleason scores  & 20,000 patient samples spanning 33 cancer types & \url{https://portal.gdc.cancer.gov/repository} \\ \midrule

PANDA (2020) \citep{bulten2022artificial} & Prostate & Whole-slide images & Gleason grading of prostate cancer & Development set: 
10,616 biopsies; Tuning set: 393; Internal validation set: 545; External validation: 1071 & \url{ https://www.kaggle.com/c/prostate-cancer-grade-assessment/data}  \\ \midrule

SCGM dataset \citep{prados2017spinal} & Spinal Cord & MRI images  & Spinal cord gray matter segmentation & Training: 40 images; Test: 40 images & \url{http://niftyweb.cs.ucl.ac.uk/program.php?p=CHALLENGE} \\ \midrule 

Montgomery (2014) \citep{jaeger2014two} & Chest & Chest X-rays & Segmentation & 138 images in two classes: normal (80)
and manifestations of TB (58) &
\url{ https://www.kaggle.com/datasets/raddar/tuberculosis-chest-xrays-montgomery} \\ \midrule

Shenzhen (2014) \citep{jaeger2014two} & Chest & Chest X-rays & Segmentation & 662 images in two classes: normal (326)
and manifestations of TB (336) &
\url{https://www.kaggle.com/datasets/raddar/tuberculosis-chest-xrays-shenzhen} \\ \midrule

NIH Chest X-ray (2017) \citep{wang2017chestx} & Chest & Chest X-rays & Classification & 112,120 X-ray images with disease labels from 30,805 unique patients. & \url{https://www.kaggle.com/datasets/nih-chest-xrays/data} \\ \midrule

ChestX-ray8 (2017) \citep{wang2017chestx} & Chest & Chest x-ray images & Classification and Localization of Common Thorax Diseases & 108,948 frontal-view X-ray images of 32,717 unique patients with the text-mined eight disease image labels & \url{https://nihcc.app.box.com/v/ChestXray-NIHCC/}  \\ \midrule

MIMIC-CXR (2019) \citep{johnson2019mimic} & Chest & Chest x-ray images & Detection & Total of 377,110 images with semi-structured free-text radiology report that describes the radiological findings of the images & \url{https://physionet.org/content/mimic-cxr/2.0.0/}  \\ \midrule

ChestX-ray14 (2019) & Chest & Chest x-ray images & Classification and Localization of Common Thorax Diseases & 112,120 frontal chest radiographs from 30,805 distinct patients with 14 binary labels & \url{https://stanfordmlgroup.github.io/competitions/chexpert/}  \\ \midrule

CC-COVID (2020) \citep{zhang2020clinically} & Chest & CT images & Lung-lesion segmentation & 532,506 CT images from NCP, common pneumonia, and normal controls & \url{https://ncov-ai.big.ac.cn/download?lang=en}  \\ \midrule

SegTHOR (2020) \citep{lambert2020segthor} & Chest & CT images & Segmentation of Thoracic Organs & Training: 40 CT scans; Test: 20 CT scans & \url{https://competitions.codalab.org/competitions/21145} \\ \midrule

VinDr-CXR (2021) \citep{nguyen2022vindr} & Chest & Chest x-ray images & Classification; Detection & Training: 15000 scans; Test: 3000 scans & \url{https://vindr.ai/datasets/cxr}  \\ \midrule

ChestXR (2021) & Chest & Chest x-ray images & Classification & 20,000+ images and 3 classes: COVID-19, Pneumonia and Normal cases & \url{https://cxr-covid19.grand-challenge.org/Dataset/} \\ \midrule

MICCAI2018 IVDM3Seg dataset & Intervertebral Disc & MRI images & Intervertebral discs (IVD) localization and segmentation & 24 3D multi-modality MRI data sets each data set contains four aligned high-resolution 3D volumes, so total 96 high-resolution 3D MRI volume data & \url{https://ivdm3seg.weebly.com/data.html}  \\ 
 \bottomrule
\end{tabular}
}
\end{center}
\end{table*}

\section{No supervision} \label{NS}

Learning with no supervision, commonly referred to as unsupervised learning, involves the challenge of obtaining supervision signals in the absence of explicit guidance. One primary technique used for this purpose is self-supervised learning (SSL). In SSL, representations are acquired by training on an auxiliary pretext task and later transferred to a target downstream task of interest. The effectiveness of SSL relies significantly on the design of well-crafted pretext tasks. These pretext tasks introduce implicit inductive biases into the model, making it crucial to select them thoughtfully to ensure their relevance to the specific domain of interest. Self-supervised learning can be divided into four broad categories: predictive, generative, contrastive, and multi self-supervision \citep{shurrab2022self}. A summary of recent methods for learning with no supervision is provided in Table~\ref{table:no sup}.

\subsection{Predictive self-supervision} \label{PSSL}

\begin{figure}[t]
\begin{center}
\includegraphics[scale=0.4]{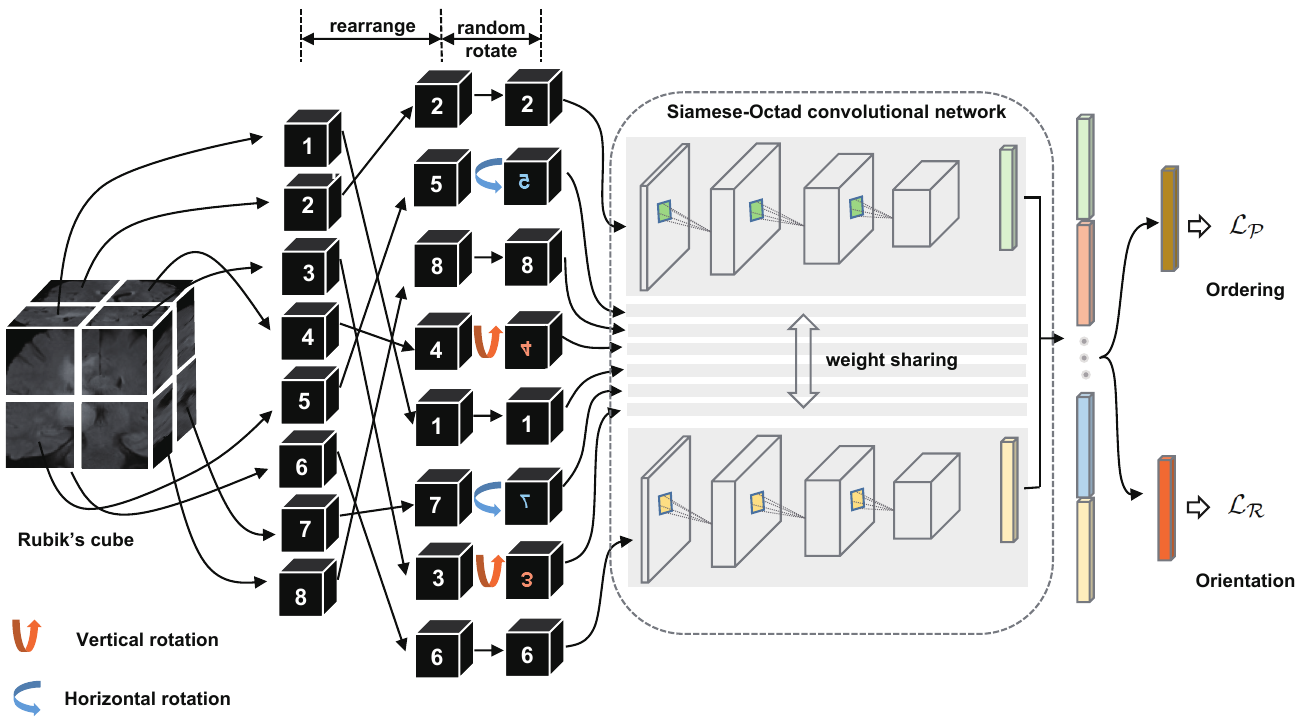}
\end{center}
\caption{Illustration of the Rubik's Cube pretext task: A Siamese network with M (representing the number of cubes) shared-weight branches, referred to as Siamese-Octad, is employed to solve the Rubik's Cube. The backbone network for each branch can be any well-known 3D CNN. The feature maps derived from the final fully-connected or convolutional layer of all branches are concatenated and used as input for separate tasks' fully-connected layers, namely cube ordering and orientation. These tasks are supervised by the permutation loss ($L_P$) and rotation loss ($L_R$), respectively (image from \citep{zhuang2019self}).}
\label{Rubik}
\end{figure}

In this section, we explore predictive self-supervision, where the pretext task is cast as either a classification or regression task. Specifically, each unlabeled image is assigned a pseudo label, which is generated directly from the data itself. These pseudo labels can take on categorical or numerical values, depending on the design specifications of the pretext task. Common transformation-based predictive tasks involve aspects such as assessing relative position \citep{doersch2015unsupervised}, solving jigsaw puzzles \citep{noroozi2016unsupervised}, and determining rotation angles \citep{gidaris2018unsupervised}, among others. These traditional pretext tasks, and their variations, have been explored in MIA and have demonstrated their effectiveness. For instance, Bai et al. \citep{bai2019self} introduced an approach for segmenting cardiac MRI scans by proposing a pretext task focused on predicting anatomical positions. This pretext task aimed to utilize the various cardiac views available in the MRI scans, such as short-axis, 2CH long-axis, and 4CH long-axis, to represent different cardiac anatomical regions, including the left and right atrium and ventricle. To accomplish this, the authors defined a series of bounding boxes corresponding to specific anatomical positions within a given view and trained their network to predict these positions. Taleb et al. \citep{taleb2021multimodal} introduced a novel approach inspired by Jigsaw puzzle-solving, which makes use of multiple imaging modalities. In this method, an input image is composed of disordered patches from different modalities, and the model's task is to reconstruct the original image by correctly assembling these patches. Their work represents a notable enhancement over the traditional Jigsaw puzzle approach.
Zhuang et al. \citep{zhuang2019self} proposed a self-supervised task called Rubik cube recovery, inspired by the early work on Jigsaw puzzle solving for 2D natural images. The task involves two operations: cube rearrangement and cube rotation, as shown in Figure~\ref{Rubik}. The Rubik cube recovery task uses 3D input, where a Rubik cube is divided into a 3D grid of 2×2×2 sub-cubes. The addition of the cube rotation task ensures learning of rotation invariant features, going beyond the original Jigsaw puzzle task, which only focuses on learning translation-invariant features. Rubik cube+ \citep{zhu2020rubik} improves upon the Rubik cube recovery pretext task by using cube masking operation along with both cube rearrangement and cube rotation operations. Nguyen et al. \citep{nguyen2020self} introduced a spatial awareness pretext task with the aim of acquiring semantic and spatial representations from volumetric images. This concept of a spatial pretext task was influenced by Chen et al.'s \citep{chen2019self} context restoration framework; however, it was formulated here into a classification problem. Recently, Zhou et al. \citep{zhou2023unified} performed multi-scale pixel restoration and siamese feature comparison within the feature pyramid. This approach effectively retains semantic, pixel-level, and scale information all at once.

\subsection{Generative self-supervision}  \label{GSSL}
\begin{figure}[t]
\begin{center}
\includegraphics[scale=0.65]{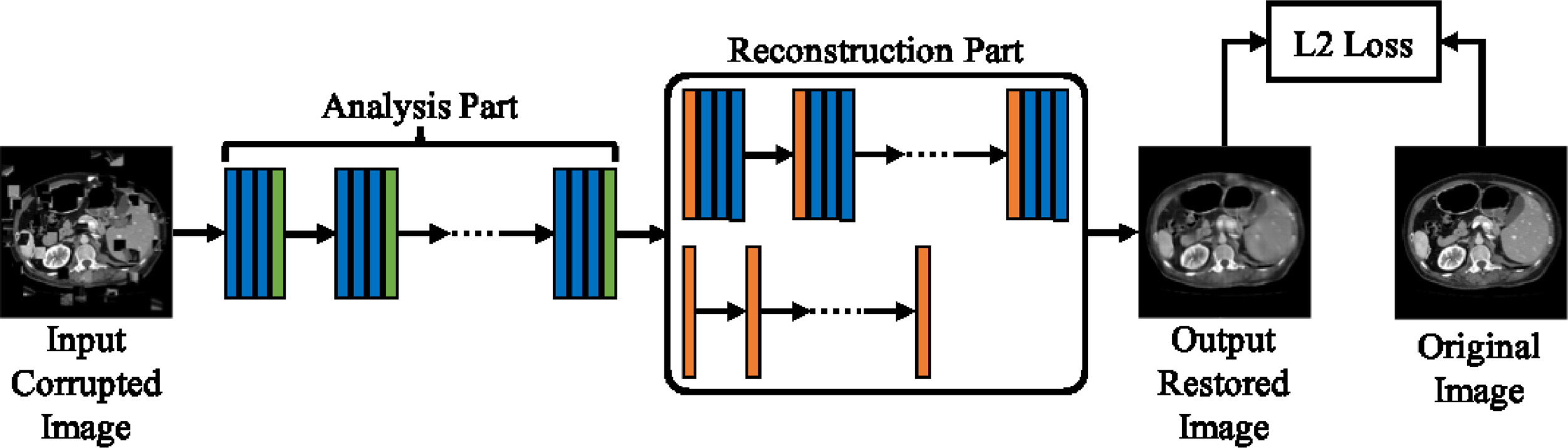}
\end{center}
\caption{CNN architecture for self-supervised context restoration learning, where blue, green, and orange strides indicate convolutional units, downsampling units, and upsampling units, respectively. The specific structure of the CNN in the reconstruction part may vary based on the subsequent task (image from \citep{chen2019self}).}
\label{SSL_GAN}
\end{figure}

The generative self-supervised learning approach seeks to learn underlying features in the input data by framing pretext tasks as generative problems \citep{shurrab2022self}. The idea behind generative pretext tasks is that the model can acquire valuable representations from unlabeled data by either learning to reconstruct the input data itself or by generating new examples that follow the same distribution as the input data. Ross et al. \citep{ross2018exploiting} utilized the image colorization pretext task to address the segmentation of endoscopic medical instruments in endoscopic video data. However, instead of using the original architecture employed in the colorization task, they opted for a conditional Generative Adversarial Network (GAN) architecture. This choice aimed to promote the generation of more realistic colored images. The authors evaluated their approach on six datasets from both medical and natural domains to assess its effectiveness in downstream tasks.
Chen et al. \citep{chen2019self} introduced a new generative pretext task that involves randomly selecting two isolated patches from an input image and swapping their positions. This swapping process is repeated iteratively, resulting in a corrupted version of the original image while preserving its overall distribution. Subsequently, a generative model is used to restore the corrupted image back to its original version (see Figure~\ref{SSL_GAN}). Building upon earlier context-restoration-based studies, Zhou et al. \citep{zhou2019models} incorporated four data transformations (non-linear transformation, local-shuffling, outer-cutout, and inner-cutout) into a cohesive reconstruction model called \emph{Model Genesis}. Harvella et al. \citep{hervella2020learning} introduced a self-supervised multi-modal reconstruction task for retinal anatomy learning. They assumed that distinct modalities of the same organ could offer complementary knowledge, leading to valuable representations for subsequent tasks.

In the medical domain, conventional pretext tasks that heavily rely on the existence of bigger objects in natural images are inadequate because disease-related features are usually found in smaller regions of the medical image. To address this, Holmberg et al. \citep{holmberg2020self} introduced a pretext task, cross-modal self-supervised retinal thickness prediction, for ophthalmic disease diagnosis. This task involves the utilization of two distinct modalities: infrared fundus images and optical coherence tomography scans (OCT). Initially, they extracted retinal thickness maps from OCT scans by training a segmentation model with the limited annotated dataset, which served as ground-truth annotations for the preliminary task. Then, a model was trained to predict the thickness maps utilizing unlabeled fundus images and the previously predicted thickness maps as labels. Other examples of generative self-supervised pretext tasks include the image denoising method proposed by Prakash et al. \citep{prakash2020leveraging} and the Rubik cube++ (introduced by Tao et al. \citep{tao2020revisiting}). In the Rubik cube++ approach, significant modifications were made to the earlier Rubik cube method \citep{zhuang2019self}. Instead of treating it as a classification task, they approached it as a generative problem using a GAN-based framework. The generator's task was to bring back the initial arrangement of the Rubik cube before applying transformations, whereas the discriminator was responsible for distinguishing between correct and incorrect arrangements of the generated cubes.

\begin{figure*}[t]
\begin{center}
\includegraphics[scale=0.3]{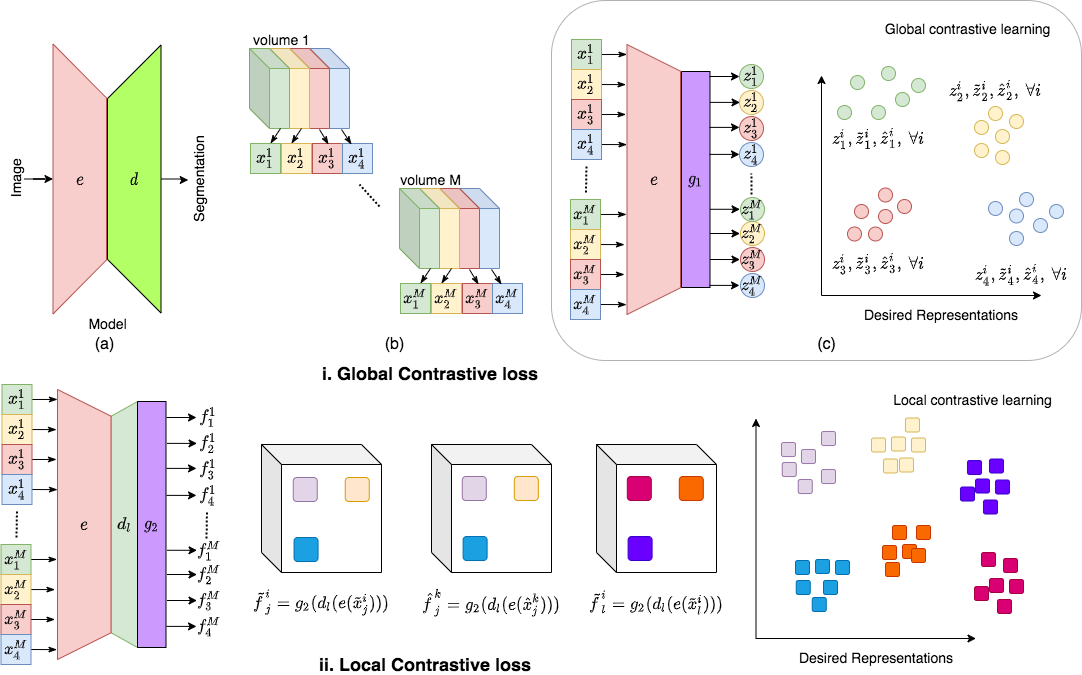}
\end{center}
\caption{Illustration of the enhanced SimCLR for 3D medical image segmentation: (i) Outline of the global contrastive loss employed for pre-training the encoder $e$ using dense layers $g_1$. (ii) Outline of the local contrastive loss utilized for pre-training the decoder $d_l$ with 1 × 1 convolutional layers $g_2$, with frozen weights of encoder $e$ obtained from the previous training stage (image from \citep{chaitanya2020contrastive}).}
\label{fig_conSSL}
\end{figure*}

\subsection{Contrastive self-supervision} \label{CSSL}
Contrastive learning is designed to maximize the mutual information between positive image pairs and, if needed, minimize the representation similarity of negative image pairs. Positive pairs consist of two augmented views of the same instance, whereas negative pairs come from different instances. This allows the network to learn discriminative representations of instances, which are beneficial for pattern recognition tasks. In contrastive learning, the effectiveness of learned representations heavily depends on the choice of positive and negative pairs. However, the conventional pair generation methods used for natural images might not be suitable for medical images with intricate semantic concepts, leading to potentially meaningless representations. To tackle this challenge, researchers have dedicated considerable effort to meticulously devising pair selection strategies within widely used contrastive learning frameworks \citep{zhang2023dive}. These strategies aim to retain the pathological semantics present in medical images, resulting in significant performance enhancements for medical datasets compared to traditional methods.

Contig \citep{taleb2022contig} employs a contrastive loss to align images and various genetic modalities within the feature space. The approach is devised to seamlessly incorporate multiple modalities from each individual into a single end-to-end model, even when the modalities available may differ among individuals. Sowrirajan et al. \citep{sowrirajan2021moco} asserted that the augmentations used in MOCO \citep{he2020momentum} are not suitable for gray-scale medical images. Specifically, blurring and random crop could potentially remove important lesions. To address this issue, they introduced MoCo-CXR, a modified version of MOCO, specifically tailored for chest X-ray images by adapting the augmentations to better suit this medical imaging context. Vu et al. \citep{vu2021medaug} introduced a SSL technique called MedAug, inspired by MoCo-CXR. In their method, positive pairs are generated from diverse images of a single patient based on their metadata. Azizi et al.  \citep{azizi2021big} presented a similar work to MedAug, which was based on the SimCLR framework \citep{chen2020simple}. They introduced a method called \emph{Multi-Instance Contrastive Learning} to create more informative positive pairs from various images of a similar patient. Chaitanya et al. \citep{chaitanya2020contrastive} enhanced SimCLR for 3D medical image segmentation (see Figure~\ref{fig_conSSL}). They introduced a novel contrasting strategy that leveraged the structural similarity of volumetric medical images. Additionally, they introduced a local contrastive loss to facilitate the learning of more detailed and fine-grained representations. Ciga et al. \citep{ciga2022self} introduce a contrastive SSL approach for digital histopathology. They conducted training on 57 unlabeled histopathology datasets. Their findings reveal that enhancing the feature quality is achievable by combining multiple multi-organ datasets with diverse staining and resolution characteristics. Some techniques leverage anatomical priors within contrastive methods to further enhance performance across various tasks \citep{9760421, he2023geometric}. Specifically, He et al. \citep{he2023geometric} introduce Geometric Visual Similarity Learning (GVSL). GVSL incorporates the concept of topological invariance into the metric, ensuring a dependable assessment of inter-image similarity. This approach aims to learn a consistent representation for equivalent semantic regions across different images.

\subsection{Multi-self supervised learning: combining multiple SSL pretext tasks into one framework} 
\label{Multi-SSL}

\begin{figure}[t]
\begin{center}
\includegraphics[scale=0.4]{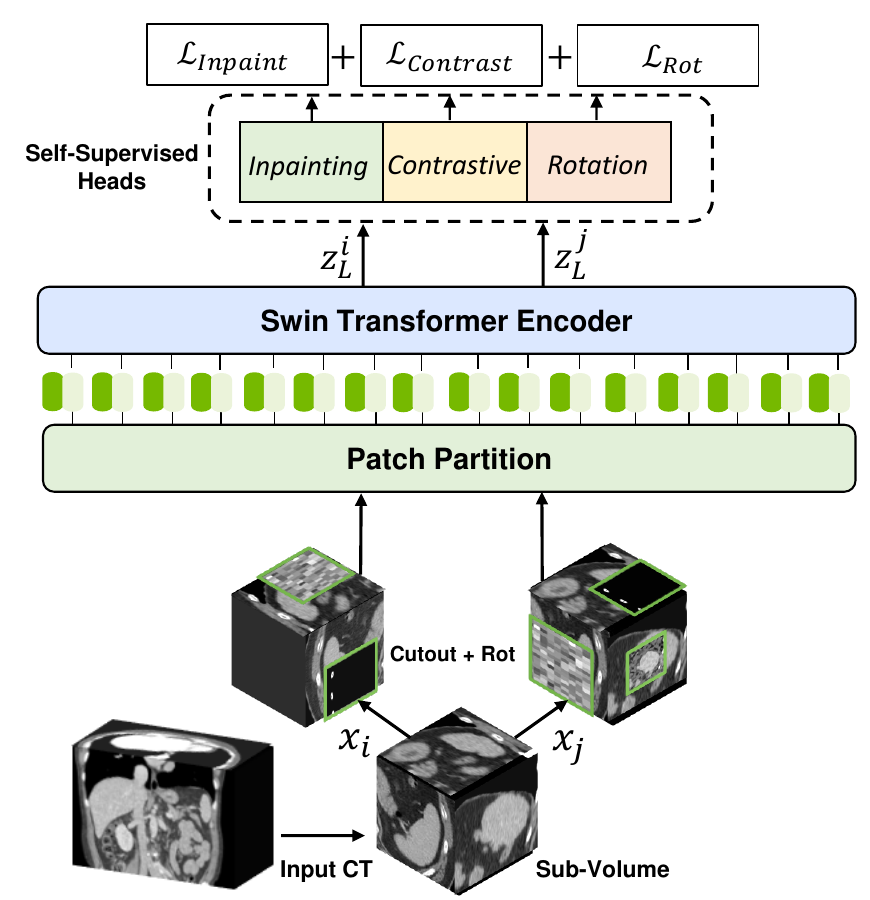}
\end{center}
\caption{The pre-training framework's \citep{tang2022self} outline begins with the random cropping of input CT images into sub-volumes, followed by the application of random inner cutout and rotation augmentations. These processed images are then utilized as input for the Swin UNETR encoder. The framework leverages masked volume inpainting, contrastive learning, and rotation prediction as proxy tasks aimed at acquiring contextual representations from the input images (image from \citep{tang2022self}).}
\label{Swin transformer}
\end{figure}

Multi-SSL integrates various types of pretext tasks, including predictive, generative, and contrastive tasks. By doing so, it aims to overcome the limitation of single pretext tasks, which might learn task-specific features. By employing different self-supervision signals during network training, multi-SSL aims to extract more robust and generalizable representations. Taleb et al. \citep{taleb20203d} proposed that medical images with a 3D nature offer the potential to learn rich representations compared to 2D images. To accommodate this, they employed five predesigned pretext tasks, namely contrastive predictive coding (CPC), exemplar CNN, rotation prediction, relative position prediction, and Jigsaw puzzle, to adapt to the characteristics of 3D medical images. Haghighi et al. \citep{haghighi2020learning} introduced Semantic Genesis, building upon the Model Genesis approach \citep{zhou2019models}. This framework comprises three modules: self-classification, self-restoration, and self-discovery, aimed at learning semantics-enriched representations. In a further extension of Model Genesis, Zhang et al. \citep{zhang2021sar} incorporated a scale-aware proxy task for predicting the input's scale. This addition allows for the learning of multi-level representations. Zhou et al. \citep{zhou2021preservational} combined generative and
contrastive SSL into a Preservational Contrastive Representation Learning (PCRL) framework, where preservational learning is introduced for the generative SSL to keep more information. Tang et al. \citep{tang2022self} introduce a novel 3D transformer-based architecture known as Swin UNEt TRansformers (Swin UNETR), with a hierarchical encoder for self-supervised pre-training. In their proposed pre-training framework, input CT images undergo random cropping into sub-volumes and are augmented with random inner cutout and rotation operations. Subsequently, they are inputted into the Swin UNETR encoder. The authors employ masked volume inpainting, contrastive learning, and rotation prediction as proxy tasks to facilitate the learning of contextual representations from input images, as shown in Figure~\ref{Swin transformer}. CS-CO \citep{yang2022cs}, designed specifically for histopathological images, combines the strengths of generative and discriminative approaches. This method comprises two self-supervised learning phases: cross-stain prediction (CS) and contrastive learning (CO). Yan et al. \citep{yan2023representation} employ Masked Autoencoders (MAE) but demonstrate that directly applying MAE is suboptimal for dense downstream prediction tasks such as multi-organ segmentation. To address this limitation, they propose a self-supervised pre-training approach on large-scale unlabeled medical datasets, leveraging both contrastive and generative modeling techniques.

Yan et al. \citep{yan2023representation} used Masked Autoencoders (MAE) but demonstrated that directly applying MAE is suboptimal for dense downstream prediction tasks, such as multi-organ segmentation. To address this limitation, they proposed a self-supervised pre-training approach on large-scale unlabeled medical datasets, leveraging both contrastive and generative modeling techniques.

\begin{table*}[t]
    \centering
    \scriptsize
    \begin{center}
    \caption{Overview of recent methods in \emph{No Supervision} category.}
    \label{table:no sup}
   \resizebox{\textwidth}{!}{
    \begin{tabular}{p{0.8cm}p{3.5cm}p{4cm}p{3.5cm}p{4.7cm}}
    \toprule
   Reference & \multicolumn{1}{l}{Task} & \multicolumn{1}{l}{Pretext task} & \multicolumn{1}{l}{Dataset} & \multicolumn{1}{l}{Result} \\ \toprule

\citep{bai2019self} & Cardiac segmentation & Anatomical Position Prediction &  Private Dataset: 3825 Subjects & DSC: 0.93 \\ \midrule

\citep{zhuang2019self} & Brain tumor segmentation Brain hemorrhage classification & Rubik’s Cube Recovery & BraTS 2018; Private Dataset: 1,486 Images & BraTS 2018: mIoU: 0.773; Private: Acc: 0.838 \\ \midrule

\citep{zhu2020rubik} & Brain tumor segmentation Brain hemorrhage classification & Rubik cube+ (cube ordering, cube orientation and masking identification) & BraTS-2018; Private Dataset: 1,486 CT volumes & BraTS 2018: Mean Dice: 81.70; Private: Acc: 87.84  \\ \midrule

\citep{chen2019self} & Fetal image classification Abdominal multi-organ localization Brain tumour segmentation & Image Context Restoration & Private Fetus Dataset: 2,694 Images; Private Multi-organ Dataset: 150 Images; BraTS 2017 & Private Fetus Dataset: F1: 0.8942; Private Multi-organ Dataset: Mean Distance: 2.90; BraTS 2017: DSC: 0.8557  \\ \midrule

\citep{hervella2020learning} & Optic disc segmentation & Multi-modal Reconstruction & Isfahan MISP & AUC: 0.818 \\ \midrule


\citep{tao2020revisiting} & Pancreas and Brain Tissue segmentation & Rubik cube ++ & NIH PCT; MRBrainS18 & NIH PCT: DSC: 0.8408; MRBrainS18: DSC: 0.7756 \\ \midrule

\citep{azizi2021big} & Chest X-ray classification Skin lesions classification & Multi-Instance Contrastive Learning (SimCLR) & Priavte Dermatology Dataset; CheXpert & Private: Top-1 Acc: 0.7002; CheXpert: AUC: 0.772 \\ \midrule

\citep{tiu2022expert} & Lung & Contrastive Learning & CheXpert & AUC: 0.889 \\ \midrule

\citep{9760421} & 2D and 3D landmark detection; 3D Lesion matching & Global and Local Contrastive Learning & DeepLesion; NIH LN; Private Dataset: 94 Patients & Mean Radial Error: 4.3; Maximum Radial Error: 16.4 \\ \midrule

\citep{haghighi2020learning} & Lung & Self-Discovery + Self-Classification + Self-Restoration & LUNA; LiTS; CAD-PE; BraTS 2018; ChestX-ray14; LIDC-IDRI; SIIM-ACR & Classification: LUNA: AUC: 0.9847; Segmentation: IoU: LiTS: 0.8560; BraTS 2018: 0.6882 \\ \midrule

\citep{taleb20203d} & Brain tumors segmentation pancreas tumor segmentation  & CPC Jigsaw puzzle Exemplar CNN
Rotation Prediction Relative position prediction & BraTS 2018; DECATHLON; DRD & BraTS 2018: DSC: 0.9080; DECATHLON: DSC $\approx$ 0.635; DRD DRD: DSC $\approx$ 0.80 \\ \midrule

\citep{yan2023representation} & Multi-organ segmentation & Masked Autoencoders + contrastive and generative modeling & Pre-training Dataset: Abdomen-1K; Fine-tuning Dataset: ABD-110; Thorax-85; HaN & ABD-110: Dice score: 84.67; Thorax-85: Dice score: 90.37; HaN: Dice score: 77.31   \\

\bottomrule
\end{tabular}
}
\end{center}
\end{table*}

\section{Inexact supervision} \label{InS}

\begin{figure}[t]
\begin{center}
\includegraphics[scale=0.4]{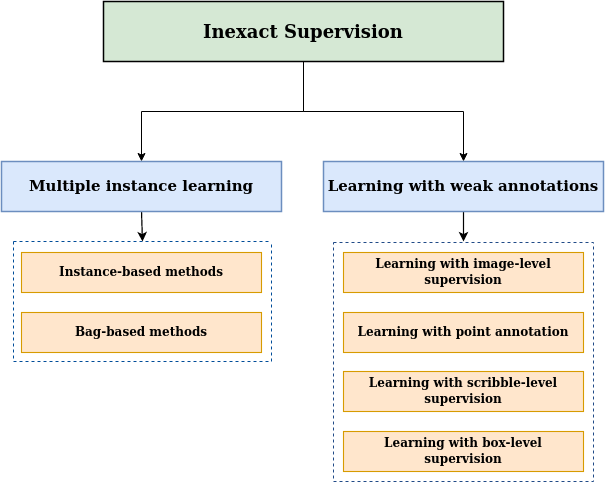}
\end{center}
\caption{Taxonomy of \emph{Inexact Supervision} methods.}
\label{fig_inexact}
\end{figure}

Inexact supervision pertains to situations where some form of supervision information is available but lacks the exactness desired for the task. In this context, we classify inexact supervision into two categories: Multiple Instance Learning (MIL) and learning with weak annotations (Figure \ref{fig_inexact}). In the MIL framework (Subsection~\ref{MIL}), each image is treated as a \emph{bag}, and the patches extracted from it are regarded as \emph{instances}. When a bag is labeled as negative, it implies that all instances within it are also considered negative. Conversely, if a bag is labeled as positive, it indicates the presence of at least one positive instance within it. This labeling strategy at the bag level significantly reduces the labeling burden compared to labeling each individual instance separately, which proves advantageous across various tasks. Learning with weak annotations (Subsection~\ref{LWA}) refers to a scenario in which the available training data is annotated with labels that are less detailed or less precise than what might be ideal for a particular task. In many medical imaging tasks, obtaining precise annotations at a fine-grained level (such as pixel-level annotations) can be highly valuable but also costly and time-consuming. Weak annotations offer an alternative approach where the labels provided for the training data are of a coarser or less specific nature, making them easier and more cost-effective to obtain. These weak annotations can take various forms, including image-level, point-level, scribble-level, or box-level. In all of these scenarios, the provided annotations are less detailed and precise compared to comprehensive pixel-level annotations. A summary of recent methods for learning with inexact supervision is provided in Table~\ref{table:inexact sup}.

\subsection{Multiple instance learning}  \label{MIL}
Multiple-instance learning (MIL) \citep{qu2022towards} arises when obtaining detailed annotations for individual pixels or patches in an image becomes impractical, time-consuming, or infeasible. Instead, global labels representing the overall image condition are more readily available. However, these global labels do not directly correspond to every pixel or patch within the image. MIL extends supervised learning to train classifiers using weakly labeled data. In MIL, every image is viewed as a \emph{bag} containing numerous patches, also referred to as \emph{instances}. If an image, or \emph{bag}, is classified as disease-positive, it implies that at least one patch, or \emph{instance}, within that image is disease-positive. Conversely, if an image is labeled as disease-negative, it signifies that all patches, or \emph{instances}, in that image are negative instances. The current approaches within deep MIL can be classified into two categories: instance-based methods and bag-based methods.

\subsubsection{Instance-based methods} 

The main concept behind the instance-based method is to train an effective instance classifier to predict the possible labels for individual instances (e.g., image patches) within each bag. Subsequently, the MIL-pooling (the aggregation process is commonly referred to as MIL-pooling) method is applied to combine the predictions of all instances within each bag, ultimately generating the bag's prediction. Given that the actual labels of individual instances are unknown, these approaches typically begin by assigning pseudo-labels to each instance based on their respective bags (i.e., all instances within a positive bag are assigned positive labels, and all instances within a negative bag are assigned negative labels). Subsequently, the instance classifier is trained using pseudo-labels in a supervised manner until it converges \citep{qu2022towards}. Various MIL pooling techniques are employed in this process, including Mean-pooling \citep{wang2018revisiting}, Max-pooling \citep{wang2018revisiting}, Average-pooling \citep{schwab2020localization} log-sum-exp-pooling \citep{ramon2000multi}, Noisy-or-pooling \citep{maron1997framework}, Noisy-and-pooling \citep{kraus2016classifying}, and Dynamic pooling \citep{yan2018deep}, among others.
Couture et al. \citep{couture2018multiple} propose an improved MIL aggregation approach that employs a quantile function as the pooling mechanism. This innovative technique allows for a comprehensive representation of the variations within each sample, leading to improved global classification accuracy. In the recent study by Qu et al. \citep{qu2023generalized}, they applied instance-level contrastive learning to aggregate various tumor features for the purpose of diagnosing pancreatic cancer.

\subsubsection{Bag-based methods}

Bag-based methods rely on shared instance-level feature extractors to capture the features of each instance within a bag. These features are then aggregated using MIL-pooling to obtain bag-level features, followed by supervised training of the bag classifier until convergence is achieved. In bag-based methods, MIL-pooling aggregates instance features rather than instance predictions, as is the case in instance-based methods. Bag-based methods excel in bag classification because they have access to true bag labels, making their training process free from noise and more accurate than instance-based methods. However, they are less suitable for localization tasks, and their instance feature aggregation lacks flexibility in showcasing the contributions of individual instances to bag classification. These methods are suitable when the target pattern is expected to be visible at the whole-bag level rather than being localized to specific instances within the bag \citep{cheplygina2019not}. 

\begin{figure*}[t]
\begin{center}
\includegraphics[scale=0.12]{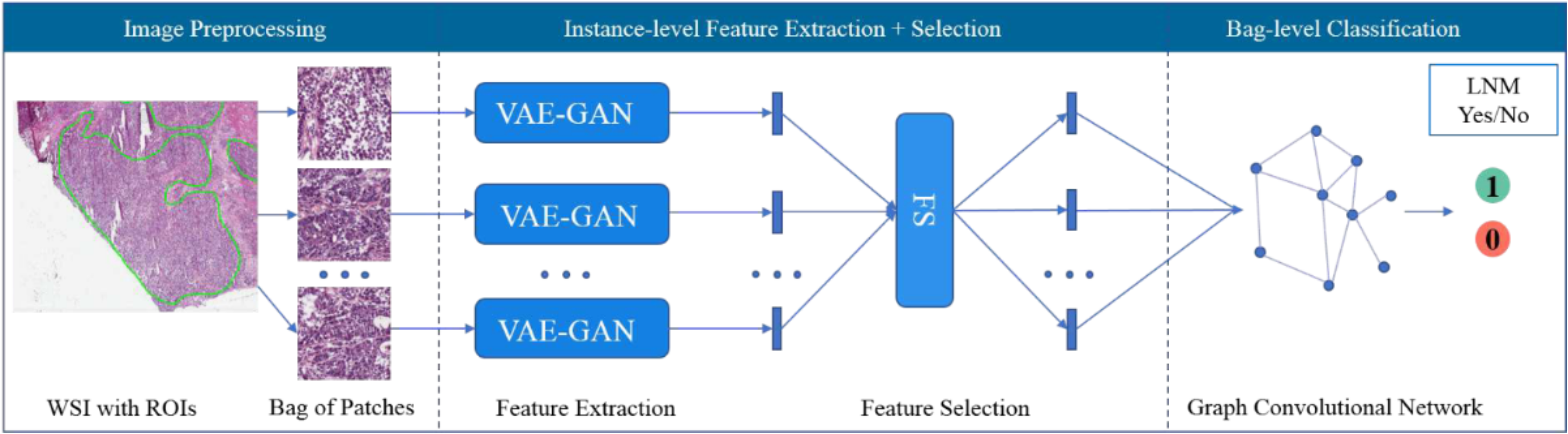}
\end{center}
\caption{Illustration of the framework from Zhao et al.'s work \citep{zhao2020predicting}: VAE-GAN functions as the instance-level feature extractor. The feature selection process identifies and selects discriminative instance-level features. A Graph Convolutional Network (GCN) is employed to synthesize the selected instance-level features, responsible for generating bag representations and performing the final classification (image adapted from \citep{zhao2020predicting}).}
\label{fig_MIL}
\end{figure*}

Bag-based methods primarily vary in three key components: the first being the instance-level feature extraction module, the second involving instance-level feature selection, and lastly, the method by which the instance features are aggregated to produce bag-level features.

Concerning the instance-level feature extractor, the majority of methods utilize CNNs to automatically extract robust features from patches or employ pre-trained models \citep{shao2021transmil}. Recently, there has been an emergence of methods that utilize unsupervised learning to extract features at the patch level. In this context, \citep{zhao2020predicting} train the feature extractor using a combination model that includes both a variational autoencoder and a generative adversarial network (VAE-GAN) as shown in Figure~\ref{fig_MIL}. Various methods employ a self-supervised contrastive learning approach to obtain instance-level feature representations. For instance, \citep{lu2019semi} uses contrastive predictive coding (CPC) from \citep{oord2018representation}, while \citep{li2021mil} utilizes SimCLR from \citep{chen2020simple}. Additionally, Chikontwe et al. \citep{chikontwe2021dual} integrate an unsupervised contrastive loss with their proposed MIL method to enhance the learning of instance-level features.

Regarding the feature selection, the high resolution of medical images poses a challenge when applying deep Multiple Instance Learning (MIL) methods since only a limited number of patches can be selected from these images for MIL. To address this, some approaches use techniques such as random patch selection \citep{raju2020graph}, intelligent sampling using weakly supervised discriminator \citep{su2022attention2majority} and discriminative patch selection \citep{adnan2020representation, zhao2020predicting}. Additionally, patch clustering methods \citep{sharma2021cluster, lu2021data, yan2023histopathological} have been employed. Patch clustering serves the purpose of ensuring the representativeness of the selected patches to a certain degree, as a few patches chosen from a cluster can approximately represent the entire cluster. Ultimately, representative clusters are utilized to make the final prediction. Sharma et al. \citep{sharma2021cluster} employ clustering and sampling on the patch features extracted through the feature extractor. Subsequently, they integrate these features using an adaptive attention mechanism to facilitate end-to-end training. To enhance the feature space learning, Lu et al. \citep{lu2021data} select instances with the highest and lowest attention scores within the current bag for clustering. To advance upon these prior techniques, Yan et al. \citep{yan2023histopathological} introduce a patch clustering approach based on unsupervised and self-supervised learning methods.

For the Bag level representation, pooling methods such as max pooling, average pooling, and log-sum-exp pooling \citep{ramon2000multi} are typically adopted in this step. However, these pooling methods are not trainable, which can restrict their usefulness. To address this limitation, Ilse et al. \citep{ilse2018attention} introduced a fully trainable approach that uses the attention mechanism to assign weights to instances, thus indicating the contribution of individual instances to bag classification. This work has spurred a wave of research into attention-based aggregation methods \citep{lu2019semi, wang2021learning, hashimoto2020multi, li2021novel, li2021mil}. Hashimoto et al. \citep{hashimoto2020multi} utilized the attention mechanism to combine instance features at various resolutions. Li et al. \citep{li2021mil} introduced a dual-stream aggregator that relies on masked non-local operations for conducting instance-level classification as well as bag-level classification. In contrast to the methods mentioned earlier, their model computes attention explicitly using a trainable distance measurement. It's not just important to consider the contribution of various instances to bag classification; the relationships among these instances should also be fully explored. To address this, several methods proposed to use Transformer to aggregate instance features \citep{wang2022lymph, shao2021transmil}. Shao et al. \citep{shao2021transmil} introduced Vision Transformer (ViT) into MIL for gigapixel Whole Slide Images (WSIs) because ViT offers significant benefits in capturing long-distance information and correlations among instances in a sequence. Wang et al. \citep{wang2022lymph} aimed to improve lymph node metastasis prediction by incorporating a pruned Transformer model into MIL. To address the issue of limited samples in the original dataset and prevent overfitting, they also developed a knowledge distillation mechanism using data from similar datasets. 
Different from the approaches mentioned above, \citep{zhao2020predicting} work builds the bag representation with a Graph Convolutional Network.

\subsection{Learning with weak annotations} \label{LWA}

\begin{figure*}[t]
\begin{center}
\includegraphics[scale=0.45]{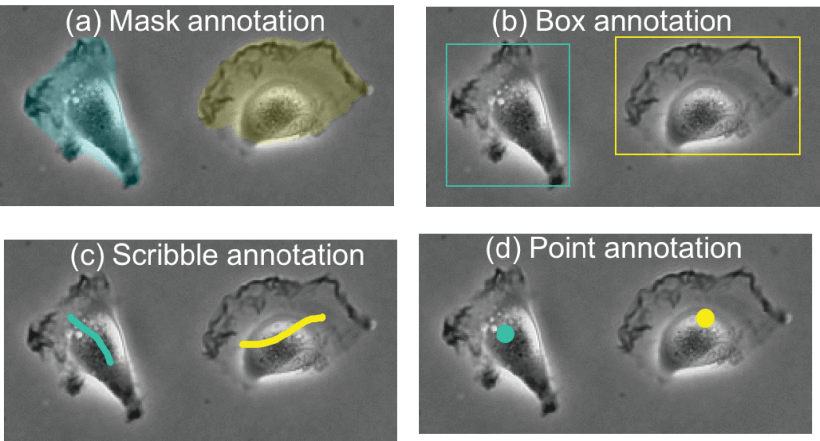}
\end{center}
\caption{Illustration of fully supervised mask annotation, weakly supervised box annotation, scribble annotation and point annotation (image from \citep{zhao2020weakly}).}
\label{fig_LWA}
\end{figure*}

Learning with weak annotations refers to a scenario where the available training data is annotated with labels that are less detailed or less precise than what might be ideal for a particular medical imaging task. In many MIA applications, obtaining precise annotations at a fine-grained level, such as pixel-level annotations, can be challenging, or expensive. Weak annotations provide a cost-effective alternative with coarser labels. These weak annotations can take various forms, including: (\ref{image}) \textit{Image-level annotations}: Only category labels are provided for each training image, lacking precise instance-level information. (\ref{point}) \textit{Point-level annotations}: A single specific location or coordinate within an image is marked to highlight a key feature. (\ref{scribble}) \textit{Scribble-level annotations}: A subset of pixels within each training image is annotated. (\ref{box}) \textit{Box-level annotations}: Object bounding boxes are annotated for each training image, offering coarse localization information but not pixel-level accuracy (see Figure~\ref{fig_LWA}). In each of these cases, the annotations are less detailed or less precise than full pixel-level annotations, which presents challenges but also reduces the labeling effort compared to exhaustive pixel-level annotation requirements.

\subsubsection{Learning with image-level supervision}  \label{image}
In this section, we examine approaches that exclusively rely on image-level supervision for tasks like image detection and segmentation. It's worth noting that image-level supervision is commonly employed to train models for image classification. The challenge here arises from the substantial gap in supervision between the high-level information provided by image-level labels and the detailed pixel-level predictions required for tasks like detection and segmentation \citep{shen2023survey}. In most cases, the Class Activation Maps (CAMs) \citep{zhou2016learning} are commonly used as the standard approach for producing initial regions of interest using classification models. Essentially, CAMs leverage prior of cross-label constraints to identify these initial regions within an image based on the information derived from a classification model. Nonetheless, the accuracy of localizing using CAMs is relatively limited. To tackle this challenge, researchers have devised multiple strategies aimed at enhancing CAMs to enable tasks such as segmentation with only image-level supervision.
For example, Li et al. \citep{li2022deep} introduce an approach named CAM-deep level set (CAM-DLS). In this method, they integrate the DLS loss into the classification loss during the training of the classification network. This DLS loss leverages CAMs to emphasize regions within breast tumors. Similarly, Chen et al. \citep{chen2022c} present a causal CAM approach for organ segmentation. This method employs the concept of causal inference, incorporating a category-causality chain and an anatomy-causality chain.

\subsubsection{Learning with point annotation}  \label{point}
Point annotation involves marking a single specific location or coordinate within an image to indicate a key feature or point of interest. Some works \citep{khan2019extreme, roth2021going, dorent2021inter} concentrate on employing extreme points as annotations for accomplishing pixel-level segmentation. Specifically, Khan et al. \citep{khan2019extreme} investigate a method designed to extract information from extreme points and create a confidence map. This map serves as a guide for neural networks to comprehend the precise object location within the boundaries set by the extreme points. Similarly, Roth et al. \citep{roth2021going} utilize a network that takes two types of input: an image channel and a point channel representing user-defined extreme points. This point channel is subsequently integrated into the network to provide additional guidance during segmentation training. Specifically, it is used as an extra input for attention gates and is incorporated into the loss function, effectively enhancing the segmentation process. Nevertheless, these methods demand annotators to identify the object's boundary, a task that remains labor-intensive in practical applications. In comparison, some methods \citep{qu2020weakly, lin2022label, zhao2020weakly} employ center point annotation to accomplish pixel-level segmentation. To achieve this, certain studies employ the Voronoi diagram \citep{kise1998segmentation} and clustering algorithms to create initial coarse pixel-level labels. Subsequently, various techniques are applied to enhance the segmentation outcomes, including iterative optimization \citep{qu2020weakly} and co-training \citep{lin2022label, lin2023nuclei}. Zhao et al. \citep{zhao2020weakly} employ a framework that combines self-training and co-training to address cell segmentation. They introduce a divergence loss to mitigate overfitting and a consistency loss to ensure agreement among multiple co-trained networks.

\begin{table*}[t]
    \centering
    \scriptsize
    \begin{center}
    \caption{Overview of recent methods in \emph{Inexact supervision} category.}
    \label{table:inexact sup}
   \resizebox{\textwidth}{!}{
    \begin{tabular}{p{0.8cm}p{3.5cm}p{4cm}p{3.5cm}p{5cm}}
    \toprule
   Reference & \multicolumn{1}{l}{Task} & \multicolumn{1}{l}{Algorithm Design} & \multicolumn{1}{l}{Dataset} & \multicolumn{1}{l}{Result} \\ \toprule

\citep{hashimoto2020multi} & Cancer subtype classification & Domain Adversarial + Multi-scale MIL & Private Dataset: 196 Images & Acc: 0.871 \\ \midrule

\citep{raju2020graph} & Colorectal cancer staging, & Graph Attention MIL & MCO & Acc: 0.811; F1: 0.798 \\ \midrule

\citep{shao2021transmil} & Whole slide image classification & Transformer-based MIL & CAMELYON 2016; TCGA-NSCLC; TCGA-RCC & Acc: CAMELYON: 0.8837; TCGA-NSCLC: 0.8835; TCGA-RCC: 0.9466 \\ \midrule

\citep{wang2022lymph} & Lymph node metastasis prediction & Transformer-based MIL + Knowledge Distillation & Private Dataset: 595 Images & AUC: 0.9835; P: 0.9482; R: 0.9151; F1: 0.9297 \\ \midrule

\citep{zhang2022dtfd} & Histopathology whole slide image classification & Double-Tier Feature Distillation MIL & CAMELYON 2016; TCGA-Lung & CAMELYON 2016: AUC: 0.946; TCGA-Lung: AUC: 0.961 \\ \midrule

\citep{schwab2020localization} & Chest X-rays classification & Jointly Classification and Localization & RSNA-Lung; MIMIC-CXR; Private Dataset: 1,003 Images & AUC: 0.93 \\ \midrule

\citep{couture2018multiple} & Breast cancer classification & Quantile Function-based MIL & CBCS3 & Acc: 0.952 \\ \midrule

\citep{ilse2018attention} & Cancer classification  & Attention-based MIL & TMA-UCSB; CRCHistoPhenotypes & TMA-UCSB: Acc: 0.755; CRCHistoPhenotypes: Acc: 0.898 \\ \midrule

\citep{wang2021learning} & Pancreatic ductal adenocarcinoma classification and segmentation & Jointly Global-level Classification and Local-level Segmentation & Private Dataset: 800 Images & DSC: 0.6029; Sens: 0.9975 \\ \midrule

\citep{campanella2019clinical} & Detection of lymph node metastases  & Hybrid MIL & MSK breast cancer & AUC: 0.965 \\ \midrule

\citep{chen2021diagnose} & Breast Cancer (HER2 scoring: negative, equivocal and positive) & Hybrid MIL & Private dataset: 1105 cases  & Accuracy: 0.8970 \\ \midrule


\citep{li2022deep} & Breast tumor segmentation & CAM + Level-Set & Private dataset: 3062 BUS images & DSC: fat 0.830 ± 0.118; mammary gland 0.843 ± 0.100; muscle 0.807 ± 0.154; thorax layers 0.910 ± 0.114 \\ \midrule

\citep{chen2022c} & Segmentation &Causal Inference; CAM & ACDC; ProMRI; CHAOS & ProMRI DSC: 0.864±0.004; ASD: 3.86±1.20; MSD: 3.85±1.33 Abdominal Organ ACDC DSC: 0.875±0.008; ASD: 1.62±0.41; MSD: 1.17±0.24
CHAOS DSC: 0.781 \\ \midrule


\citep{khan2019extreme} & Multi-organ segmentation & Confidence Map Supervision & SegTHOR & DSC Aorta: 0.9441 ± 0.0187; Esophagus 0.8983 ± 0.0416  \\ \midrule

\citep{roth2021going} & Multi-organ segmentation & Random Walker + Iterative Training & BTCV; MSD; CT-ORG & MO-Liver 0.956 ± 0.010; MO-Pancreas 0.747 ± 0.082; DSC: MSD-spleen 0.958 ± 0.007; MO-Spleen 0.954 ± 0.027 \\ \midrule

\citep{dorent2021inter} & Brain tumor segmentation & CNN + CRF & Vestibular-Schwannoma-SEG & DSC: 0.819±0.080; HD95: 3.7±7.4; P: 0.929±0.059 \\ \midrule

\citep{lin2022label} & Multi-organ segmentation & Co-/Self-Training & MoNuSeg; CPM & MoNuSeg DSC: 0.7441; AJI: 0.5620; CPM DSC: 0.7337; AJI: 0.5132 \\ \midrule

\citep{zhao2020weakly} & Cell segmentation & Self-/Co-/Hybrid-Training & PHC; Phase100 & DSC PHC: 0.871; Phase 100: 0.811 \\

\bottomrule
\end{tabular}
}
\end{center}
\end{table*}

\subsubsection{Learning with scribble-level supervision}  \label{scribble}

In this section, we examine techniques related to scribble-based supervision, where annotations are given for a limited number of pixels, often in the form of manually drawn scribbles. These scribbles essentially act as seed regions. The key challenge is to extend semantic information from these sparsely annotated scribbles to all other pixels that lack labels. Some approaches address this challenge by aiming to expand the scribbles or reconstruct the complete mask for model training \citep{bai2018recurrent, ji2019scribble, chen2022scribble2d5}. Nevertheless, the iterative training necessary for the pixel-relabeling process is time-consuming and susceptible to the introduction of noisy labels. To eliminate the necessity for relabeling, several approaches have utilized conditional random fields for refining segmentation results, either in post-processing \citep{can2018learning} or as a trainable layer \citep{tang2018regularized}. Specifically, Can et al. \citep{can2018learning} use region growing to create seed areas. They apply a random walk-based segmentation method that generates per-pixel probability maps for each label, assigning values only when the probability exceeds a specific threshold. However, these methods failed to provide more effective guidance for model training. Conversely, alternative techniques \citep{luo2022scribble, valvano2021learning} introduced new modules to assess the quality of segmentation masks, thereby encouraging the generation of realistic predictions. For instance, Gabriele et al. \citep{valvano2021learning} proposed an adversarial training and an attention gating mechanism to produce segmentation masks, leading to enhanced object localization across multiple resolutions, while Zhang et al. \citep{zhang2020accl} leveraged the PatchGAN discriminator to incorporate shape priors. However, these methods required additional
data source of complete masks. On the other hand, Zhang et al. \citep{zhang2022cyclemix} utilize mix augmentation and cycle consistency within the Scribble-Pixel approach. This demonstrates enhancements in both weakly and fully supervised segmentation methodologies. Several studies utilize consistency learning for scribble-based supervision \citep{gao2022segmentation, zhang2022shapepu, lee2020scribble2label}. Scribble2Label \citep{lee2020scribble2label} combines guidance signals from scribble annotations and pseudo labels using exponential moving averages for cell segmentation. Based on the teacher-student framework, Gao et al. \citep{gao2022segmentation} propose SOUSA, where the student model receives weak supervision through scribbles and a Geodesic distance map created from those scribbles. Simultaneously, a substantial volume of unlabeled data containing different forms of perturbations is provided to both the student and teacher models. The alignment of their output predictions is enforced using a combination of Mean Square Error (MSE) loss and a Multi-angle Projection Reconstruction (MPR) loss.

\subsubsection{Learning with box-level supervision}  \label{box}

In this section, we evaluate approaches for semantic segmentation guided by box-level supervision. Utilizing box-level supervision proves to be a more robust substitute for image-level guidance, as it inherently reduces the exploration area for object detection. For object segmentation, Rajchl et al. \citep{rajchl2016deepcut} recover pixel-wise annotations given a database of images with corresponding bounding boxes. To achieve this goal, they devise an iterative energy minimization problem within a densely connected conditional random field framework to adjust and refine the parameters of a CNN model throughout the iterative process. Wang et al. \citep{wang2021bounding} utilize MIL and a smooth maximum approximation method based on the concept of bounding box tightness. In this context, bounding box tightness implies that an object instance should have contact with all four sides of its bounding box. Consequently, if there is a vertical or horizontal crossing line within the box, it results in a positive bag classification because it covers at least one foreground pixel. In the work presented by \citep{wei2022boxpolyp}, they introduce a fusion filter sampling (FFS) module designed to create pixel-level pseudo labels from box annotations while minimizing noise.


\section{Incomplete supervision} \label{Incomplete}
\begin{figure}[t]
\begin{center}
\includegraphics[scale=0.26]{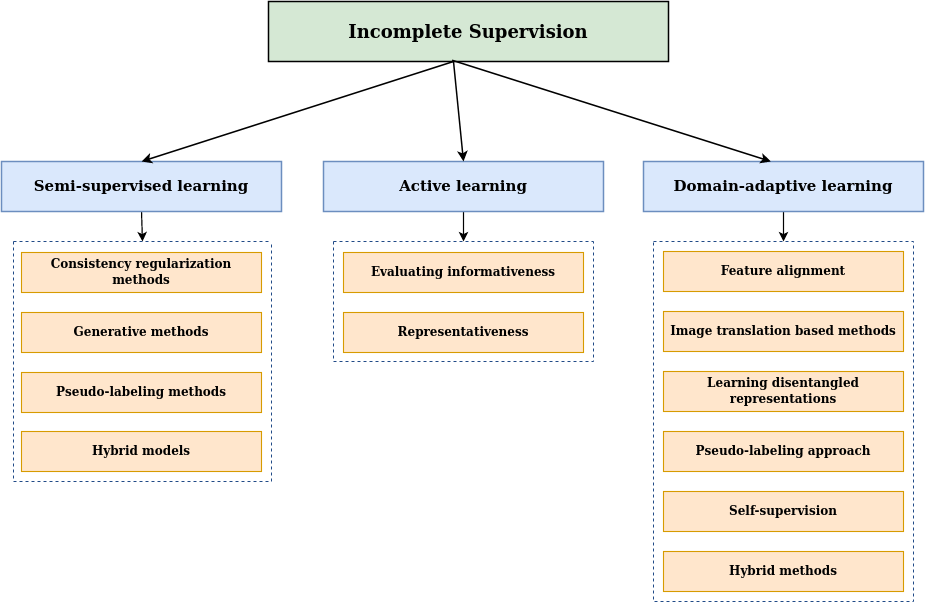}
\end{center}
\caption{Taxonomy of \emph{Incomplete Supervision} methods.}
\label{fig_incomplete}
\end{figure}

Incomplete supervision refers to a scenario where we have access to a limited quantity of labeled data, which is inadequate for training an effective learner, while there exists a large pool of unlabeled data. We categorize incomplete supervision into three broad subcategories: Semi-supervised Learning, Active Learning, and Domain-adaptive Learning (Figure \ref{fig_incomplete}). Semi-supervised learning aims to enhance learning performance by leveraging both labeled and unlabeled data automatically. In Domain-adaptive Learning, a domain shift occurs between labeled and unlabeled data. Conversely, Active learning operates on the assumption that there is an \emph{oracle}, like a human expert, who can be consulted to obtain ground-truth labels for specific unlabeled instances. A summary of recent methods for learning with incomplete supervision is provided in Table~\ref{table:incomplete sup}.

\subsection{Semi-supervised learning} \label{semi}

In this section, we will examine techniques used in semi-supervised learning (Semi-SL). In this approach, only a small portion of the training images have annotations, while the majority of training images remain unannotated. The goal of semi-supervised learning is to incorporate the vast number of unlabeled training images into the training process in order to enhance model performance \citep{jiao2022learning, 9941371}. Semi-supervised Learning can be categorized into Consistency regularization, Generative, Pseudo-labeling, and Hybrid methods.

\subsubsection{Consistency regularization methods}

Consistency regularization methods rely on the concept of smoothness or manifold assumption, suggesting that perturbing data points should not alter the model's predictions. Importantly, this approach does not rely on label information, making it an effective constraint for learning from unlabeled data. Within this framework, various perturbations are available and can be classified into two categories: input perturbations and feature map perturbations. These perturbations must be relevant and meaningful for the specific task at hand. Commonly employed input perturbations encompass random rotation, Gaussian blurring, Gaussian noise, contrast variations, and scaling. Notably, Bortsova et al. \citep{bortsova2019semi} and Li et al. \citep{li2020transformation} employ consistency learning by applying different transformations to input images. Another widely adopted form of consistency is mix-up consistency \citep{berthelot2019mixmatch, basak2022embarrassingly}, where the segmentation of interpolation of two inputs is encouraged to remain consistent with the interpolation of segmentation results for those inputs. Moreover, recent investigations by \citep{zheng2022double} and \citep{li2021dual} delve into perturbations at the feature map level. Zheng et al. \citep{zheng2022double} propose a method that introduces random noise into the parameter calculations of the teacher model. Li et al. \citep{li2021dual} introduce seven distinct feature perturbations, each associated with an additional decoder, all conditioned on maintaining consistency with the primary decoder. Furthermore, there are studies that simultaneously apply perturbations at both the input and feature map levels \citep{xu2021shadow, shu2022cross}.
 
\begin{figure}[t]
\begin{center}
\includegraphics[scale=0.26]{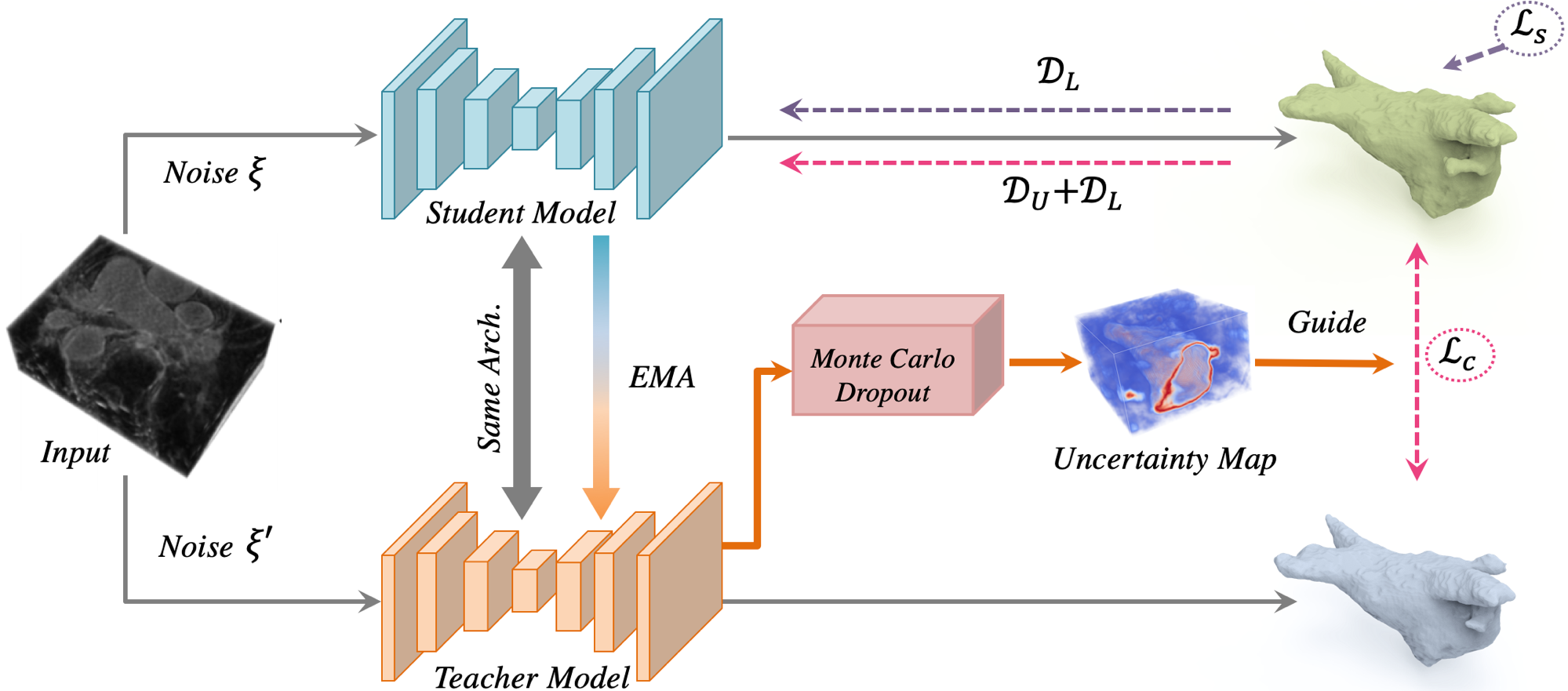}
\end{center}
\caption{Illustration of the Uncertainty-Aware Mean Teacher (UA-MT) framework. The student model is trained by minimizing the supervised loss $L_s$ on labeled data and the consistency loss $L_c$ on both unlabeled and labeled data. The teacher model's estimated uncertainty is used to instruct the student in learning from the more dependable teacher-provided targets (image courtesy of Yu et al. \citep{yu2019uncertainty}).}
\label{UAMT}
\end{figure}

In contrast to incorporating perturbations, alternative consistency learning techniques are also available. For instance, the $\pi$-model \citep{sajjadi2016regularization} is a straightforward yet powerful approach that utilizes a shared encoder to generate various views of the input sample through augmentation. It enforces the classifier to provide consistent predictions for different augmentations of the same input. Simultaneously, the training process incorporates label information to enhance the classifier's overall performance. Li et al. \citep{li2018semi} developed a semi-supervised algorithm for skin lesion segmentation based on the $\pi$-model approach. Temporal ensembling \citep{laine2016temporal} was created with the aim of enhancing the prediction stability of the $\pi$-model. This is achieved by incorporating an exponentially moving average module to update predictions. Several researchers have adopted this module to tackle MIA related challenges \citep{cao2020uncertainty, luo2020deep}. To achieve precise breast mass segmentation, Cao et al. \citep{cao2020uncertainty} incorporate uncertainty into the temporal ensembling model. They utilize uncertainty maps as guidance for the neural network to ensure the reliability of the generated predictions. Likewise, Luo et al. \citep{luo2020deep} suggest an uncertainty-aware temporal ensembling method for chest X-ray disease screening. In the training process of temporal ensembling, the activation of each training sample is updated only once in one epoch. Mean teacher (MT) \citep{tarvainen2017mean} overcomes this limitation by applying exponentially moving average on model parameters instead of network activations. Several methods enhance the MT framework for its application in MIA contexts \citep{yu2019uncertainty, xu2023ambiguity, wang2021tripled, zhu2023hybrid}. To enhance the performance of the MT, Yu et al. \citep{yu2019uncertainty} introduced the Uncertainty-Aware Mean Teacher (UA-MT) framework (see Figure~\ref{UAMT}) for 3D left atrium segmentation. In this approach, the teacher model, in addition to producing target outputs, also assesses the uncertainty associated with each target prediction using Monte Carlo sampling. This allows the removal of unreliable predictions, retaining only those with low uncertainty for consistency loss calculations. This process offers more reliable guidance to the student model, promoting the teacher model to produce higher-quality target predictions. Wang et al. \citep{wang2021tripled} incorporated multi-task learning into the mean teacher framework including segmentation, reconstruction, and SDF prediction tasks to enhance data, model, and task consistency. Additionally, they introduced an uncertainty-weighted integration (UWI) approach to assess uncertainty across all tasks and created a triple-uncertainty method to guide the student model to learn reliable information from the teacher. 

Recently, Xu et al. \citep{xu2023dual} present a dual uncertainty-guided mixing consistency network for precise 3D semi-supervised segmentation, emphasizing the consideration of context information at the volume level. To segment surgical images, Lou et al. \citep{lou2023min} propose a Min-Max Similarity (MMS) method. This approach adopts a dual-view training strategy, utilizing classifiers and projectors to construct pairs of all-negative features and positive/negative feature pairs. This formulation transforms the learning process into solving an MMS problem.

\subsubsection{Generative methods}

\begin{figure*}[t]
\begin{center}
\includegraphics[scale=0.1]{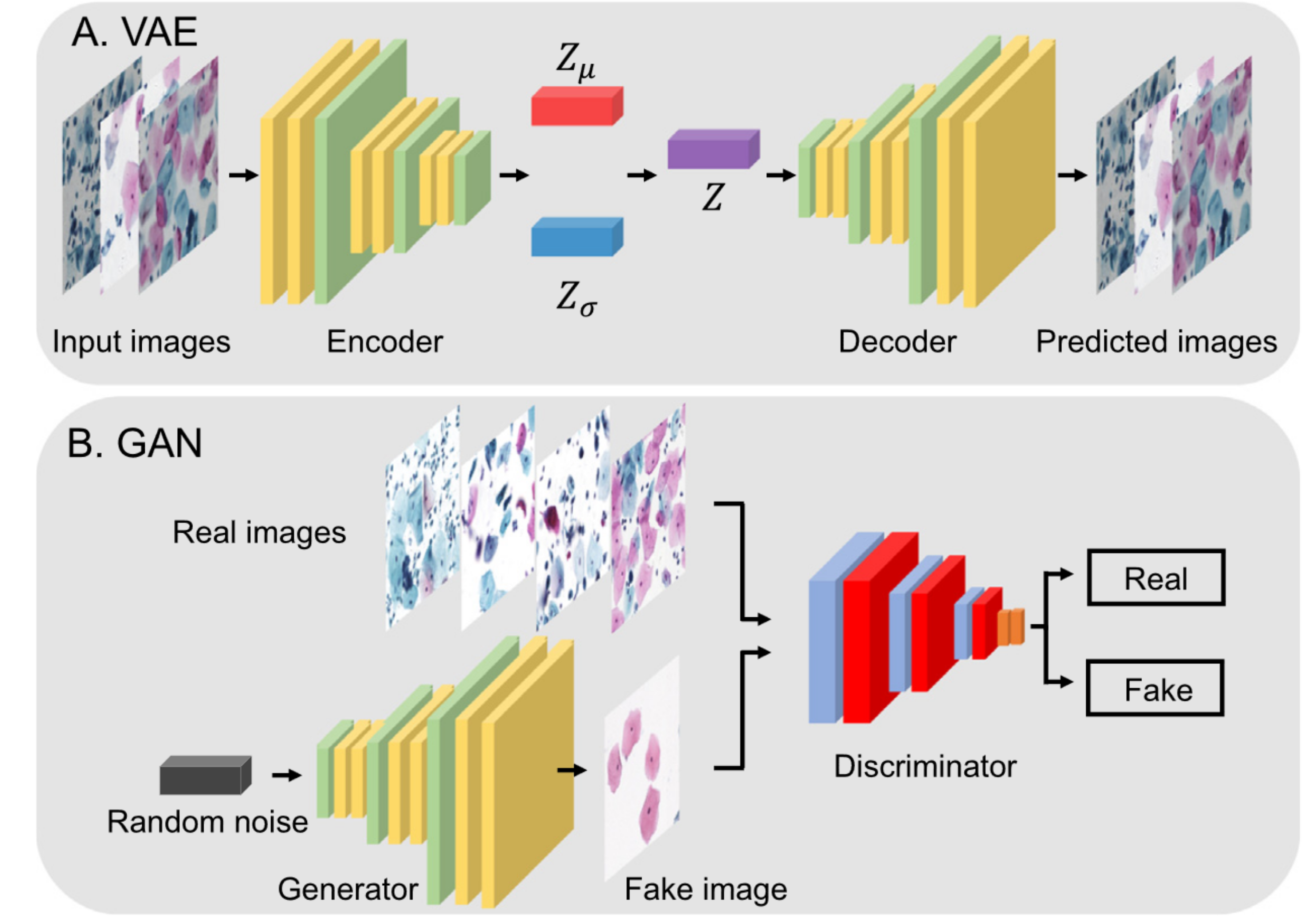}
\end{center}
\caption{Illustration of VAE and GAN architectures: (A) In the VAE architecture, there is an encoder-decoder structure. Here, $Z_\mu$ represents the mean vector, $Z_\sigma$ denotes the standard deviation vector, and $Z$ is the sampled latent vector. (B) In the GAN architecture, there are both a generator and a discriminator (image courtesy of \citep{jiang2022deep}).}
\label{generative}
\end{figure*}

The generative adversarial network (GAN) has shown potential performance on semi-supervised learning \citep{sedai2017semi, wu2021collaborative, wang2023cat}. GANs consist of two main parts: a generator and a discriminator. The generator's goal is to deceive the discriminator by producing fake data that appears real, while the discriminator aims to distinguish between real and synthetic data (see Figure~\ref{generative}(B)). These two networks engage in a zero-sum game, where any gain made by one network comes at the expense of the other. There are different ways to use GANs in Semi-SL settings. One such approach involves employing adversarial techniques to encourage the outputs of unlabeled images to closely resemble those of the labeled images \citep{zhang2017deep, peiris2021duo}. Peiris et al. \citep{peiris2021duo} incorporate a critic network into their segmentation architecture. This network engages in a min-max game by distinguishing between the predicted masks and the actual ground truth masks. The outcomes of their experiments indicate that this approach can enhance the definition of boundaries in the prediction masks. Additionally, the discriminator can be employed to generate pixel-wise confidence maps, facilitating the selection of reliable pixel predictions for consistency learning. The study by Wu et al. \citep{wu2021collaborative} introduces a pair of discriminators to anticipate confidence maps and differentiate between segmentation outcomes originating from labeled or unlabeled data. Constrained Adversarial Training (CAT) \citep{wang2023cat} focuses on generating anatomically accurate segmentations. This method incorporates unlabeled samples into an adversarial training framework, which serves to regularize the network and facilitate constraint learning.

Hou et al. \citep{hou2022semi} use a GAN-based framework with three enhancements: First, a U-Net style network is employed as the discriminator. Second, a \emph{polluted discriminator} is introduced, incorporating auxiliary \emph{leaking links} from the generator to encourage the generation of moderate, though unrealistic, samples, thereby enhancing semi-supervised learning. Third, the discriminator undergoes regularization via the mean-teacher mechanism, enhancing segmentation generalization through input and weight perturbations. Certain approaches employ GANs as a method for data augmentation within the context of Semi-SL. For instance, Chaitanya et al. \citep{chaitanya2021semi} integrate unlabeled data directly into GAN's adversarial training process to enhance the generator's performance for improving medical data augmentation. They assert that incorporating unlabeled samples enables greater diversity in terms of shape and intensity, thereby enhancing the model's robustness and guiding the optimization process.

A Variational Autoencoder (VAE) \citep{kingma2013auto} consists of two main components: an encoder that transforms input data into a latent representation and a decoder that reconstructs the latent representation into the original data space. In order to regularize the encoder of the VAE, a prior over the latent distribution is commonly introduced (see Figure~\ref{generative}(A)). As one of the initial attempts to apply VAE to semi-supervised segmentation tasks, Sedai et al. \citep{sedai2017semi} employed a dual-VAE approach for segmenting the optic cup in retinal fundus images. This method involved two VAEs, where one VAE learned the data distribution from unlabeled data and transferred its acquired knowledge to the other VAE responsible for segmentation using labeled data. Wang et al. \citep{wang2022rethinking} extended the VAE architecture to 3D medical image segmentation by introducing a mean vector and covariance matrix to account for correlations across different slices within an input volume.

\subsubsection{Pseudo-labeling methods}

In pseudo-labeling, a model is trained on the available labeled data. It then predicts labels for unlabeled samples with high confidence, effectively creating pseudo-labels. Finally, the model is retrained using both the labeled data and these newly generated pseudo-labeled samples, improving its performance through the utilization of additional unlabeled data. Pseudo-labeling methods can be mainly categorized into two sub-categories: Self-training methods and Co-training learning methods.

\begin{figure*}[t]
\begin{center}
\includegraphics[scale=0.35]{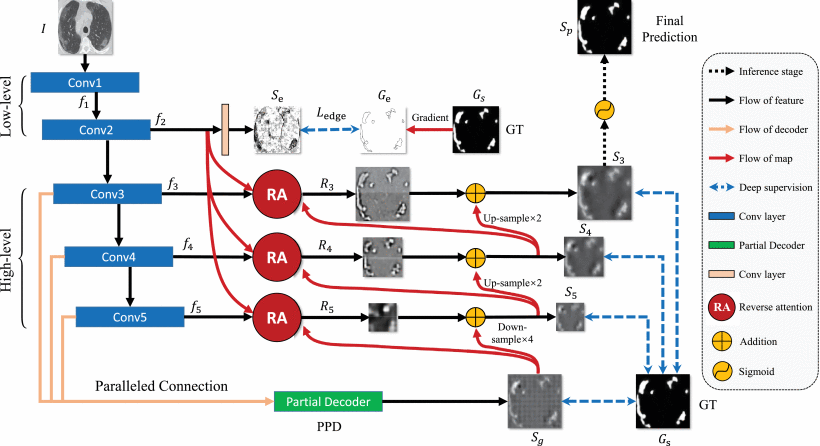}
\end{center}
\caption{Illustration of the Inf-net framework: CT images are initially processed through two convolutional layers for the extraction of high-resolution (i.e., low-level) features. An edge attention module is incorporated to enhance the representation of boundaries within the region of interest. Subsequently, the obtained low-level features, denoted as $f_2$, undergo three convolutional layers to extract high-level features. These high-level features serve two primary purposes. Firstly, they are used to feed a parallel partial decoder (PPD), which aggregates these features and generates a global map denoted as $S_g$. This global map aids in the coarse localization of lung infections. Secondly, these high-level features, along with $f_2$, are directed through multiple cascaded reverse attention (RA) modules under the guidance of $S_g$. The RA module $R_4$ depends on the output of another RA module, $R_5$. Finally, the output of the last RA module, denoted as $S_3$, is passed through a sigmoid activation function for the final prediction of lung infection regions (image from \citep{fan2020inf}).}
\label{infnet}
\end{figure*}

\textbf{Self-training models:}
In the self-training framework, an initial model is trained using limited labeled data. Then, this initial model is utilized to generate pseudo labels for the unlabeled data. Subsequently, the labeled dataset is combined with the pseudo-labeled dataset to update the initial model. The training process iteratively alternates between these two steps until a predetermined number of iterations is reached. Self-training approaches primarily vary in terms of model initialization, pseudo label generation, and their strategies for addressing pseudo label noise. According to the study by \citep{lee2013pseudo}, pseudo labels with higher confidence tend to be more effective. Consequently, various methods that take into account confidence or uncertainty in pseudo labels have been introduced to generate more consistent and reliable pseudo labels, such as refining pseudo labels through conditional random fields \citep{bai2017semi}, uncertainty-aware confidence evaluation \citep{wang2021semi}. Similarly, Ke et al. \citep{ke2022three} proposed a three-stage self-training framework to refine pseudo labels in a stage-wise manner. It reduces the uncertainty in the predicted probability for the pseudo-masks using a multi-task model. Inf-net \citep{fan2020inf} addresses the shortage of well-annotated data for segmentation of COVID-19 lung infections in CT images. Further, a parallel partial decoder (PPD), reverse attention (RA), and edge attention were further added to improve the performance of the model, as shown in Figure~\ref{infnet}. In contrast to conventional pseudo-labeling techniques, which rely on a threshold to pick confidently classified samples, Liu et al. \citep{liu2022acpl} propose the Anti-Curriculum Pseudo-labeling (ACPL) method. ACPL utilizes a mechanism known as \emph{cross-distribution sample informativeness} to identify highly informative unlabeled samples for pseudo-labeling. It also employs an ensemble of classifiers to generate precise pseudo-labels. This approach enables ACPL to effectively handle multi-class and multi-label imbalanced classification issues in the field of MIA.

Recently, Chen et al. \citep{chen2023magicnet} introduced a teacher-student framework for multi-organ segmentation in CT scans. They proposed a learning paradigm involving $N^3$ small cubes extracted from each CT scan, called magic-cubes. Two data augmentation strategies were designed. First, labeled and unlabeled data cubes were mixed to teach unlabeled data organ semantics in their relative positions. Second, for smaller organs, data cubes were shuffled and fed into the student network. Finally, the original magic-cubes were reconstructed to align with the ground-truth or teacher's supervision. Further, the teacher network's predicted pseudo labels are improved by blending them with the learned representations of the small cubes. This blending strategy considers local attributes like texture, luster, and boundary smoothness, addressing the lower performance observed for smaller organs.

\textbf{Co-training models:}
In the Co-training framework \citep{blum1998combining}, a model is trained on a dataset with two or more views or representations of the data. These views are typically different but complementary. The key idea is that if each view provides unique information about the data, the model can learn more effectively from the combined knowledge of all views. In contrast to the self-training framework, which expands the labeled dataset based on a single model's confidence, co-training iteratively selects instances on which the model is confident based on different views, expanding the labeled dataset with complementary information. The essence of co-training lies in the process of creating two or more deep models that can effectively capture distinct and nearly independent perspectives. These approaches typically involve utilizing diverse data sources, implementing various network architectures, and applying specialized training techniques to acquire a range of diverse deep models \citep{jiao2022learning}. In the context of medical images, data can originate from various modalities or medical centers, resulting in distinct distributions. In this regard, \citep{zhu2021semi} and \citep{chen2022mass} make use of different views derived from diverse modalities within the co-training framework. Some approaches employ different network architectures as distinct views. For instance, Luo et al. \citep{luo2022semi} propose cross-teaching between CNN and Transformer models, which implicitly promotes consistency and complementarity between these distinct networks. Peng et al. \citep{peng2020deep} generate adversarial examples as an alternative view. Similarly, for 3D images, Zhao et al. \citep{zhao2022mmgl} utilize coronal, sagittal, and axial views of images as diverse input views. Recently, Wang et al. \citep{wang2023dhc} address the issue of imbalanced class distribution in Semi-SL methods using the Dual-debiased Heterogeneous Co-training (DHC) framework. They introduce two loss weighting techniques called Distribution-aware Debiased Weighting (DistDW) and Difficulty-aware Debiased Weighting (DiffDW). These strategies utilize pseudo labels dynamically to help the model address data and learning biases effectively.

\subsubsection{Hybrid models} 
An emerging area of research in Semi-SL involves integrating the previously mentioned methods into a unified framework to achieve improved performance. These combined approaches are referred to as hybrid methods \citep{wang2021deep, wang2020focalmix, zhang2022boostmis}. Several studies have explored the combination of pseudo-labeling and contrastive learning methods \citep{chaitanya2023local, basak2023pseudo, zhang2023multi, zeng2023ss} for different tasks. Specifically, both Chaitanya et al. \citep{chaitanya2023local} and Basak et al. \citep{basak2023pseudo} introduce a self-training method based on local contrastive learning, guided by pseudo-labels, and demonstrate its effectiveness across various medical segmentation datasets. For COVID-19 Screening and Lesion Segmentation, Zeng et al. \citep{zeng2023ss} present a double-threshold pseudo-labeling approach and a novel inter-slice consistency regularization technique designed specifically for CT images. Wang et al. \citep{wang2021deep} utilize self-training with consistency regularization to efficiently extract valuable information from unlabeled data, and they incorporate virtual adversarial training to enhance the model's generalization capability. ASE-Net \citep{lei2022semi} comprises segmentation networks and a discriminator network. The segmentation network is constructed using the MT framework, while the discriminator network employs an adversarial consistency training strategy (ACTS) with two discriminators focused on consistency learning. This strategy helps establish prior relationships between labeled and unlabeled data.

\subsection{Active learning} \label{active}

\begin{figure}[t]
\begin{center}
\includegraphics[scale=0.4]{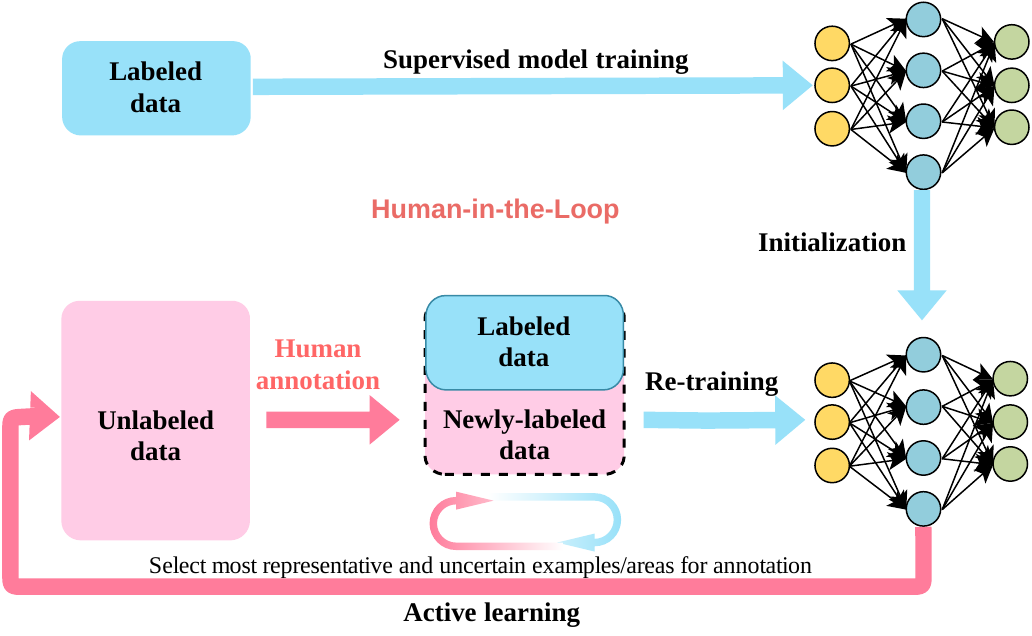}
\end{center}
\caption{Overview of the active learning paradigm: In a cycle, a deep learning model is trained on a labeled medical dataset. Then, active sampling strategies are implemented to select the data that is most valuable to the model from an unlabeled medical dataset. Finally, oracles are used to annotate the selected data. Image courtesy of Peng and Wang \citep{9363892}.}
\label{fig_AL}
\end{figure}

Active learning (AL) \citep{budd2021survey} operates on the assumption that the ground-truth labels of unlabeled instances can be obtained by querying an expert annotators (see Figure~\ref{fig_AL}). Assuming that the labeling cost is solely based on the number of queries, the objective of active learning is to minimize the number of queries required while still achieving effective model training with minimized labeling costs. In situations where there is a limited set of labeled data but an abundance of unlabeled data, active learning aims to identify the most valuable unlabeled instance for querying. There are two commonly used selection criteria: informativeness and representativeness. Informativeness assesses how effectively an unlabeled instance reduces the uncertainty of a statistical model, while representativeness calculates how well an instance represents the structure of input patterns.

\subsubsection{Evaluating informativeness}
The primary category of informativeness measures revolves around calculating uncertainty. The key idea is that including the ground truth for samples with higher uncertainty in the training set can provide more valuable information. In the deep learning area, uncertainty-based sampling has seen widespread usage in recent active learning methods \citep{liu2020deep, wen2018comparison, wu2021covid}. Specifically, Wen et al. \citep{wen2018comparison} introduce an active learning approach that employs uncertainty sampling to facilitate quality control of nucleus segmentation in pathology images. Wu et al. \citep{wu2021covid} use both the network loss and diversity condition as the uncertainty metric for sampling from a loss prediction network. They apply this method to the COVID-19 classification task. Zhou et al. \citep{zhou2021active} introduce the concept of active selection policies, where the highest confidence is determined based on the entropy and diversity of the sampled data in the mean prediction outcomes. For classifying radiology images, Balram et al. \citep{balaram2022consistency} introduce an integrated end-to-end solution that merges consistency-driven semi-supervised learning with uncertainty-guided active learning, aiming to alleviate the need for extensive manual annotations. Another prevalent approach for estimating informativeness involves assessing the agreement among various models executing the same task. The reasoning is that greater disagreement observed between predictions on similar data points indicates a higher degree of uncertainty. These techniques are commonly employed in situations where ensembling is utilized to enhance performance. Ensembling involves training multiple models to execute the same task with slight variations in parameters or settings \citep{kuo2018cost, beluch2018power}. Beluch Bcai et al. \citep{beluch2018power} showcase the effectiveness of ensembles in active learning and compare them to alternative approaches. Kuo et al. \citep{kuo2018cost} employed an ensemble technique to assess uncertainty in the context of intracranial hemorrhage segmentation, utilizing the Jensen-Shannon divergence. Additionally, they made an effort to predict the time required for manual delineation using a log-linear model. Their approach involved selecting examples for manual segmentation based on maximizing the cumulative uncertainty within a specified time constraint. Atzeni et al. \citep{atzeni2022deep} adopt an iterative method, requesting manual delineation for a single Region of Interest (ROI) on a single slice per iteration rather than labeling all structures within a slice or volume. They update a segmentation CNN that generates dense segmentations for all slices using mixed-cross entropy loss, effectively utilizing partially annotated images. Similar to Kuo et al. \citep{kuo2018cost}, they use tracing time, based on boundary length, as a practical measure of effort. However, in contrast to Kuo et al. \citep{kuo2018cost}, they also account for multiple ROIs and their spatial relationships.

Bayesian neural networks have gained significant interest due to their capacity to represent and propagate the probability of deep learning models. Gal et al. \citep{gal2017deep} introduce the concept of using Bayesian CNNs for AL, specifically employing \emph{Bayesian Active Learning by Disagreement} (BALD). Their study demonstrates the superior performance of Bayesian CNNs compared to deterministic CNNs within the context of AL. Mahapatra et al. \citep{mahapatra2018efficient} employ a conditional GAN to generate chest X-ray images based on a real image. Additionally, they use a Bayesian neural network to assess the informativeness of each generated sample, determining whether it should be utilized as training data. If selected, the sample is used to fine-tune the network. Their study demonstrates that this method achieves comparable performance to training on fully labeled data, even when working with a dataset where only 33 \% of the pixels in the training set have annotations. This provides significant time, effort, and cost savings for annotators. Dai et al. \citep{dai2020suggestive} proposed a distinctive method for brain tumor segmentation. Instead of traditional approaches, they adopted a novel strategy to select the most informative example. This involved moving through the image space along the gradient direction of the Dice loss and identifying the nearest neighbor of this image within a lower-dimensional latent space, which was learned using a variational autoencoder. Certain studies address the challenge of the cold start problem in Active Learning, which pertains to the initial selection of images for labeling when no labeled data is available as a starting point \citep{nath2022warm, li2023hal}. Nath et al. \citep{nath2022warm} address the issue of cold start by introducing a proxy task and subsequently leveraging the uncertainty generated from this proxy task to prioritize the annotation of unlabeled data.

\subsubsection{Representativeness}

These approaches go beyond relying solely on uncertainty-based methods and instead focus on evaluating the diversity within chosen samples to minimize repetitive annotations. By introducing a representativeness measure, these strategies aim to promote the selection of samples from various areas of the distribution, leading to greater sample diversity and ultimately enhancing the performance of AL. To this end, Yang et al. \citep{yang2017suggestive} introduce Suggestive Annotation, a deep active learning framework designed for medical image segmentation. This framework utilizes a different approach to uncertainty sampling and incorporates a form of representativeness density weighting. The method involves training multiple models, each of which excludes a portion of the training data. These models are then leveraged to calculate an ensemble-based uncertainty measure. Ozdemir et al. \citep{ozdemir2021active} create a Bayesian network and utilize Monte Carlo dropout to derive variance information as a measure of model uncertainty. In addition, they employ infoVAE \citep{zhao2017infovae} to build a representativeness metric, which aids in the selection of samples through maximum likelihood sampling within the latent space. Li et al. \citep{li2021pathal} adopt k-means clustering and curriculum classification (CC) techniques, leveraging CurriculumNet \citep{guo2018curriculumnet}, to estimate uncertainty and representativeness in their approach. Li et al. \citep{li2023hal} tackle the challenge of the cold start problem by employing representativeness sampling that relies on the distance matrix to choose an initial dataset that is representative. They also introduce a hybrid sample selection approach that incorporates pixel entropy, region consistency, and image diversity scores to filter the samples. These three scores reflect informativeness at different levels: pixel, region, and image. This strategy, which combines these three levels of scores, proves to be more effective in selecting the most valuable samples compared to using a simple pixel uncertainty score alone. Wang et al. \citep{wang2021annotation} utilize model ensembles to guide user labeling, focusing on cells that optimize a blend of uncertainty, diversity (evaluated using a clustering algorithm), and representativeness assessed through cosine similarity of features.

\subsection{Domain-adaptive learning}  \label{UDA}

Domain adaptive learning, also known as domain adaptation \citep{suru2023deep}, is a learning paradigm focused on improving the performance of a model on a target domain by leveraging knowledge learned from a source domain. In this context, a domain refers to a specific distribution of data, which can vary in terms of characteristics like data collection settings, sensor types, lighting conditions, or other factors that affect the data's distribution. The main challenge addressed by domain adaptive learning is the domain shift problem. This problem arises when there is a mismatch between the source domain (where the model is trained) and the target domain (where the model needs to perform well). Due to this mismatch, a model trained on one domain may not generalize effectively to another domain. Domain adaptive learning methods aim to bridge this gap by adapting the model to the target domain. Unsupervised Domain Adaptation (UDA) is a specific case of domain adaptation where you only have labeled data in the source domain and no labeled data in the target domain. The adaptation process is entirely unsupervised, meaning it relies solely on unlabeled data in the target domain. These methods encompass various approaches, including feature alignment, image translation-based methods, learning disentangled representations, pseudo-labeling approaches, self-supervision, and hybrid methods.

\subsubsection{Feature alignment}
The fundamental idea behind feature alignment in UDA is to lessen the distinction between the source and target domains by learning domain-invariant representations. Various UDA approaches map images from both domains onto a common latent space to mitigate disparities. This can be accomplished directly by reducing a disparity measure that quantifies domain dissimilarities. Alternatively, it can be realized implicitly through adversarial learning techniques. The objective is to align the feature distributions of both the source and target domains, ensuring that the learned representations can be smoothly transferred and effectively utilized in diverse domains.

\textbf{Explicit discrepancy minimization:} Methods focused on explicitly minimizing discrepancies usually create a measure or loss function that calculates how different the source and target distributions are from each other. This measure is then reduced during training to encourage the development of features that work well in both domains. Different measures, like Maximum Mean Discrepancy (MMD) \citep{yu2022domain, fang2023unsupervised}, Kullback-Leibler (KL) divergence \citep{hu2022synthesis}, and Contrastive Loss (CL) \citep{sahu2020endo, gomariz2022unsupervised}, can be employed for this purpose. Specifically, Yu et al.
\citep{yu2022domain} use two separate feature encoders for both the target and source domains. They integrate an attention technique to focus on particular brain regions and employ MMD to acquire features that work well across domains for the prediction of subjective cognitive decline. Another explicit measurement used in UDA is the Characteristic Function (CF) distance \citep{wu2020cf}. This metric calculates the distinction between the distributions of latent features in the frequency domain instead of the spatial domain.

\begin{figure}[t]
\begin{center}
\includegraphics[scale=0.65]{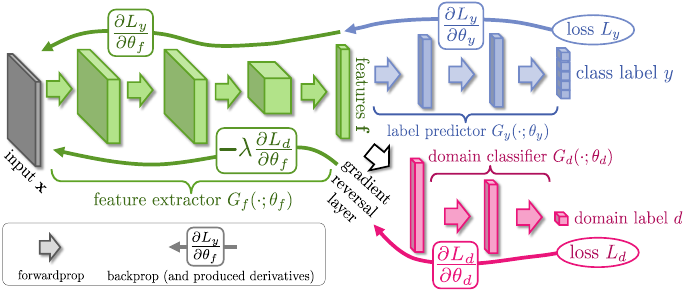}
\end{center}
\caption{The figure demonstrates the Domain Adversarial Neural Network (DANN) framework, a classic and effective model designed for learning domain-invariant features using adversarial training (image courtesy of Ganin \citep{ganin2016domain}).}
\label{fig_7}
\end{figure}

\textbf{Implicit discrepancy minimization:} Implicit methods for reducing differences in UDA mainly rely on the concepts of adversarial learning. To ensure that feature distributions are comparable between different domains, a technique called domain-adversarial neural network (DANN) \citep{ganin2016domain} is used. This approach involves incorporating a gradient reversal layer (GRL) into the framework of Generative Adversarial Networks (GANs), as illustrated in Figure~\ref{fig_7}. The network comprises two classifiers and shared feature extraction layers. With the help of GRL, DANN aims to maximize the loss due to domain confusion while minimizing the loss associated with label prediction for source samples and domain confusion loss for all samples. DANN serves as a foundational model for different UDA methods that are built upon adversarial learning principles. 

Different research studies employ implicit techniques for a range of classification issues. For instance, Ren et al. \citep{ren2018adversarial} utilize an adversarial loss along with siamese architecture for whole slide images. Zhang et al. \citep{zhang2019whole} leverage adversarial learning and introduce focal loss to tackle the problem of class imbalance in histopathology images. More recently, Feng et al. \citep{feng2023contrastive} engage in binary and multi-class classification tasks related to diagnosing pneumonia. They make use of a conditional domain adversarial network to narrow the domain discrepancy and implement a contrastive loss to address the challenge of limited data in the target domain. Certain investigations have combined self-training and adversarial learning for the task of medical image segmentation \citep{bian2020uncertainty, liu2021s, huang2022domain}. Specifically, Liu et al. \citep{liu2021s} proposed the Self-cleansing UDA (S-cuda) technique, which is specifically designed to address the issue of domain shift and handle noisy labels in the source domain. This method utilizes self-training to produce accurate pseudo-labels for both the noisy source and unlabeled target domains. Beyond image classification and segmentation, various other applications also make use of implicit discrepancy methods. For instance, these methods are applied in bronchoscopic depth estimation \citep{karaoglu2021adversarial}, reconstructing precise high-resolution (HR) representations from low-resolution (LR) OCTA images \citep{hao2022sparse}, and automating sleep staging \citep{yoo2021transferring}.

\subsubsection{Image translation based methods}
\begin{figure}[t]
\begin{center}
\includegraphics[scale=0.25]{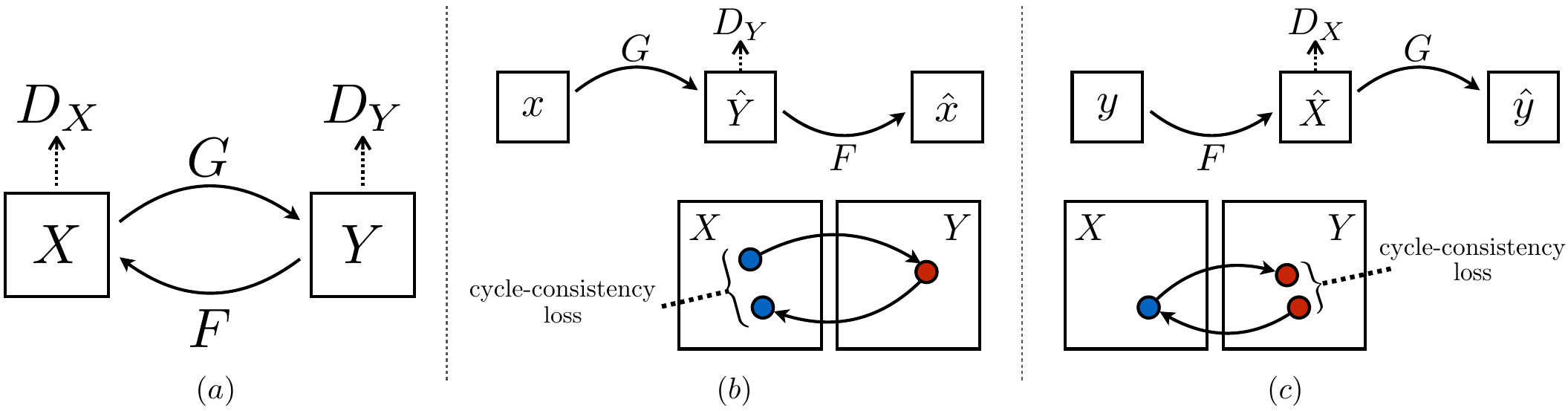}
\end{center}
\caption{Illustration of CycleGAN framework: (a) CycleGAN comprises two mapping functions $G: X \rightarrow Y$ and $F: Y \rightarrow X$, accompanied by adversarial discriminators $D_Y$ and $D_X$. $D_Y$ encourages $G$ to translate $X$ into outputs indistinguishable from domain $Y$, while $D_X$ performs the reverse task.
(b) The forward cycle-consistency loss is represented as: $x \rightarrow G(x) \rightarrow F(G(x)) \approx x$. (c) The backward cycle-consistency loss is represented as: $y \rightarrow F(y) \rightarrow G(F(y)) \approx y$ (image courtesy of Zhu et al. \citep{zhu2017unpaired}).}
\label{fig_cyclegan}
\end{figure}

Image translation techniques achieve domain alignment by altering the pixel-level appearance of source data to match the characteristics of a target domain. Generative Adversarial Networks (GANs) are often used for tasks involving direct mapping between pixels for image translation. A widely used approach in this category is CycleGAN \citep{zhu2017unpaired}, which operates as an image-to-image translation architecture (see Figure~\ref{fig_cyclegan}). It transforms features from one image domain into another without relying on paired training examples. In the medical field, several approaches apply CycleGAN for unsupervised domain adaptation (UDA). However, CycleGAN's emphasis on pixel-level mapping might not consistently ensure the preservation of semantic information in medical images. To overcome this limitation, multiple studies have integrated semantic understanding into the framework. Various works \citep{tang2019tuna, jiang2018tumor, jiang2020psigan} have incorporated task-specific losses within the UDA context. These task-specific losses are designed to enhance the UDA procedure by introducing extra constraints aligned with the unique requirements of the task at hand.

Certain works employ attention mechanisms to capture distant relationships \citep{tomar2021self, kapil2021domain}. In the context of cross-modality domain adaptation, Tomar et al. \citep{tomar2021self} employ a dual cycle consistency loss to maintain semantic content while performing image translation. They propose a self-attentive spatial adaptive normalization technique that comprises two components: the synthesis module and the attention module. The synthesis module's intermediate layers receive semantic layout information from the attention module, aiding in the learning of the translation process. Certain studies exploring UDA in image detection also employ image translation techniques. For instance, Xing et al. \citep{xing2019adversarial} delve into UDA for cell detection across different data modalities. They leverage the CycleGAN framework to adjust source images to align with the target domain. Their methodology involves training a structured regression-based object detector using these adapted source images. Furthermore, they refine the detector by incorporating pseudo-labels derived from the target training dataset. Extending their earlier study, Xing et al. \citep{xing2020bidirectional} enhance their method by introducing bidirectional mapping. This involves translating images both from the source to the target and vice versa. They also expand this framework to address the semi-supervised scenario. In a subsequent extension of their research, Xing and Cornish \citep{xing2022low} tackle not just the UDA challenges in cell/nucleus detection but also address the challenge of having scarce training data in the target domain.

\subsubsection{Learning disentangled representations}
Rather than imposing the demanding requirement of making the entire model or features domain-invariant, an alternative approach is to ease this constraint by permitting certain components to be domain-specific \citep{zhou2022domain}. This essentially involves acquiring disentangled representations. The key idea of disentangled representation is to differentiate between the content and style of an image. The underlying premise is that the content, which refers to anatomical information, remains uniform across domains, whereas the style, encompassing attributes like texture and lighting, is specific to each domain. In the process of achieving disentangled representation \citep{yang2019unsupervised}, initial steps involve extracting style and content codes from both the source and target images using specialized encoders. Subsequently, generators are employed to create images in the opposite domains by combining content codes from one domain with style codes from the other. This interplay of generators aims to deceive discriminators by generating images that confuse the domains' distinguishing features, leading to the desired disentangled representation. Wang et al. \citep{wang2022cycmis} incorporated the segmentation stage and diverse image translation stage into a cohesive end-to-end approach. 

Sun et al. \citep{sun2022attention} employ a combination of the attention mechanism and disentanglement to further mitigate the disparities between domains. Specifically, they adopt a preliminary alignment phase to address issues like variations in brightness between MRI and CT images. Following this, they introduce an improved approach to disentanglement that leverages the Hilbert-Schmidt independence criterion to encourage independence and complementary characteristics between content and style attributes. Lastly, they integrate an attention bias mechanism to emphasize the alignment of regions relevant to the task of cardiac segmentation. Several studies enhance disentanglement learning by employing various approaches. For example, Xie et al. \citep{xie2022unsupervised} utilize a zero loss to ensure that the domain-specific encoder only captures information from its corresponding domain. Similarly, Yang et al. \citep{yang2021mutual} implement a coarse-to-fine prototype alignment process before feature disentanglement to enhance the separation of features.

\subsubsection{Pseudo-labeling approach}
Pseudo-labeling is a widely used strategy in UDA to make use of unlabeled data in the target domain. This method involves generating pseudo-labels to unlabeled data in the target domain using a model trained on labeled data from the source domain. Nevertheless, these pseudo-labels can be inaccurate due to the domain gap, leading to noise. Thus, a crucial aspect of pseudo-labeling is how various networks reduce the uncertainty and eliminate noise from the pseudo-labels to enhance their precision. To reduce the uncertainty of pseudo labels, Wu et al. \citep{wu2021uncertainty} introduce an uncertainty-aware model that integrates Monte Carlo dropout layers into a U-Net architecture. Likewise, the Strudel approach \citep{groger2021strudel} involves incorporating uncertainty details into the training process through an uncertainty-guided loss function. This aids in eliminating labels with low level of certainty. Some studies adopt a curriculum learning strategy, beginning with simpler instances to facilitate the model's learning process and gradually introducing more complex cases over time \citep{qi2020curriculum, cho2022effective}. When facing a situation where classes are imbalanced, pseudo-labels frequently demonstrate an uneven distribution because the model tends to have greater confidence in dominant or less complex classes. To address this, Mottaghi et al. \citep{mottaghi2022adaptation} introduce a new strategy for pseudo-label selection. This involves using a subset of pseudo-labels based on the reciprocal of class frequency, favoring less common or challenging classes. This technique effectively addresses label distribution imbalance, boosting the surgical activity recognition model's reliability and performance. 

\subsubsection{Self-supervision}
Certain studies address UDA by employing the self-supervision strategy. In this approach, alignment is achieved by concurrently conducting auxiliary self-supervised tasks in both domains. Each self-supervised task aims to bring the domains closer by focusing on relevant directions. Successfully training these self-supervised tasks alongside the primary task in the source domain has proven effective in generalizing to the unlabeled target domain \citep{sun2019unsupervised}. Various auxiliary self-supervised tasks are available, but not all are suitable for UDA. Consequently, the primary challenge in the self-supervision method is to identify an appropriate self-supervised task that enables the model to learn valuable representations from the data and promote alignment between the domains. Koohbanani et al. \citep{koohbanani2021self} present a method called Self-Path for histology image classification. In this approach, they propose three innovative domain-specific self-supervision tasks. These tasks involve predicting the magnification level, solving a magnification jigsaw puzzle, and predicting the Hematoxylin channel. These tasks are strategically designed to utilize the contextual, multi-resolution, and semantic features inherent in histopathology images. The Self-rule to multi-adapt (SRMA) technique \citep{abbet2022self} is applied in the detection of cancer tissue. This method uses a limited set of labeled images from the source domain and integrates structural details from both domains by identifying visual similarities using self-supervision within each domain and across domains. Additional self-supervised tasks include the jigsaw puzzle auxiliary task, where the spatial correlation in an image is learned by reconstructing a CT scan from shuffled patches \citep{fu2020domain}, and an auxiliary task focused on edge generation \citep{xue2020dual}.

\subsubsection{Hybrid methods}
Various works employ feature alignment and image translation methods together to enhance the performance of UDA. These are called hybrid methods. Hybrid methods encompass a two-step procedure: initially, image transformation modifies the source images to align them with the target domain's appearance, and subsequently, feature adaptation is applied to narrow the remaining disparity between the generated target-like images and the real target images \citep{chen2020unsupervised}. The benefit of employing hybrid techniques lies in their ability to retain pixel-level, feature-level, and semantic information. The Cycle-Consistent Adversarial Domain Adaptation technique (CyCADA), introduced by Hoffman et al. \citep{hoffman2018cycada}, is a hybrid learning method developed for natural images. It consists of two stages: image adaptation and feature adaptation, both of which undergo sequential training without direct interactions. CyCADA has found extensive application as a fundamental model in different medical imaging scenarios \citep{jia2019cone, liu2020pdam}.

In contrast to CyCADA, Chen et al. \citep{chen2019synergistic, chen2020unsupervised} introduce an alternative technique known as Synergistic Image and Feature Alignment (SIFA), which facilitates concurrent image and feature translation. In particular, the feature encoder is shared, enabling it to simultaneously alter the image's appearance and extract domain-invariant representations for the segmentation task. To enhance domain adaptation accuracy further, certain research studies incorporate attention mechanisms alongside image and feature alignment techniques \citep{cui2021bidirectional, chen2022dual}. Chen et al. \citep{chen2022dual} employ the same framework as SIFA. However, in their approach, the alignment of the feature space is directed by the dual adversarial attention mechanism. This mechanism concentrates on specific regions identified by the spatial and class attention mechanisms rather than treating all semantic feature components uniformly. Label-efficient UDA (LE-UDA) \citep{zhao2023uda} tackles both domain shift and source label scarcity. The approach utilizes a hybrid method to handle domain shift, while for source label scarcity, it incorporates two teacher models. These models leverage information within domains as well as across domains from diverse datasets. In a recent study, Li et al. \citep{li2023self} introduce a self-training adversarial learning framework for retinal OCT fluid segmentation tasks that utilize a hybrid approach.

\begin{table*}[]
    \centering
    \scriptsize
    \begin{center}
    \caption{Overview of recent methods in \emph{Incomplete Supervision} category.}
    \label{table:incomplete sup}
   \resizebox{\textwidth}{!}{
    \begin{tabular}{p{0.8cm}p{3.5cm}p{4cm}p{3.5cm}p{4.7cm}}
    \toprule
   Reference & \multicolumn{1}{l}{Task} & \multicolumn{1}{l}{Algorithm Design} & \multicolumn{1}{l}{Dataset} & \multicolumn{1}{l}{Result} \\ \midrule 

\citep{zhang2017deep} & Gland segmentation & Deep adversarial network & 2015 MICCAI Gland Challenge dataset & F1: 0.916; ObjectDice: 0.903 \\ \midrule

\citep{wu2021collaborative} & Polyp segmentation & Adversarial learning & Kvasir-SEG;
CVC-Clinic DB & Kvasir-SEG: Dice: 15\% label: 0.7676, 30\% label: 0.8095; CVC-Clinic DB: Dice: 15\% label: 0.8218, 30\% label: 0.8929   \\ \midrule

\citep{chaitanya2021semi}& Heart; Prostate; Pancreas segmentation & Semi-supervised GAN & ACDC; DECATHLON &ACDC: DSC (Dice coefficient): 0.834; DECATHLON: DSC: 0.529 \\ \midrule
\citep{hou2022semi} & Fundus segmentation & Leaking GAN & DRIVE, STARE, CHASE DB1 & DRIVE: Acc: 95.74 Sp: 86.72 Se: 97.50; STARE: Acc: 95.65 Sp: 91.86 Se: 91.02; CHASE DB1: Acc: 96.83 Sp:92.21 Se:94.72   \\ \midrule
\citep{sedai2017semi} & Optic cup segmentation & Teacher-student VAE & DRD &DSC: 0.80 \\ \midrule
\citep{wang2022rethinking} & Kidney; Heart; Liver & Generative Bayesian Deep Learning & KiTS; ASG; DECATHLON & DSC: KiTS: 0.898; ASG: 0.884; DECATHLON: 0.935 \\ \midrule
\citep{li2018semi} &  Skin lesion segmentation  & $\Pi$-model & ISIC 2017 & DSC: 0.874; Acc: 0.943 \\ \midrule

\citep{bortsova2019semi} & Chest X-ray segmentation & Elastic deformations perturbations for CL & JSRT dataset & MeanIOU: 5 labeled samples: 85.0 $\pm$ 2.8; 10 labeled samples: 87.9 $\pm$ 0.8   \\ \midrule

\citep{cao2020uncertainty} & Breast & Uncertainty-aware Temporal Ensembling & Private Dataset: 170 Volumes; ISIC 2017 & Private Dataset: DSC: 0.7287; ISIC 2017: DSC: 0.8178 \\ \midrule

\citep{xu2023dual} & Brain Tumor and Left Atrial Segmentation& Dual Uncertainty-Guided Mixing Consistency & BraTS2020; LA2018 & BraTS: Dice: 85.94 \%; LA2018: Dice: 89.28 \% \\ \midrule


\citep{bai2017semi} & Heart & CRF-based Self-training & Private Dataset: 8050 & Images DSC: 0.920 \\ \midrule
\citep{liu2022acpl} &  Thorax Disease and Skin lesion classification & Anti-Curriculum Pseudo-labeling & Chest X-Ray14; ISIC 2018 & Chest X-Ray14: AUC: 81.77; ISIC 2018: AUC: 94.36  Sensitivity: 72.14 F1: 62.23 \\ \midrule

\citep{chen2022mass} & Multi-organ abdominal segmentation & Co-training using different modalities & BTCV; CHAOS & BTCV: Mean Dice score: 10 \% labels: 81.3; CHAOS: Mean Dice score: 10 \% labels: 82.1             \\ \midrule
\citep{luo2022semi} & Cardiac segmentation & Co-training using different network architectures &  ACDC dataset     &  Mean DSC: 0.848 (0.085); Mean HD95: 7.6 (10.8)  \\ \midrule
\citep{zhao2022mmgl}& Cardiac segmentation & Co-training using different transformations & MM-WHS dataset & 10\% labeled data: Dice: 0.743, mIOU: 0.601, PixAcc: 0.973; 20\% labeled data: Dice: 0.828, mIOU: 0.714, PixAcc: 0.979; 40\% labeled data: Dice: 0.849, mIOU: 0.746, PixAcc: 0.985;
\\ \midrule
\citep{wang2021deep} & Breast; Retina & Self-training + Virtual Adversarial Training & RetinalOCT; Private Dataset: 39,904 Images & Acc: 0.9513; Macro-R (Macro-Recall): 0.9330 \\ \midrule

\citep{wang2020focalmix} & Lung detection & MixMatch + Focal Loss & LUNA; NLST & LUNA: CPM: 0.872 \\ \midrule

\citep{zhang2022boostmis} & Metastatic epidural spinal cord classification & Consistency Regularization + Pseudo-labeling + Active Learning & Private Dataset: 7,295 Images; & Acc: 0.9582; Macro-P (Macro-Precision): 0.8609 \\ \midrule

\citep{chaitanya2023local} & Cardiac and Prostate segmentation & Self-training + Contrastive loss & ACDC; Prostate; MMWHS dataset & ACDC: DSC: 0.881; Prostate: DSC: 0.693; MMWHS: DSC: 0.803 \\ \midrule

\citep{basak2023pseudo} & Cardiac, Tumour and histopathology images segmentation & Self-training + Contrastive loss & ACDC; KiTS19; CRAG & ACDC: DSC: 0.891; KiTS19: DSC: 0.919; CRAG: 0.882 \\ \midrule


\citep{zhou2021active} & Colon & Traditional Data Augmentation Entropy + Diversity & Private Dataset: 6 colonoscopy videos 38 polyp videos + 121 CTPA datasets & Classification: 4 \% input: AUC: 0.9204;
Detection: 2.04 \% input: AUC: 0.9615 \\ \midrule

\citep{wu2021covid} & Lung Classification& Loss Prediction Network & CC-CCII Dataset & 42 \% Chest X-Ray input: Acc: 86.6\%
\\ \midrule

\citep{gal2017deep} & Skin disease classification & BALD + KL-divergence & ISIC 2016 & 22 \% image input: AUC: 0.75 \\ \midrule

\citep{mahapatra2018efficient} & Chest & Bayesian Neural Network + cGAN Data Augmentation & JSRT Database; ChestX-ray8 & Classification: 35 \% input: AUC: 0.953; Segmentation: 35 \% input: DSC: 0.910 \\ \midrule

\citep{yang2017suggestive} & Gland; Lymph & Cosine Similarity + Bootstrapping + FCN & GlaS 2015; Private Dataset: 80 US images & MICCAI 2015: 50 \% input: F1: 0.921; Private Dataset: 50 \% input: F1: 0.871 \\ \midrule

 \citep{ozdemir2021active}& Shoulder & BNN + MMD Divergence & Private Dataset: 36 Volume of MRIs & 48 \% MRI input: DSC $\approx$ 0.85 \\ \midrule



\citep{fang2023unsupervised} &  Major depressive order identification & Feature (MMD) & REST-meta-MDD Consortium & Site/Hospital - 20 → Site/Hospital - 1: ACC (\%): 59.73 $\pm$ 1.63; AUC (\%): 62.50 $\pm$ 2.50; SEN (\%): 69.46 $\pm$ 6.43; SPE (\%): 50.00 $\pm$ 9.63; PRE (\%): 58.49 $\pm$ 2.58  \\ \midrule

\citep{hu2022synthesis} & 3D Medical Image Synthesis & KL divergence & BraTS 2019 dataset (2 subsets are used CBICA and TCIA) & CBICA → TCIA: Dice: 0.773; TCIA → CBICA: Dice: 0.874     \\ \midrule

\citep{gomariz2022unsupervised} & Segmentation of retinal fluids in 3D OCT images & Contrastive and supervised loss & Two large OCT datasets (Spectralis and Cirrus) & Spectralis → Cirrus: Dice: 62.33; UVD: 10.88  \\ \bottomrule

\multicolumn{5}{r}{\footnotesize\textit{continued on the next page}}\\

\end{tabular}
}
\end{center}
\end{table*}

\begin{table*}[]
\ContinuedFloat
    \centering
    \scriptsize
    \begin{center}
    \caption{Overview of recent methods in \emph{Incomplete Supervision} category (continued).}
    
   \resizebox{\textwidth}{!}{
    \begin{tabular}{p{0.8cm}p{3.5cm}p{4cm}p{3.5cm}p{4.7cm}}
    \toprule
   Reference & \multicolumn{1}{l}{Task} & \multicolumn{1}{l}{Algorithm Design} & \multicolumn{1}{l}{Dataset} & \multicolumn{1}{l}{Result} \\ \toprule

\citep{zhang2019whole} & Histopathology cancer classification & Adversarial learning +
Entropy loss + Focal loss & Private cross-modality dataset & WSI → Microscopy images(MSIs): Accuracy: 90.48; Precision: 90.67; Recall: 90.35; F1-measure: 90.50  \\ \midrule

\citep{feng2023contrastive} & Automated pneumonia diagnosis & Conditional domain adversarial network + Contrastive loss & RSNA dataset (Stage I); Child X-ray dataset & RSNA dataset → Child X-ray: AUC score: 90.57; RSNA + COVID → TTSH dataset: weighted AUC: 88.27  \\ \midrule


\citep{karaoglu2021adversarial} & Bronchoscopic Depth Estimation & Adversarial learning & Synthetic dataset and human pulmonary dataset & Mean abs. rel. diff: 0.379; RMSE: 7.532; Accuracy: 0.856 \\ \midrule

\citep{jiang2018tumor} &  Lung Cancer Segmentation & CycleGan + Tumor-aware loss & The Cancer Imaging Archive (TCIA) CT dataset  and Private MRI dataset & CT → MRI: Validation set (Unsupervised): DSC: 0.62 $\pm$ 0.26 HD95: 7.47 $\pm$ 4.66; Test set (Unsupervised): DSC: 0.74 $\pm$ 0.15 HD95: 8.88 $\pm$ 4.8   \\ \midrule

\citep{tomar2021self} & Brain tumor and cardiac segmentation & Dual CycleGan + Self-attentive spatial adaptive normalization & MM-WHS challenge; BraTS & MM-WHS: MRI → CT: Mean Dice: 0.78 $\pm$ 0.10, Mean ASSD: 4.9 $\pm$ 1.5  CT → MR: Mean Dice: 0.70 $\pm$ 0.11, Mean ASSD: 9.5 $\pm$ 3.2; BraTS: MRI-T2 → MRI-T1: 0.50 $\pm$ 0.06  \\ \midrule

\citep{yang2019unsupervised} & Liver segmentation & Disentangled Representation & LiTS challenge 2017 dataset (CT slices) and multi-phasic (MRI slices) & CT → MR: Dice: 0.81         \\ \midrule

\citep{wang2022cycmis} & Cardiac (LV and MYO) segmentation & Disentangled Representation + Semantic consistency loss & MS-CMRSeg; MM-WHS challenge & MS-CMRSeg: bSSFP CMR → LGE CMR images: DSC: 79.08, ASSD: 1.68; MM-WHS: CT → MRI: DSC: 84.51, ASSD: 1.00, MRI → CT: DSC: 84.77 ASSD: 0.98         \\ \midrule

\citep{sun2022attention} & Cardiac segmentation & Disentangled Representation + HSIC + Attention bias & MMWHS challenge 2017 dataset & MRI → CT: Dice: 80.2, ASD: 5.1;  CT → MRI: Dice: 66.3, ASD: 4.9             \\ \midrule

\citep{groger2021strudel} & White Matter Hyperintensity
Segmentation & Self training + Uncertainty-guided loss & WMH; ADNI-2 & WMH → ADNI-2: DSC: 0.69 $\pm$ 0.18, H95: 11.2 $\pm$ 14.5    \\ \midrule

\citep{qi2020curriculum} & Epithelial-stroma (ES) classi-
fication & Curriculum learning & Netherland Cancer Institute’s (NKI) dataset, Vancouver General Hospital’s (VGH) and IHC
dataset & Accuracy: VGH → NKI: 91.50, IHC → NKI: 82.51,  NKI → VGH: 92.62, IHC → VGH: 80.49, VGH → IHC: 88.15, NKI → IHC: 81.90   \\ \midrule

\citep{mottaghi2022adaptation} & Pseudo labeling  & Surgical activity recognition models across operating rooms & Dataset of full-length surgery videos from two robotic ORs (OR1 and OR2) & OR1 → OR2: Accuracy: 70.76 mAP: 83.71, OR2 → OR1: Accuracy: 73.53, mAP: 89.96 \\ \midrule

\citep{koohbanani2021self} & Classification of Pathology Images & Prediction of magnification level and Hematoxylin channel + Solving jigsaw puzzle & WSIs: Camelyon16 and In house dataset(LNM-OSCC) & Camelyon16: AUC-ROC: 93.7 \% LNM-OSCC: 97.4 \%\\ \midrule

\citep{abbet2022self} & Colorectal tissue type classification; Multi-source patch classification & Intra-domain and cross-domain self-supervision & Kather-16 [242], Kather-19 [243], Colorectal cancer tissue phenotyping dataset (CRC-TP)[244] and In-house dataset & Kather-19 → Kather-16: overall weighted F1 score: 87.7; Kather-19 + Kather-16 → CRC-TP: weighted F1 score: 83.6  \\ \midrule

\citep{xue2020dual} & Cardiac segmentation & Edge generation task +
Adversarial learning & MM-WHS challenge dataset (2017) & MRI $\rightarrow$ CT: Dice: 76.98, ASD: 4.6 \\ \midrule

\citep{jia2019cone} & Cone-beam computed tomography (CBCT) segmentation & Image and feature alignment & Private Dataset: 90 patients & CT $\rightarrow$ CBCT: DSC: 83.6 \%  \\ \midrule

\citep{chen2019synergistic} & Cardiac segmentation & Synergistic image and feature alignment (SIFA) & MM-WHS challenge dataset & MRI $\rightarrow$ CT: Dice: 73.0, ASD: 8.1  \\ \midrule

\citep{chen2022dual} & Skull segmentation and Cardiac segmentation & CycleGan + Feature
space alignment is led by the dual adversarial attention mechanism & CQ500; ADNI; MM-WHS challenge & Skull: CT $\rightarrow$ MRI: DSC: 84.07 \%, ASSD: 1.18; Cardiac: MRI $\rightarrow$ CT: DSC: 76.7 \%, ASSD: 5.1  \\ \midrule

\citep{zhao2023uda} & Multi-organ and Cardiac segmentation & CycleGan and feature alignment & MICCAI 2015 Multi-Atlas Abdomen Labeling(CT images), ISBI 2019 CHAOS Challenge (MR images); MM-WHS & 
Cardiac: MRI $\rightarrow$ CT: Dice: 70.8, ASD: 9.6; CT $\rightarrow$ MRI: Dice: 66.5, ASD: 4.0; Multi-organ: MRI $\rightarrow$ CT: Dice: 82.8, ASD: 2.3; CT $\rightarrow$ MRI: Dice: 87.7, ASD: 1.0  \\ 

\bottomrule
\end{tabular}
}
\end{center}
\end{table*}

\section{Inaccurate supervision}  \label{Inaccurate}

\begin{figure}[t]
\begin{center}
\includegraphics[scale=0.1]{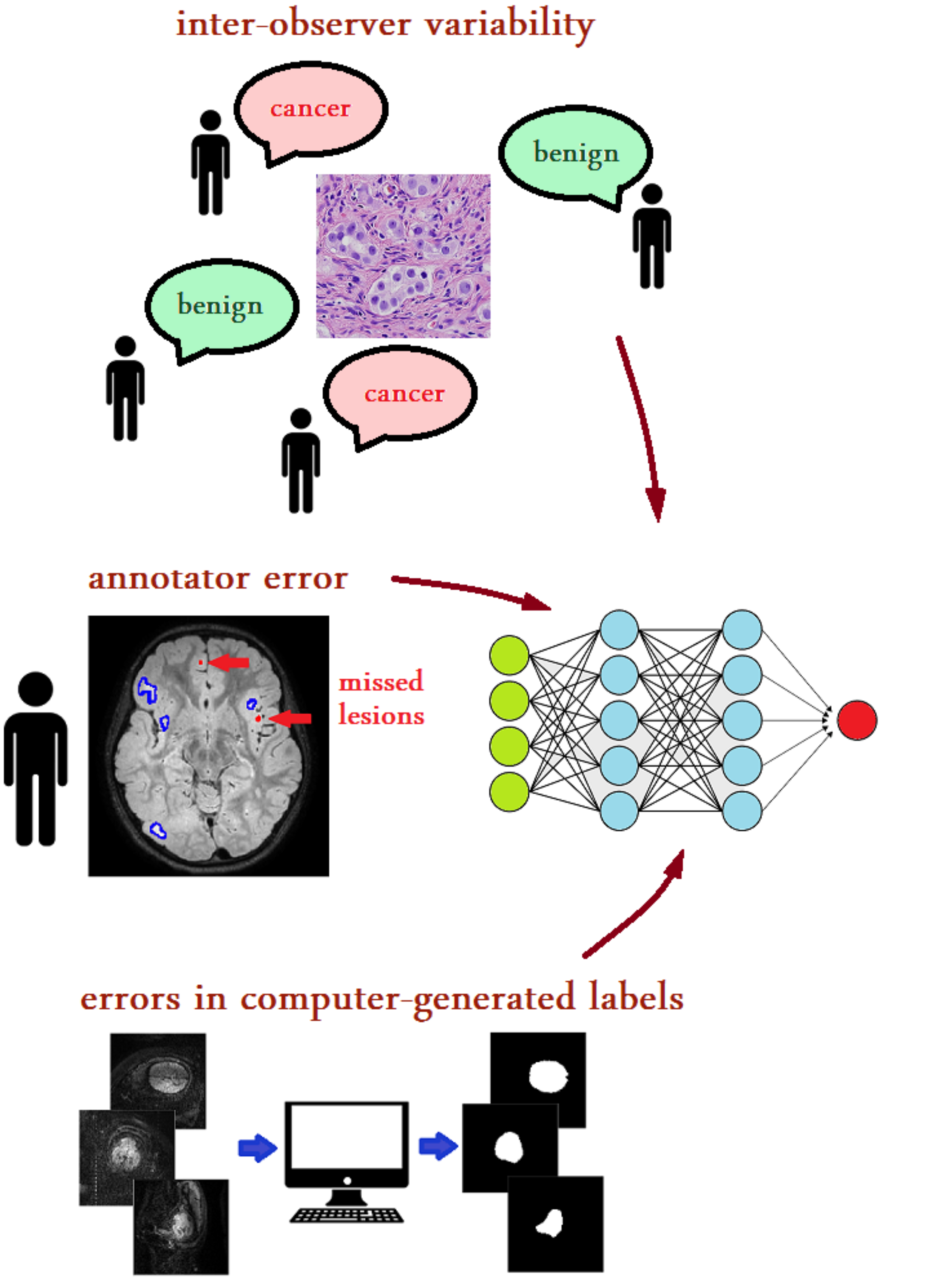}
\end{center}
\caption{The major sources of label noise encompass variations among different observers, mistakes made by human annotators, and inaccuracies in computer-generated labels. The impact of label noise in medical datasets is expected to grow as larger datasets are curated for deep learning purposes (image courtesy of Karimi \citep{karimi2020deep}).}
\label{fig_Noisy}
\end{figure}

Inaccurate supervision refers to a scenario where the provided supervision information isn't always entirely accurate, meaning that errors may be present in some of the label information. Such noisy labels \citep{karimi2020deep} can originate from various sources, including human errors in the labeling process, inter-observer variability among medical experts, or reliance on non-experts or automated systems for data labeling (see Figure~\ref{fig_Noisy}). Since noisy labels can significantly harm the generalization capabilities of deep neural networks, it is imperative to develop robust techniques to handle and mitigate the impact of noisy labels. This is particularly vital in the field of MIA, where precision and accuracy are critical for medical diagnosis and treatment. A summary of recent approaches for learning with inaccurate supervision is provided in Table~\ref{table:inaccurate sup}. We categorize inaccurate supervision approaches into three broad groups: Robust loss Design, data re-weighting and training procedures.

\subsection{Robust loss design} \label{RLD}

Certain investigations modify the loss function as a strategy to mitigate the impact of noisy labels. Researchers have introduced novel loss functions like Mean Absolute Error (MAE) \citep{ghosh2017robust} and Generalized Cross Entropy \citep{zhang2018generalized} to tackle the issue of label noise in natural images. In addition, some works choose to adapt loss functions specifically for tasks in medical imaging \citep{matuszewski2018minimal, wang2020noise, chen2022adaptive}. For COVID-19 Pneumonia Lesion segmentation, Wang et al. \citep{wang2020noise} proposed an enhanced Dice loss that addresses noise-related challenges. This improved loss function is an extended version of the traditional Dice loss, tailored for segmentation tasks, and incorporates the Mean Absolute Error (MAE) loss to enhance its robustness against noisy data. Chen et al. \citep{chen2022adaptive} introduce a new and versatile loss function called Adaptive Cross Entropy (ACE), designed to handle noise in labels without requiring hyperparameter fine-tuning during training. They provide both theoretical and practical evaluations of the ACE loss and demonstrate its efficacy across various publicly available datasets. Previous segmentation methods that handle noisy labels have typically focused on preserving semantics in a pixel-wise manner, which involves actions like pixel-wise label correction. However, they often overlook the potential benefits of considering pairwise relationships between pixels \citep{guo2022joint}. Notably, it has been observed that capturing these pairwise affinities can significantly decrease label noise. Building on this insight, Guo et al. \citep{guo2022joint} introduce a joint class-affinity segmentation model that takes into account both pixel-wise label correction and pairwise pixel relationships in order to reduce label noise. To further reduce the impact of label noise, they introduce a strategy called class-affinity loss correction (CALC), which includes class-level and affinity-level loss correction.

\subsection{Data re-weighting}  \label{DR}

Broadly speaking, these methods aim at down-weighting those training samples that are more likely to have incorrect labels. In this context, Xue et al. \citep{xue2019robust} introduced an approach for classifying skin lesions with noisy labels. Their method involved a data re-weighting technique, effectively excluding data samples with significant loss values in each training batch.
To predict pancreatic cancer regions in whole-slide images (WSIs), Le et al. \citep{le2019pancreatic} utilized a noisy label classification technique. This approach incorporates a limited set of clean training samples and dynamically assigns weights to training samples to address sample noise. These weights are assigned in real time to align the network loss with that of the clean samples. For multi-organ segmentation, Zhu et al. \citep{zhu2019pick} introduced an approach known as \emph{pick and learn}. In this method, a deep learning model is trained to identify incorrect labels and assign weights to each sample within a training batch. The goal is to reduce the impact of samples with inaccurate labels. Simultaneously, the primary segmentation model is trained in parallel, incorporating these weights into its loss function.

Strategies involving resampling and reweighting at the pixel level are intended to focus the segmentation model on learning from reliable pixels. For example, Mirikharaji et al. \citep{mirikharaji2019learning} introduced a method for skin lesion segmentation that incorporates pixel-wise weighting. This approach learns weight maps that adapt spatially and adjusts the influence of each pixel using a meta-reweighting framework. The Tri-network approach by Zhang et al. \citep{zhang2020robust} employs three cooperating networks and dynamically identifies informative samples based on the consensus among predictions generated by these distinct networks. Meanwhile, Wang et al. \citep{wang2020meta} employ meta-learning techniques to automatically estimate an importance map, allowing them to extract reliable information from crucial pixels.

\subsection{Training procedures}  \label{TP}

The methods in this category are very diverse. Several approaches in this category rely on Multi-network Learning, which frequently employs techniques like collaborative learning and co-training to train multiple networks simultaneously. Meanwhile, others follow the Multi-round Learning paradigm, which iteratively enhances the chosen set of clean examples without the need for maintaining additional DNNs. This improvement is achieved by repeating the training process in multiple rounds. Min et al. \citep{min2019two} adapted concepts from Malach and Shalev-Shwartz \citep{malach2017decoupling} to create label-noise-resistant techniques for medical image segmentation. They simultaneously trained two distinct models and exclusively updated these models using data samples where the predictions of the two models disagreed. Rather than solely relying on final layer predictions, Min et al. \citep{min2019two} incorporated attention modules at different network depths, allowing them to utilize gradient information from various feature maps to identify and reduce the influence of samples with incorrect labels. They demonstrated encouraging outcomes in MRI-based cardiac and glioma segmentation tasks. For medical image classification, Xue et al. \citep{xue2022robust} utilize a self-ensemble model along with a noisy label filter to effectively identify clean and noisy samples. Subsequently, they employ a collaborative training approach to train the clean samples, aiming to mitigate the impact of imperfect labels. Additionally, they introduce an innovative global and local representation learning scheme, which serves as an implicit regularization method for enabling the networks to make use of noisy samples in a self-supervised fashion. For COVID-19 pneumonia lesion segmentation, Yang et al. \citep{yang2022learning} propose a dual-branch network that learns from both accurate and noisy annotations separately. They introduce the Divergence-Aware Selective Training (DAST) strategy to distinguish between severely noisy and slightly noisy annotations. For severely noisy samples, they apply regularization through dual-branch consistency between predictions from the two branches. Additionally, they refine slightly noisy samples and incorporate them as supplementary data for the clean branch to prevent overfitting. Li et al. \citep{li2022learning} focus on selecting training pixels with reliable annotations from pixels with uncertain network predictions. They introduce the online prototypical soft label correction (PSLC) method to estimate pseudo-labels for label-unreliable pixels. They then calibrate the total segmentation loss using the segmentation loss of label-reliable and label-unreliable pixels. For hepatic vessel segmentation, Xu et al. \citep{xu2021noisy, xu2022anti} utilized a small set of accurately labeled data alongside a larger set of noisily labeled data. They employed confident learning with the help of a weight-averaged teacher model. This strategy involved progressively refining the noisy labels in the low-quality dataset through pixel-wise soft correction. Shi et al. \citep{shi2021distilling} present a framework designed to address noisy labels by extracting valuable supervision information from both pixel-level and image-level sources. Specifically, they make explicit estimations of pixel-wise uncertainty, treating it as a measure of noise at the pixel level. They then propose a robust learning approach at the pixel level, utilizing both the original labels and pseudo-labels. Additionally, they present a complementary image-level robust learning method to incorporate more information alongside pixel-level learning.

\begin{table*}[t]
    \centering
    \scriptsize
    \begin{center}
    \caption{Overview of recent methods in \emph{Inaccurate Supervision} category.}
    \label{table:inaccurate sup}
   \resizebox{\textwidth}{!}{
    \begin{tabular}{p{0.8cm}p{3.5cm}p{4cm}p{3.5cm}p{5cm}}
    \toprule
   Reference & \multicolumn{1}{l}{Task} & \multicolumn{1}{l}{Algorithm Design} & \multicolumn{1}{l}{Dataset} & \multicolumn{1}{l}{Result} \\ \toprule

\citep{zhu2019pick} & Multi-organ segmentation  & Loss Re-weighting  & JSRT & 25 \% noise: Dice: 0.895; 50 \% noise: Dice: 0.898; 75 \% noise: Dice: 0.895 \\ \midrule

\citep{mirikharaji2019learning} & Skin Lesions segmentation & Example reweighting & ISIC 2017 & Unsupervised noise: Dice: 73.55 \% \\ \midrule

\citep{xue2019robust} & Skin lesion classification & Sample reweighting + Online Uncertainty Sample Mining &  ISIC 2017 & 5\% noise: Acc: 84.5; 10\%: Acc: 83.6; 20\%: ;Acc: 80.7; 40\%: Acc: 80.7;  \\ \midrule

\citep{zhang2020robust} & Clinical stroke lesion and multi organ segmentation  & Tri-teaching network & Private clinical stroke dataset; JSRT & Clinical stroke: Dice: 68.12 \%; JSRT: Dice: 80.43  \\ \midrule

\citep{min2019two} & Cardiac and Brain tumour segmentation  & Two-Stream Mutual Attention Network &  HVSMR 2016; BRATS 2015 & HVSMR 2016: Myocardium: Dice: 0.820 ADB: 0.824 HDD: 4.73, Blood Pool: Dice: 0.926 ADB: 0.957 HDD: 8.81; BRATS 2015: Mean Dice: 0.792 \\ \midrule

\citep{xu2021noisy} & Hepatic Vessel Segmentation & Mean-Teacher-assisted Confident Learning & 3DIRCADb; MSD8 (Used for training) & 3DIRCADb: Dice: 0.7245 PRE: 0.7570 ASD: 1.1718 HD: 7.2111 \\ \midrule

\citep{shi2021distilling} & Left Atrial(LA) and cervical cancer Segmentation & Pixel-wise and Image-level Noise Tolerant learning & Left Atrial(LA); Private dataset & LA: 25 \% Noise : Dice(\%): ASD: 1.60 50 \% Noise: Dice(\%): 89.04 ASD: 1.92 75 \% Noise: Dice(\%): 76.25 ASD: 4.56; Private dataset: Dice(\%): 75.31 ASD: 1.76  \\ \midrule

\citep{guo2022joint} & Surgical instrument segmentation & Pixel-wise label correction and pairwise pixel relationships in order to reduce label noise.& Endovis18 & Average: Ellipse noise: Dice (\%): 71.384 Jac (\%): 58.452; Symmetric noise: Dice (\%): 74.058 Jac (\%): 62.667; Asymmetric noise: Dice (\%): 74.410 Jac (\%): 63.029\\

\bottomrule
\end{tabular}
}
\end{center}
\end{table*}

\section{Only limited supervision}  \label{Only}

\begin{figure}[t]
\begin{center}
\includegraphics[scale=0.26]{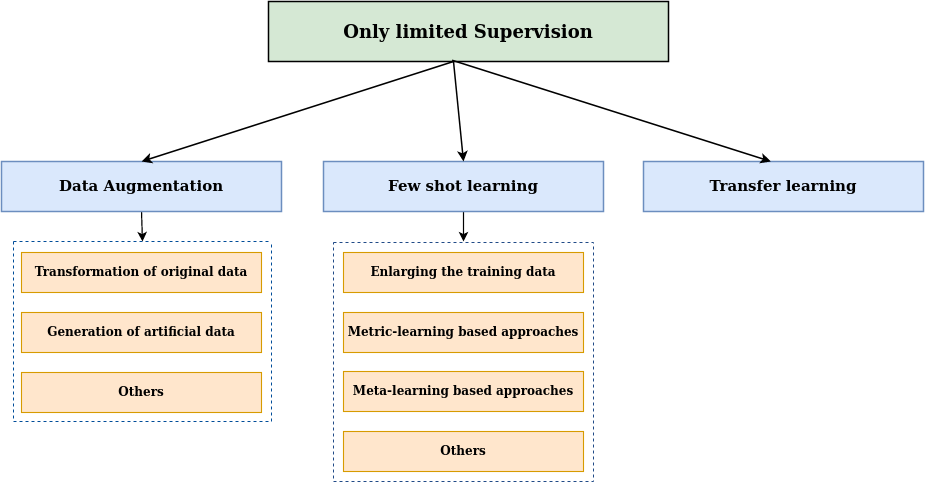}
\end{center}
\caption{Taxonomy of \emph{Only Limited Supervision} methods.}
\label{fig_only}
\end{figure}

\emph{Only Limited Supervision} refers to a set of methods in which the available supervision or labeling for training data is constrained or limited in nature. Furthermore, unlabeled data is also unavailable. These methods are typically applied in scenarios where acquiring extensive or detailed annotations is challenging or resource-intensive. Instead, they employ alternative strategies such as few-shot learning, transfer learning, and data augmentation to maximize the utility of the limited available supervision (Figure \ref{fig_only}). These approaches enhance model performance and facilitate tasks like segmentation, classification, or detection with minimal labeled data. A summary of recent approaches for learning with only limited supervision is provided in Table~\ref{table:only limited sup}.

\subsection{Data Augmentation} \label{DA}
Data augmentation offers a means to significantly increase the quantity and diversity of training data, all while avoiding the need for additional sample collection. These augmentation techniques encompass both straightforward yet remarkably impactful transformations like cropping, padding, and flipping, as well as more intricate generative models \citep{garcea2022data}. The efficacy of data augmentation strategies varies based on factors like input nature and visual tasks. Therefore, the field of medical imaging might necessitate distinct augmentation approaches that yield plausible data instances and effectively enhance the regularization of deep neural networks. Moreover, data augmentation can also address the issue of underrepresented classes by generating additional instances, such as generating synthetic lesions. Following earlier surveys \citep{garcea2022data, nalepa2019data}, we have categorized data augmentation methods into three broad groups: transformation of original data, generation of artificial data and other categories.

\subsubsection{Transformation of original data}

The first data augmentation category involves applying various image manipulation techniques to existing samples. This can be divided into three subcategories: (1) \textit{Affine Transformations} are geometric changes that retain lines and parallelism, though not necessarily distances and angles. This is ensured by transformation constraints, typically preserving the image's aspect ratio along axes of symmetry. The transformations include translation, rotation, flipping, scaling, cropping, and shearing \citep{pereira2016brain, liu2017deep}. (2) \textit{Elastic transformations} involve applying a spatial deformation field to an image. Unlike affine transformations, they don't enforce the preservation of collinearity or aspect ratio. As a result, elastic transformations can introduce shape variations and can be employed to enhance the robustness of segmentation algorithms \citep{nalepa2019data}. and (3) \textit{Pixel-Level Transformations} alter pixel values to modify characteristics like saturation, contrast, noise, and brightness \citep{galdran2017data}. Given that medical imaging is often grayscale, color-based changes are rare. Pixel-level transformations aid deep neural networks' robustness across different scanners and protocols that might affect pixel distribution.

The majority of transformations within this category are quite straightforward to apply and are either readily available in deep learning frameworks or can be easily incorporated using versatile libraries \citep{buslaev2020albumentations}. Recently, there has been a surge in frameworks and libraries tailored for the medical field, like the Medical Open Network for AI (MONAI) \citep{perez2021torchio}. However, it's important to note that since these techniques rely on altering the original samples, they can't enhance the network's ability to generalize beyond its initial training data. They often generate samples that are highly correlated.

\subsubsection{Generation of artificial data}
Creating artificial or synthesized samples can provide a broader range of diverse and complex examples, effectively addressing the limitations of methods based on transformations. The prevalent method for medical image synthesis is through generative networks, particularly generative adversarial networks (GANs) \citep{goodfellow2014generative}. However, generating synthetic images can also involve techniques like combining features or employing specialized modeling approaches designed for specific medical imaging tasks or modalities. While these approaches offer increased diversity, they often require higher computational resources and introduce complexity. Additionally, artificially generated samples might not fully capture the visual attributes or distribution of genuine data instances.

Depending on the specific domain, dataset, and task at hand, certain families of GANs may be more suitable, while others may be entirely impractical. Translation-based GANs, which encompass models like CGAN \citep{mirza2014conditional}, pix2pix \citep{isola2017image}, CycleGan \citep{zhu2017unpaired}, and SPADE \citep{park2019semantic}, specialize in learning how to transform various types of images. For instance, they can convert a segmentation mask into a newly synthesized input or transform a non-contrast CT-scan into a contrast CT-scan. In contrast, noise-based generation models like DC-GAN \citep{radford2015unsupervised}, StyleGAN2 \citep{karras2020analyzing}, and PGAN \citep{karras2018progressive} offer greater flexibility. However, noise-based generation techniques may encounter challenges when dealing with small training datasets, necessitating mitigation strategies such as patch extractions and traditional data augmentation. Although translation based models are known for producing images of exceptionally high quality, they are restricted in terms of the quantity of images they can create because they rely on the use of segmentation masks or different image modalities as input requirements. Noise-based approaches, on the other hand, do not face such limitations but often yield images with lower visual quality and run the risk of reproducing artifacts (such as vignettes or rulers) that could reinforce biases present in the dataset \citep{bissoto2021gan}. These frameworks and other extended versions of GANs have been widely employed for augmenting various types of organ images, including liver \citep{frid2018synthetic, frid2018gan}, skin \citep{rashid2019skin}, chest \citep{waheed2020covidgan}, eye \citep{zhao2018synthesizing}, lung \citep{nishio2020attribute}, breast \citep{chen2021enhanced, muramatsu2020improving}, brain \citep{guo2020lesion, rezaei2018conditional}, and more. Readers interested in a comprehensive review of GANs for medical image augmentation can refer to the work of Chen et al. \citep{chen2022generative}.

Other than GANs, Copy-paste methods have been utilized to generate artificial data. Copy-paste is a simple yet effective data augmentation technique and, it has demonstrated the potential to amplify the generalization power of deep neural networks. In essence, copy-paste involves copying portions of one image and pasting them onto another. Notably, the mix-up technique by \citep{zhang2018mixup}, and CutMix \citep{yun2019cutmix} are well-known approaches for mixing entire images and mixing image crops, respectively. Several studies have extended these methods to address specific objectives in MIA. For example, TumorCP  \citep{yang2021tumorcp} employs lesion masks to extract lesions from scans and paste them onto another scan at appropriate locations, guided by the lesion masks in the target scan. In the context of nuclei segmentation, InsMix \citep{lin2022insmix} follows a Copy-Smooth-Paste principle and conducts morphology-constrained generative instance augmentation. SelfMix \citep{zhu2022selfmix} leverages both tumor and non-tumor information for lesion segmentation. Given a pair of annotated training images, CarveMix \citep{zhang2023carvemix} combines a region of interest (ROI) based on lesion location and geometry, replacing the corresponding voxels in other labeled images. TensorMixup \citep{wang2022data} is a method that merges two image patches using a tensor and has been utilized to enhance the precision of tumor segmentation. Certain alternative approaches concentrate on medical datasets that frequently exhibit skewness towards negative cases. These methods encompass the creation and incorporation of fabricated lesions into individuals who are otherwise healthy. For example, these methods have been applied to simulate lesions resembling multiple sclerosis in brain MR images \citep{salem2019multiple} or to introduce cancer indicators into breast mammography images \citep{cha2020evaluation}.

\subsubsection{Others}
Apart from the aforementioned categories, there exist augmentation techniques designed for specific purposes. Notably, modalities such as CT and MRI possess a volumetric nature. Leveraging 3D convolutions presents the advantage of incorporating information from neighboring slices, leading to enhanced performance given a large dataset. While several image transformations, particularly affine transformations, are well-established for 2D images, extending them to 3D settings may pose challenges in terms of computational efficiency. Examples of augmentations designed with 3D data augmentation techniques include 3D GANs \citep{sun2020mm}, 3D affine transformations \citep{onishi2020multiplanar}, and multiplanar image synthesis \citep{chen2020fully}. 

Learnable data augmentation, also known as neural data augmentation, is an advanced technique in deep learning where the augmentation parameters are learned by the neural network during the training process. Unlike traditional data augmentation methods that apply fixed transformations to input data, learnable data augmentation allows the model to adaptively determine the augmentation parameters based on the data and the task at hand. Methods following this strategy involve training two networks simultaneously: one network learns to solve a specific task, while the second network learns how to augment the data for the first one.  One of the most common approaches for learnable data augmentation is autoaugment \citep{cubuk2018autoaugment}. This method aims to optimize network performance by identifying the most effective combination of established transformations (like affine transformations, pixel-level modifications, etc.). The ideal augmentation policy is determined through a neural network, which can be trained using adversarial training \citep{chen2022enhancing}, evolutionary algorithms \citep{fujita2018data}, or reinforcement learning \citep{cubuk2018autoaugment}.

\subsection{Few shot learning}  \label{FSL}

Few-shot learning (FSL) takes inspiration from human-like robust reasoning and analytical abilities. Wang et al. \citep{wang2020generalizing} provided a standard definition for machine learning based on experience (E), task (T), and performance (P): A computer program is considered to learn from experience E with respect to certain classes of task T and performance measure P if its performance can enhance with E on T as measured by P. It's important to note that E, the experience in FSL, is quite limited. Formally, within each few-shot task, we are given three sets: a support set denoted as S, a query set referred to as Q, and an auxiliary set labeled as A. The support set S encompasses C distinct categories, with each category comprising K training samples, essentially forming a C-way K-shot configuration. The query set Q comprises unlabeled query data. We categorize the current deep FSL methods into enlarging the training data, metric-learning based methods, meta-learning based methods and others.

\textbf{Enlarging the training data:} These approaches augment the training data to enlarge the number of samples, enabling the utilization of standard deep learning models and algorithms on the augmented dataset to attain a more accurate model. For multi-modal medical image segmentation, Mondal et al. \citep{mondal2018few} expand the training set through the use of GANs. On the other hand, Zhao et al. \citep{zhao2019data} present a learning-based technique for data augmentation. Specifically, their approach starts with a single labeled image and a set of unlabeled examples. By employing learning-based registration methods, they model the spatial and appearance transformations between the labeled and unlabeled examples. These transformations encompass effects such as non-linear deformations and variations in imaging intensity. Subsequently, they generate new labeled examples by sampling these transformations and applying them to the labeled example, resulting in a diverse range of realistic images. These synthesized examples are then used to train a supervised segmentation model. For the segmentation of new WM tracts in a few-shot scenario, \citep{lu2022transfer, liu2022one} have proposed efficient data augmentation techniques. Specifically, Lu et al. \citep{lu2022transfer} introduce an efficient data augmentation technique that creates synthetic annotated images through tract-aware image mixing. Additionally, they employ a transfer learning method for few-shot segmentation. \citep{fischer2023self}

\textbf{Metric-learning based approaches:} These approaches offer a straightforward and adaptable framework where they directly assess the similarities or dissimilarities between query images in the query set and labeled images in the support set. A classical metric-learning technique, known as the Prototypical Network (ProtoNet) by Snell et al. \citep{snell2017prototypical}, illustrates this concept. ProtoNet computes prototype representations for each base class by averaging the feature vectors and subsequently measures the distances between these prototype representations and each query image. Importantly, metric-learning based methods do not involve data-independent parameters in their classifiers, which means fine-tuning is unnecessary during the testing phase. Moreover, some researchers, such as Ali et al. \citep{ali2020additive}, have introduced an innovative additive angular margin metric to enhance the original ProtoNet's ability to classify challenging samples, especially in scenarios involving multi-center, underrepresented, and difficult-to-classify endoscopy data. To enhance prototype-based few-shot segmentation model for abdominal organs, Wang et al. \citep{wang2022few} introduce a regularization technique. This enhancement involves two key elements: self-reference and contrastive learning. Self-reference regularization ensures that a class prototype accurately represents the entire organ within a support image. Contrastive learning aids in the understanding of similarity between foreground and background features. 

\begin{figure*}[t]
\begin{center}
\includegraphics[scale=0.07]{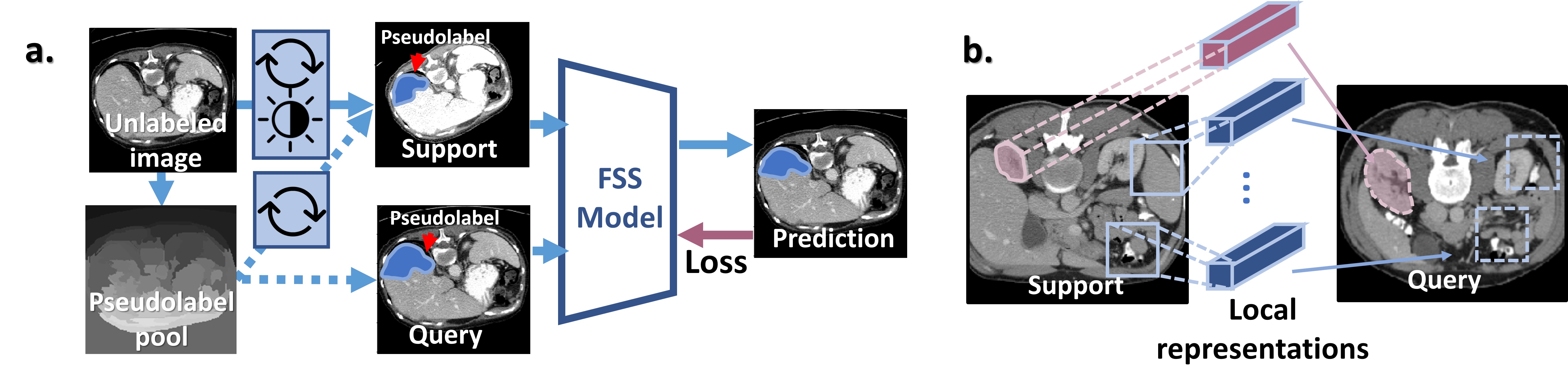}
\end{center}
\caption{Illustration of the SSL-ALPNet framework \citep{ouyang2020self}: (a) The self-supervision task involves segmenting the pseudolabel on the query image with reference to the support image, despite the applied transformations (shown in blue boxes). (b) The ALPNet method addresses the challenge of class imbalance by adaptively extracting multiple local representations of the large background class (in blue), each of which represents a distinct local background region (image from \citep{ouyang2020self}).}
\label{fig_FSL}
\end{figure*}

In contrast to natural images, there is a lack of extensive publicly available datasets for pre-training medical image segmentation models. Consequently, some self-supervised learning approaches have emerged in the domain of medical image few-shot segmentation, relying on unlabeled data. To this end, Ouyang et al. \citep{ouyang2020self, ouyang2022self} presented SSL-ALPNet, a self-supervised learning method based on superpixels. In this approach, for every unlabeled image, pseudolabels are created at the superpixel level. During each training iteration, a randomly chosen pseudolabel, along with the original image, is used as both the support and query. Random transformations are introduced between the support and query images. The primary objective of this self-supervision task is to segment the pseudolabel on the query image, using the support image as a reference, despite the applied transformations. Additionally, they incorporated an adaptive local prototype pooling module into prototypical networks to address the prevalent issue of foreground-background class imbalance in medical image segmentation, as shown in Figure~\ref{fig_FSL}. Hansen et al. \citep{hansen2022anomaly} expanded on this concept by extending the self-supervision task to supervoxels, effectively incorporating 3D information from image volumes. They introduced ADNet, a prototypical segmentation network inspired by anomaly detection, which avoids modeling the large and diverse background class with prototypes. Building on this work, their recent contribution, ADNet++ \citep{hansen2023adnet++}, presents a one-step multi-class medical image segmentation framework. The model notably enhances the current 3D FSS model for MRI and CT-based abdominal organ and cardiac segmentation.

In medical images, a significant imbalance exists between the foreground and background. Medical images typically feature a diverse background comprising numerous tissues and organs, whereas the foreground is typically uniform and occupies a relatively small area. Applying the same global operation, like masked average pooling, directly to both foreground and background, as is commonly done in the processing of natural images, can result in the loss of local information. To address this issue, recent studies in the field of prototypical FSS have introduced more adaptive prototype extraction modules to mitigate the impact of complex backgrounds. Specifically, these studies have incorporated additional priors, such as spatial location \citep{yu2021location}, and neighborhood correlations \citep{tang2021recurrent}, into the prototypes to preserve spatial and shape information.

\textbf{Meta-learning based approaches:} Meta-learning methods typically employ a meta-training approach, where they train a model on a sequence of few-shot tasks derived from the base classes during the training phase. The objective is to equip the pre-trained model with the capability to quickly adapt to entirely new tasks during the testing phase. One well-known representative of meta-learning methods is Model-Agnostic Meta-Learning (MAML) \citep{finn2017model}. MAML achieves this adaptation by pre-training the initial model parameters using second-order gradients, enabling the model to rapidly adapt to new tasks with only a limited number of gradient steps. Several studies in medical imaging employ MAML and its extensions for various few-shot learning tasks. These tasks include rare disease classification \citep{singh2021metamed}, classifying whole-genome doubling (WGD) across 17 cancer types using digitized histopathology slide images \citep{chao2021generalizing}, brain tumor segmentation \citep{achmamad2022few}, and more. Further, To address issues related to vanishing high-order meta-gradients in MAML, Khadka et al. \citep{khadka2022meta} utilize the Implicit Model Agnostic Meta-Learning (iMAML) optimization strategy \citep{rajeswaran2019meta} for few-shot lesion segmentation. In this approach, inner optimization focuses on computing weights using a CNN model, while an analytic solution is used to estimate the outer meta-gradients.

Zhao et al. \citep{zhao2022meta} suggest that meta-learning can enhance the generator's ability to \emph{learn to hallucinate} meaningful images, leading to improved segmentation models in few-shot unsupervised domain adaptation. In this regard, they introduce a meta-hallucinator to generate valuable samples, enhancing model adaptability on the target domain with limited source annotations. For diagnosis of glioblastoma multiforme progression, Song et al. \citep{song2022diagnosis} propose an interpretable structure-constrained graph neural network (ISGNN). The ISGNN used a meta-learning strategy for aggregating class-specific graph nodes to enhance classification performance on small-scale datasets while maintaining interpretability. Recently, Gao et al. \citep{gao2023discriminative} introduced a discriminative ensemble meta-learning approach for the diagnosis of rare fundus diseases. They introduced a co-regulation loss during the pre-training of the meta-learning backbone. Subsequently, ensemble-learning techniques were employed to improve performance, taking advantage of the hierarchical features within the backbone network. They explored three ensemble strategies: uniform averaging, majority voting, and stacking, to identify low-shot rare fundus diseases.

\textbf{Others:} One approach to accomplishing few-shot learning involves utilizing the encoder-decoder framework to explore the connection between a query and a support set. In this regard, Roy et al. \citep{roy2020squeeze} introduce the Squeeze and Excitep framework, which was the first implementation of a two-branch architecture for medical image few-shot segmentation. One branch, referred to as the conditioner arm, focuses on extracting foreground information from the support set. The other branch, known as the segmenter arm, engages with the conditioner arm through the spatial SE module to rectify the query feature. MprNet \citep{feng2021interactive}, as an enhancement of SENet, introduces a fusion module based on cosine similarity to facilitate information exchange between these two branches. Similarly, Kim et al. \citep{kim2021bidirectional} introduce a U-Net like network tailored for segmentation tasks. This network is designed to predict segmentation by capturing the relationship between 2D slices from the support data and a query image. It incorporates a bidirectional gated recurrent unit (GRU) to learn the coherence of encoded features among adjacent slices. Recently, Feng et al. \citep{feng2023learning} employ a hybrid method in which they introduce a segmenter built upon the encoder-decoder architecture. They incorporate spatial and prototypical priors as extra sources of supervisory information. Experimental results in multi-modalities and multi-organs segmentation showcase that the method they propose significantly surpasses previous state-of-the-art techniques. FSL is typically trained using the episode training method. Zhu et al. \citep{zhu2020FSL} introduced a Query-Relative (QR) loss, which is more effective when combined with the episode training approach than the Cross-Entropy loss for FSL.

\subsection{Transfer learning}  \label{TL}

To address the challenge of limited training data and enhance model performance, another common approach known as transfer learning is frequently employed. In this scenario, the goal is to harness knowledge acquired from similar learning tasks. The supervised transfer learning technique \citep{tajbakhsh2016convolutional} has proven valuable in addressing various issues in medical image analysis. These approaches typically involve initial pre-training of standard architectures like ResNet \citep{he2016deep} or VGG \citep{simonyan2014very} on a source domain containing abundant data, such as natural images from sources like ImageNet \citep{deng2009imagenet} or medical images. Subsequently, these pre-trained models are transferred to the target domain and fine-tuned using a significantly smaller set of training examples. Tajbakhsh et al. \citep{tajbakhsh2016convolutional} demonstrated that pre-trained CNNs, when appropriately fine-tuned, achieved performance levels at least comparable to CNNs trained entirely from the beginning. This has established transfer learning as a fundamental technique for image classification tasks across diverse modalities, spanning CT \citep{shin2016deep}, mammography \citep{huynh2016digital} MRI \citep{yuan2019prostate}, X-ray \citep{minaee2020deep}, and more.

Aside from classification, the use of transfer learning in addressing different medical image challenges, such as image segmentation and localization, has been limited \citep{tajbakhsh2020embracing, kora2022transfer, 9363892}. This trend can be attributed, in part, to the inherent 3D characteristics of medical images, which present challenges when adapting 2D models trained on natural images. Additionally, it is influenced by the effective performance of shallower segmentation networks in medical imaging, which may not gain significant advantages from fine-tuning in contrast to deep models. However, certain studies have attempted transfer learning for image segmentation. For instance, Ma et al. \citep{ma2019neural} performed fine-tuning on an autoencoder that was originally pre-trained for image segmentation tasks in natural images. Similarly, other researchers like Qin et al. \citep{qin2019transfer} utilized an encoder pre-trained for the task of image classification in natural images and added a randomly initialized decoder to it to address the task of prostate MRI segmentation. Liu et al. \citep{liu2021covid} perform a two-stage transfer learning framework for segmenting COVID-19 lung infections from CT images. Nguyen et al. \citep{nguyen2022tatl} perform task agnostic transfer learning for skin attribute detection.

Some studies have attempted to transfer knowledge from 2D models pre-trained on natural images to models intended for 3D medical applications. For example, Yu et al. \citep{yu2018recurrent} adapted models trained on natural scene videos, treating the third dimension of medical scans as a temporal axis. However, this approach may not effectively capture the 3D context of medical scans. In contrast, Liu et al. \citep{liu20183d} proposed a method to transform a 2D model into a 3D network by expanding 2D convolution filters into 3D separable anisotropic filters. Recently, Messaoudi et al. \citep{messaoudi2023cross}  introduced two transfer learning strategies. Firstly, they introduced weight transfer learning, an effective method for leveraging the weights of a pre-trained 2D classifier network by incorporating it into a network of the same or higher dimension. The second approach they proposed is dimensional transfer learning, which relies on extrapolating 3D weights from a pre-trained 2D network. Empirical evidence demonstrates that their methods outperform current state-of-the-art techniques.


\begin{table*}[t]
    \centering
    \scriptsize
    \begin{center}
    \caption{Overview of recent methods in \emph{Only Limited Supervision} category.}
    \label{table:only limited sup}
   \resizebox{\textwidth}{!}{
    \begin{tabular}{p{0.8cm}p{3.5cm}p{4cm}p{3cm}p{5.5cm}}
    \toprule
   Reference & \multicolumn{1}{l}{Task} & \multicolumn{1}{l}{Algorithm Design} & \multicolumn{1}{l}{Dataset} & \multicolumn{1}{l}{Result} \\ \toprule

\citep{yang2021tumorcp} & kidney tumor segmentation & Data Augmentation & KiTS19 & Mean Dice: 77.44 \\ \midrule

\citep{zhao2019data} & Brain tumour segmentation & Learning-based technique for data augmentation &   T1-weighted MRI brain scans described in \citep{balakrishnan2018unsupervised} & Dice score: 0.815         \\ \midrule

\citep{ali2020additive} & Clinical endoscopy image classification & Angular margin metric to ProtoNet & miniEndoGI classification datase & 5-way: 1-shot: 58.76 $\pm$ 1.64, 5-shot: 66.72 $\pm$ 1.35; 3-way: 1-shot: 75.06 $\pm$ 1.87, 5-shot: 81.20 $\pm$ 1.72; 2-way: 1-shot: 85.60 $\pm$ 2.21, 5-shot: 90.60 $\pm$ 1.70   \\ \midrule

\citep{ouyang2022self} & Cardiac and organ segmentation & Prototype-based network + Self-supervision          & Abdominal CT; Abdominal T2-SPIR MRI; Cardiac bSSFP MRI & Abdominal CT: 1-shot: Dice: 67.62, 5-shot: Dice: 75.91; Abdominal MRI: 1-shot: Dice: 76.81, 5-shot: Dice: 80.16; Cardiac: 1-shot: Dice: 77.94, 5-shot: Dice: 81.66   \\ \midrule

\citep{hansen2022anomaly} & Abdomen and cardiac segmentation & Anomaly detection-inspired FS + Self-supervision & MS-CMRSeg; CHAOS & CHAOS: Mean DSC: 72.41; MS-CMRSeg: Mean DSC: 69.62 \\ \midrule

\citep{hansen2023adnet++} & Abdomen and cardiac segmentation & Prototype-based network + Self-supervision & MS-CMRSeg; CHAOS; BTCV & CHAOS: Mean 95 HD: 12.5; Mean DSC: 80.99; BTCV: Mean 95 HD: 23.60; Mean DSC: 60.94; MS-CMRSeg: Mean 95 HD: 6.08; Mean DSC: 69.68 \\ \midrule

\citep{tang2021recurrent} & Abdomen segmentation & Context relation encoder + Recurrent mask refine-
ment module + Prototypical network & ABD-110; BTCV; CHAOS  & ABD-110: DSC: 81.91; BTCV: DSC: 72.48; CHAOS: DSC: 79.26  \\ \midrule

\citep{zhu2020FSL} & Skin Disease Classification & FSL with Query-Relative loss & Dermatology images & 5-way 1-shot: ACC\%: 52.41 Precision\%: 53.21 F1\%: 49.52;  5-way 5-shot: ACC\%: 71.99 Precision\%: 74.23 F1\%: 70.30  \\ \midrule

\citep{achmamad2022few} & Brain tumor segmentation & Meta-learning & BraTS2021 & 1-way 1-shot: DSC($\mu$ ± std)\% : 0.57 ± 0.19; 1-way 1-shot: DSC($\mu$ ± std)\% : 0.63 ± 0.16; 1-way 1-shot: DSC($\mu$ ± std)\% : 0.65 ± 0.17  \\ \midrule

\citep{chao2021generalizing} & Classification of whole-genome doubling across 17 cancer types & Model-Agnostic Meta-Learning & TCGA & AUC (average ± 1 standard deviation): 0.6944 ± 0.0773 \\ \midrule

\citep{song2022diagnosis} & Tumour classification & Interpretable structure-
constrained graph neural network & Private dataset: 150 patients &  ACC: 83.3, AUC: 81.9, SEN: 67.2 and SPE: 85.7  \\ \midrule

\citep{zhao2022meta} & Cardiac segmentation  & Gradient-based meta-hallucination learning & MM-WHS 2017 & 4-shots: Average Dice: 75.6, Average ASD: 4.8; 1-shots: Average Dice: 51.8, Average ASD: 14.1  \\ \midrule

\citep{gao2023discriminative} & Rare fundus diseases diagnosis & Meta learning + Co-regularization loss  + Ensemble-learning & FundusData-FS \citep{gao2023discriminative} & Accuracy(\%): 2-way: 1-shot: 71.53, 3-shot: 78.20, 5-shot: 81.47; Accuracy(\%): 3-way: 1-shot: 56.69, 3-shot: 62.62, 5-shot: 66.78; Accuracy(\%): 4-way: 1-shot: 48.17, 3-shot: 56.65, 5-shot: 58.60 \\ \midrule

\citep{roy2020squeeze} & Multi-organ segmentation & Squeeze and excitep framework & Visceral dataset & Mean Dice score on validation set: 0.567 \\ \midrule

\citep{feng2023learning} & Multi-organ segmentation & Encoder-decoder architecture + Spatial and prototypical priors & CHAOS & Left atrium (LA): 1-shot: Mean DSC: 86.37; 5-shot: Mean DSC: 88.02  Left ventricle (LV): 1-shot: Mean DSC: 87.06; 5-shot: Mean DSC: 87.87  \\

\bottomrule
\end{tabular}
}
\end{center}
\end{table*}

\section{Future research scope}  \label{Future}

\subsection{Continual/lifelong learning}

In healthcare, most intelligent diagnosis systems are limited in their scope, often capable of diagnosing only a few diseases. Expanding their capabilities after deployment is challenging, preventing them from achieving the breadth of diagnoses that medical specialists can. Collecting data for all diseases poses significant challenges due to privacy concerns and data sharing constraints. Consequently, training a single system to diagnose all diseases simultaneously is impractical. One potential solution is to make the system with the ability for continual learning. This would allow the system to progressively acquire the capacity to diagnose more diseases over time without needing extensive new data for previously learned diseases. Continual learning, also known as lifelong learning or incremental learning, is a learning paradigm in which a model learns and adapts to new information and tasks over time without forgetting previously acquired knowledge \citep{nguyen2018variational}. Unlike traditional deep learning approaches that assume a fixed dataset and task, continual learning addresses scenarios where data arrives continuously, and the nature of tasks can evolve over time, including the possibility of introducing new classes.

Despite its potential, there has been a limited exploration of continual learning in medical contexts. Current research has primarily focused on this paradigm in specific areas such as image segmentation \citep{zheng2021continual, zhang2021comprehensive}, disease classification \citep{li2020continual, derakhshani2022lifelonger, yang2023continual, bayasi2021culprit}, and domain adaptation \citep{lenga2020continual, chen2023generative}. Moreover, there is currently no unified framework in continual learning capable of accommodating diverse types of annotations in medical applications. We look forward to the development of an integrated framework for continual learning that can encompass the various settings and challenges highlighted in this paper. Such a framework would significantly advance the application of continual learning in the medical field, providing a more comprehensive approach to managing evolving datasets and diverse annotations.

\subsection{Incorporating domain knowledge}

Incorporating additional information beyond the existing medical datasets has emerged as a more promising strategy to tackle the issue of limited-sized medical datasets. Within this context, domain knowledge plays a vital role in guiding the development of effective deep learning algorithms for MIA. While many models used in medical vision are adapted from those designed for natural images, it's worth noting that medical images typically present more complex challenges, such as high inter-class similarity, a scarcity of labeled data, and label noise. When applied effectively, domain knowledge can mitigate these challenges with reduced time and computational requirements. Integrating domain knowledge into deep learning algorithms can be achieved by leveraging anatomical details from MRI and CT images \citep{zhou2019models}, exploiting multi-instance data from the same patient \citep{azizi2021big}, incorporating patient metadata \citep{vu2021medaug}, utilizing radiomic features, and considering textual reports that accompany the images \citep{zhang2022contrastive}. 

While the utilization of medical domain knowledge in deep learning models is a prevalent practice, it is not without its challenges. These challenges involve the selection, representation, and integration methods for medical domain knowledge \citep{xie2021survey}. Identifying such knowledge is a complex task primarily because the experiences of medical professionals tend to be subjective and ambiguous. It's often difficult for medical practitioners to provide precise and objective descriptions of the experiences they draw upon to complete specific tasks. Currently, the identification of medical domain knowledge relies on manual processes, and there is no existing method for automatically identifying medical domain knowledge within a given field. Medical professionals typically draw from various types of domain knowledge simultaneously. Furthermore, Most existing approaches, however, incorporate only a single type or a few types of medical domain knowledge, often from the same modality. Consequently, simultaneously integrating multiple forms of medical domain knowledge has the potential to provide more robust support for deep learning models across various medical applications.

\subsection{Label-efficient learning by vision transformers}

Current label-efficient segmentation techniques primarily rely on convolutional neural networks (CNNs). However, there has been a recent transformation in computer vision, driven by the introduction of the transformer module \citep{vaswani2017attention}. This innovation has given rise to vision transformers (ViT) \citep{dosovitskiy2020image} and their adaptations \citep{fan2021multiscale}, leading to significant advancements in numerous medical applications, including segmentation, detection, and classification tasks. Transformer-based models can achieve higher performance when trained on extensive datasets, but their effectiveness diminishes when data or annotations are scarce. To overcome this challenge, self-supervised transformers offer a promising solution. By utilizing unlabeled data and employing proxy tasks like contrastive learning and reconstruction, these transformers can enhance their representation learning capabilities \citep{li2023transforming}. For instance, the Self-Supervised SwinUNETR \citep{tang2022self} and unified pre-training \citep{xie2021unified} frameworks in the medical domain demonstrate that training with large-scale unlabeled 2D or 3D images is advantageous for fine-tuning models with smaller datasets. However, it's worth noting that the utilization of pre-training can be computationally demanding. Future research directions may aim to simplify and assess the efficiency of the pre-training framework, especially regarding its applicability to smaller datasets.

\subsection{Flexible target model design}
To develop better architectures for data-efficient deep learning architectures, one promising avenue is the field of automated architectural engineering. Presently, the architectures predominantly in use are crafted by human experts through iterative processes that are susceptible to errors. To circumvent the need for manual design, researchers have put forward the concept of automating architectural engineering, with one relevant domain being neural architecture search (NAS), introduced by Zoph and Le \citep{zoph2016neural}. However, it's essential to note that the majority of NAS investigations have been concentrated on image classification tasks \citep{elsken2019neural}. Regrettably, this focus has yet to yield truly transformative models capable of instigating fundamental shifts \citep{vision}. Nevertheless, the exploration of NAS for data-efficient learning in MIA remains a promising avenue.

\subsection{Federated learning}

Modern healthcare systems collect significant amounts of medical data, yet the complete utilization of this data by deep learning is hindered. This limitation stems from the data being isolated within silos and privacy concerns that limit data access \citep{rajpurkar2022ai}. To tackle this challenge, federated learning (FL) emerges as a learning paradigm seeking to address the problem of data governance and privacy. It achieves this by training algorithms collaboratively without the need to exchange the data itself \citep{larson2020ethics}.
FL preserves data privacy while collectively improving model efficiency, making it a valuable tool for data-efficient deep learning in MIA. FL has produced valuable outcomes in the MIA domain \citep{rieke2020future, antunes2022federated, sheller2020federated}. Nevertheless, existing FL algorithms are predominantly trained using supervised methods. When implementing FL in real-world MIA situations, a critical issue arises: \emph{label scarcity} can occur in local healthcare datasets. Different medical centers may have varying degrees of missing labels, or the label granularity may differ. A potential avenue for research is the development of label-efficient federated learning techniques to tackle this challenge.

\subsection{Data-efficient learning with text supervision}

Text supervision involves using textual descriptions or labels as additional sources of information during training. This can include utilizing clinical reports, medical terminology, or textual metadata associated with images or patient records. By incorporating text supervision, MIA models can learn to associate medical text with visual patterns in images, facilitating improved generalization and a deeper understanding of medical data. Some studies have explored this approach \citep{zhang2022contrastive, li2023lvit}. We encourage future investigations to further explore and expand upon this area of study.

\section{Conclusion \label{conclusion}}

The challenge of acquiring high-quality labels remains a significant hurdle for supervised learning in Medical Image Analysis (MIA). This challenge has fueled interest in alternative approaches that enhance labeling efficiency to reduce the labeled data requirement. In recent years, extensive research efforts have been dedicated to advancing data-efficient learning within the realm of medical images, resulting in the development of numerous techniques applicable across diverse application domains. In this paper, we have provided a comprehensive review that explores the recent progress in data-efficient deep learning for MIA. Specifically, we conducted a thorough examination of deep learning-based data-efficient methodologies and categorized them into five distinct groups. These categorizations are rooted in the varying degrees of supervision they depend on, covering a spectrum from scenarios with no supervision to those involving inexact, incomplete, inaccurate, and only limited supervision. Finally, we highlight several potential future directions for research and development in this area. We hope that this survey serves as a valuable resource, offering insights into the current state of data-efficient deep learning in medical imaging and inspiring further progress in this domain.

\bibliographystyle{elsarticle-num}
\bibliography{refs}

\end{document}

%% file: defs.tex
\def\x{{\bf x}}

\def\0{{\bf 0}}
\def\1{{\bf 1}}

